\documentclass[sigconf]{acmart}

\pdfoutput=1

\usepackage[detect-all=true,mode=text,binary-units=true]{siunitx}
\usepackage{colortbl}
\usepackage{enumitem}
\usepackage{booktabs}
\usepackage{multirow}
\usepackage{multicol}
\usepackage{subcaption}
\usepackage{arydshln}
\usepackage[pdftex,outline]{contour}

\usepackage{tikz}
\usetikzlibrary{calc,tikzmark}

\AtBeginDocument{%
  \providecommand\BibTeX{{%
    \normalfont B\kern-0.5em{\scshape i\kern-0.25em b}\kern-0.8em\TeX}}}

\copyrightyear{2022}
\acmYear{2022}
\setcopyright{acmlicensed}
\acmConference[ASIA CCS '22] {Proceedings of the 2022 ACM Asia Conference on Computer and Communications Security}{May 30--June 3, 2022}{Nagasaki, Japan.}
\acmBooktitle{Proceedings of the 2022 ACM Asia Conference on Computer and Communications Security (ASIA CCS '22), May 30--June 3, 2022, Nagasaki, Japan}
\acmPrice{15.00}
\acmDOI{10.1145/3488932.3497762}
\acmISBN{978-1-4503-9140-5/22/05}

\makeatletter
\if@ACM@anonymous
\else
  \g@addto@macro\@subtitlenotes{\global\let\authors=\cleanauthors}
\fi
\makeatother

\DeclareSIUnit\permille{\text{\textperthousand}}
\definecolor{perfecthost}{HTML}{481568}
\definecolor{sensiblehost}{HTML}{3E8A8D}
\definecolor{insecurehost}{HTML}{DCE319}

\definecolor{limegreen}{HTML}{32CD32}

\newcommand{\DefineRemark}[2]{%
  \expandafter\newcommand\csname rmk-#1\endcsname{#2}%
}
\newcommand{\Remark}[1]{\csname rmk-#1\endcsname}
\DefineRemark{102102siemensdate}{05-11}
\DefineRemark{102102siemenspctconnsuccess}{0.9}
\DefineRemark{102102siemenspcttlsvalidoftransport}{0.0}
\DefineRemark{102102siemenspctvalidtls}{0.0}
\DefineRemark{102102siemenstotalas}{806}
\DefineRemark{102102siemenstotalcertcns}{0}
\DefineRemark{102102siemenstotalcerts}{0}
\DefineRemark{102102siemenstotalconnerr}{2764639}
\DefineRemark{102102siemenstotalconnsuccess}{25957}
\DefineRemark{102102siemenstotalhosts}{2790596}
\DefineRemark{102102siemenstotalsuccess}{6507}
\DefineRemark{102102siemenstotalsystemid}{1160}
\DefineRemark{102102siemenstotaltls}{0}
\DefineRemark{102102siemenstotaltlsauth}{0}
\DefineRemark{102102siemenstotaltlssuccess}{0}
\DefineRemark{102102tlssiemensdate}{05-11}
\DefineRemark{102102tlssiemenspctconnsuccess}{0.9}
\DefineRemark{102102tlssiemenspcttlsvalidoftransport}{24}
\DefineRemark{102102tlssiemenspctvalidtls}{0.0}
\DefineRemark{102102tlssiemenstotalas}{0}
\DefineRemark{102102tlssiemenstotalcertcns}{0}
\DefineRemark{102102tlssiemenstotalcerts}{0}
\DefineRemark{102102tlssiemenstotalconnerr}{2764639}
\DefineRemark{102102tlssiemenstotalconnsuccess}{25957}
\DefineRemark{102102tlssiemenstotalhosts}{2790596}
\DefineRemark{102102tlssiemenstotalsuccess}{0}
\DefineRemark{102102tlssiemenstotalsystemid}{0}
\DefineRemark{102102tlssiemenstotaltls}{6125}
\DefineRemark{102102tlssiemenstotaltlsauth}{6125}
\DefineRemark{102102tlssiemenstotaltlssuccess}{6060}
\DefineRemark{18831883mqttdate}{07-06}
\DefineRemark{18831883mqttpctconnsuccess}{13}
\DefineRemark{18831883mqttpcttlsvalidoftransport}{0.0}
\DefineRemark{18831883mqttpctvalidtls}{12}
\DefineRemark{18831883mqtttotalas}{5285}
\DefineRemark{18831883mqtttotalcertcns}{0}
\DefineRemark{18831883mqtttotalcerts}{0}
\DefineRemark{18831883mqtttotalconnerr}{3189989}
\DefineRemark{18831883mqtttotalconnsuccess}{488626}
\DefineRemark{18831883mqtttotalhosts}{3678615}
\DefineRemark{18831883mqtttotalsuccess}{291344}
\DefineRemark{18831883mqtttotalsystemid}{0}
\DefineRemark{18831883mqtttotaltls}{0}
\DefineRemark{18831883mqtttotaltlsauth}{0}
\DefineRemark{18831883mqtttotaltlssuccess}{0}
\DefineRemark{18831883tlsmqttdate}{07-06}
\DefineRemark{18831883tlsmqttpctconnsuccess}{13}
\DefineRemark{18831883tlsmqttpcttlsvalidoftransport}{35}
\DefineRemark{18831883tlsmqttpctvalidtls}{12}
\DefineRemark{18831883tlsmqtttotalas}{283}
\DefineRemark{18831883tlsmqtttotalcertcns}{2625}
\DefineRemark{18831883tlsmqtttotalcerts}{3256}
\DefineRemark{18831883tlsmqtttotalconnerr}{3189989}
\DefineRemark{18831883tlsmqtttotalconnsuccess}{488626}
\DefineRemark{18831883tlsmqtttotalhosts}{3678615}
\DefineRemark{18831883tlsmqtttotalsuccess}{3420}
\DefineRemark{18831883tlsmqtttotalsystemid}{0}
\DefineRemark{18831883tlsmqtttotaltls}{171912}
\DefineRemark{18831883tlsmqtttotaltlsauth}{171905}
\DefineRemark{18831883tlsmqtttotaltlssuccess}{27849}
\DefineRemark{1999819998iec104date}{06-07}
\DefineRemark{1999819998iec104pctconnsuccess}{4.0}
\DefineRemark{1999819998iec104pcttlsvalidoftransport}{0.0}
\DefineRemark{1999819998iec104pctvalidtls}{0.0}
\DefineRemark{1999819998iec104totalas}{0}
\DefineRemark{1999819998iec104totalcertcns}{0}
\DefineRemark{1999819998iec104totalcerts}{0}
\DefineRemark{1999819998iec104totalconnerr}{3316445}
\DefineRemark{1999819998iec104totalconnsuccess}{139390}
\DefineRemark{1999819998iec104totalhosts}{3455835}
\DefineRemark{1999819998iec104totalsuccess}{0}
\DefineRemark{1999819998iec104totalsystemid}{0}
\DefineRemark{1999819998iec104totaltls}{0}
\DefineRemark{1999819998iec104totaltlsauth}{0}
\DefineRemark{1999819998iec104totaltlssuccess}{0}
\DefineRemark{1999819998tlsiec104date}{06-07}
\DefineRemark{1999819998tlsiec104pctconnsuccess}{4.0}
\DefineRemark{1999819998tlsiec104pcttlsvalidoftransport}{6.5}
\DefineRemark{1999819998tlsiec104pctvalidtls}{0.0}
\DefineRemark{1999819998tlsiec104totalas}{1}
\DefineRemark{1999819998tlsiec104totalcertcns}{1}
\DefineRemark{1999819998tlsiec104totalcerts}{1}
\DefineRemark{1999819998tlsiec104totalconnerr}{3316445}
\DefineRemark{1999819998tlsiec104totalconnsuccess}{139390}
\DefineRemark{1999819998tlsiec104totalhosts}{3455835}
\DefineRemark{1999819998tlsiec104totalsuccess}{1}
\DefineRemark{1999819998tlsiec104totalsystemid}{0}
\DefineRemark{1999819998tlsiec104totaltls}{8999}
\DefineRemark{1999819998tlsiec104totaltlsauth}{8994}
\DefineRemark{1999819998tlsiec104totaltlssuccess}{8920}
\DefineRemark{1999919999dnp3date}{07-18}
\DefineRemark{1999919999dnp3pctconnsuccess}{7.9}
\DefineRemark{1999919999dnp3pcttlsvalidoftransport}{0.0}
\DefineRemark{1999919999dnp3pctvalidtls}{1.0}
\DefineRemark{1999919999dnp3totalas}{0}
\DefineRemark{1999919999dnp3totalcertcns}{0}
\DefineRemark{1999919999dnp3totalcerts}{0}
\DefineRemark{1999919999dnp3totalconnerr}{2838073}
\DefineRemark{1999919999dnp3totalconnsuccess}{242876}
\DefineRemark{1999919999dnp3totalhosts}{3080949}
\DefineRemark{1999919999dnp3totalsuccess}{0}
\DefineRemark{1999919999dnp3totalsystemid}{0}
\DefineRemark{1999919999dnp3totaltls}{0}
\DefineRemark{1999919999dnp3totaltlsauth}{0}
\DefineRemark{1999919999dnp3totaltlssuccess}{0}
\DefineRemark{1999919999tlsdnp3date}{07-18}
\DefineRemark{1999919999tlsdnp3pctconnsuccess}{7.9}
\DefineRemark{1999919999tlsdnp3pcttlsvalidoftransport}{17}
\DefineRemark{1999919999tlsdnp3pctvalidtls}{1.0}
\DefineRemark{1999919999tlsdnp3totalas}{2}
\DefineRemark{1999919999tlsdnp3totalcertcns}{2}
\DefineRemark{1999919999tlsdnp3totalcerts}{2}
\DefineRemark{1999919999tlsdnp3totalconnerr}{2838073}
\DefineRemark{1999919999tlsdnp3totalconnsuccess}{242876}
\DefineRemark{1999919999tlsdnp3totalhosts}{3080949}
\DefineRemark{1999919999tlsdnp3totalsuccess}{2}
\DefineRemark{1999919999tlsdnp3totalsystemid}{0}
\DefineRemark{1999919999tlsdnp3totaltls}{41415}
\DefineRemark{1999919999tlsdnp3totaltlsauth}{41408}
\DefineRemark{1999919999tlsdnp3totaltlssuccess}{536}
\DefineRemark{2000020000dnp3date}{07-19}
\DefineRemark{2000020000dnp3pctconnsuccess}{5.9}
\DefineRemark{2000020000dnp3pcttlsvalidoftransport}{0.0}
\DefineRemark{2000020000dnp3pctvalidtls}{1.0}
\DefineRemark{2000020000dnp3totalas}{91}
\DefineRemark{2000020000dnp3totalcertcns}{0}
\DefineRemark{2000020000dnp3totalcerts}{0}
\DefineRemark{2000020000dnp3totalconnerr}{3753870}
\DefineRemark{2000020000dnp3totalconnsuccess}{235653}
\DefineRemark{2000020000dnp3totalhosts}{3989523}
\DefineRemark{2000020000dnp3totalsuccess}{503}
\DefineRemark{2000020000dnp3totalsystemid}{0}
\DefineRemark{2000020000dnp3totaltls}{0}
\DefineRemark{2000020000dnp3totaltlsauth}{0}
\DefineRemark{2000020000dnp3totaltlssuccess}{0}
\DefineRemark{2000020000tlsdnp3date}{07-19}
\DefineRemark{2000020000tlsdnp3pctconnsuccess}{5.9}
\DefineRemark{2000020000tlsdnp3pcttlsvalidoftransport}{18}
\DefineRemark{2000020000tlsdnp3pctvalidtls}{1.0}
\DefineRemark{2000020000tlsdnp3totalas}{2}
\DefineRemark{2000020000tlsdnp3totalcertcns}{2}
\DefineRemark{2000020000tlsdnp3totalcerts}{2}
\DefineRemark{2000020000tlsdnp3totalconnerr}{3753870}
\DefineRemark{2000020000tlsdnp3totalconnsuccess}{235653}
\DefineRemark{2000020000tlsdnp3totalhosts}{3989523}
\DefineRemark{2000020000tlsdnp3totalsuccess}{3}
\DefineRemark{2000020000tlsdnp3totalsystemid}{0}
\DefineRemark{2000020000tlsdnp3totaltls}{43134}
\DefineRemark{2000020000tlsdnp3totaltlsauth}{41697}
\DefineRemark{2000020000tlsdnp3totaltlssuccess}{1836}
\DefineRemark{22212221tlsenipdate}{04-26}
\DefineRemark{22212221tlsenippctconnsuccess}{0.2}
\DefineRemark{22212221tlsenippcttlsvalidoftransport}{100}
\DefineRemark{22212221tlsenippctvalidtls}{6.4}
\DefineRemark{22212221tlseniptotalas}{56}
\DefineRemark{22212221tlseniptotalcertcns}{1}
\DefineRemark{22212221tlseniptotalcerts}{94}
\DefineRemark{22212221tlseniptotalconnerr}{4278801}
\DefineRemark{22212221tlseniptotalconnsuccess}{6904}
\DefineRemark{22212221tlseniptotalhosts}{4285705}
\DefineRemark{22212221tlseniptotalsuccess}{111}
\DefineRemark{22212221tlseniptotalsystemid}{94}
\DefineRemark{22212221tlseniptotaltls}{6904}
\DefineRemark{22212221tlseniptotaltlsauth}{6904}
\DefineRemark{22212221tlseniptotaltlssuccess}{6904}
\DefineRemark{24042404iec104date}{06-08}
\DefineRemark{24042404iec104pctconnsuccess}{0.5}
\DefineRemark{24042404iec104pcttlsvalidoftransport}{0.0}
\DefineRemark{24042404iec104pctvalidtls}{0.0}
\DefineRemark{24042404iec104totalas}{383}
\DefineRemark{24042404iec104totalcertcns}{0}
\DefineRemark{24042404iec104totalcerts}{0}
\DefineRemark{24042404iec104totalconnerr}{3620606}
\DefineRemark{24042404iec104totalconnsuccess}{18760}
\DefineRemark{24042404iec104totalhosts}{3639366}
\DefineRemark{24042404iec104totalsuccess}{4484}
\DefineRemark{24042404iec104totalsystemid}{0}
\DefineRemark{24042404iec104totaltls}{0}
\DefineRemark{24042404iec104totaltlsauth}{0}
\DefineRemark{24042404iec104totaltlssuccess}{0}
\DefineRemark{24042404tlsiec104date}{06-08}
\DefineRemark{24042404tlsiec104pctconnsuccess}{0.5}
\DefineRemark{24042404tlsiec104pcttlsvalidoftransport}{32}
\DefineRemark{24042404tlsiec104pctvalidtls}{0.0}
\DefineRemark{24042404tlsiec104totalas}{0}
\DefineRemark{24042404tlsiec104totalcertcns}{0}
\DefineRemark{24042404tlsiec104totalcerts}{0}
\DefineRemark{24042404tlsiec104totalconnerr}{3620606}
\DefineRemark{24042404tlsiec104totalconnsuccess}{18760}
\DefineRemark{24042404tlsiec104totalhosts}{3639366}
\DefineRemark{24042404tlsiec104totalsuccess}{0}
\DefineRemark{24042404tlsiec104totalsystemid}{0}
\DefineRemark{24042404tlsiec104totaltls}{5958}
\DefineRemark{24042404tlsiec104totaltlsauth}{5954}
\DefineRemark{24042404tlsiec104totaltlssuccess}{5878}
\DefineRemark{30111911foxdate}{05-17}
\DefineRemark{30111911foxpctconnsuccess}{0.7}
\DefineRemark{30111911foxpcttlsvalidoftransport}{0.0}
\DefineRemark{30111911foxpctvalidtls}{0.5}
\DefineRemark{30111911foxtotalas}{1494}
\DefineRemark{30111911foxtotalcertcns}{0}
\DefineRemark{30111911foxtotalcerts}{0}
\DefineRemark{30111911foxtotalconnerr}{4347639}
\DefineRemark{30111911foxtotalconnsuccess}{32723}
\DefineRemark{30111911foxtotalhosts}{4380362}
\DefineRemark{30111911foxtotalsuccess}{21995}
\DefineRemark{30111911foxtotalsystemid}{12600}
\DefineRemark{30111911foxtotaltls}{0}
\DefineRemark{30111911foxtotaltlsauth}{0}
\DefineRemark{30111911foxtotaltlssuccess}{0}
\DefineRemark{30111911tlsfoxdate}{05-17}
\DefineRemark{30111911tlsfoxpctconnsuccess}{0.7}
\DefineRemark{30111911tlsfoxpcttlsvalidoftransport}{10}
\DefineRemark{30111911tlsfoxpctvalidtls}{0.5}
\DefineRemark{30111911tlsfoxtotalas}{0}
\DefineRemark{30111911tlsfoxtotalcertcns}{0}
\DefineRemark{30111911tlsfoxtotalcerts}{0}
\DefineRemark{30111911tlsfoxtotalconnerr}{4347639}
\DefineRemark{30111911tlsfoxtotalconnsuccess}{32723}
\DefineRemark{30111911tlsfoxtotalhosts}{4380362}
\DefineRemark{30111911tlsfoxtotalsuccess}{0}
\DefineRemark{30111911tlsfoxtotalsystemid}{0}
\DefineRemark{30111911tlsfoxtotaltls}{3404}
\DefineRemark{30111911tlsfoxtotaltlsauth}{3388}
\DefineRemark{30111911tlsfoxtotaltlssuccess}{3394}
\DefineRemark{30113011httpdate}{05-17}
\DefineRemark{30113011httppctconnsuccess}{4.3}
\DefineRemark{30113011httppcttlsvalidoftransport}{0.0}
\DefineRemark{30113011httppctvalidtls}{0.3}
\DefineRemark{30113011httptotalas}{1558}
\DefineRemark{30113011httptotalcertcns}{0}
\DefineRemark{30113011httptotalcerts}{0}
\DefineRemark{30113011httptotalconnerr}{4190742}
\DefineRemark{30113011httptotalconnsuccess}{189620}
\DefineRemark{30113011httptotalhosts}{4380362}
\DefineRemark{30113011httptotalsuccess}{23055}
\DefineRemark{30113011httptotalsystemid}{21722}
\DefineRemark{30113011httptotaltls}{0}
\DefineRemark{30113011httptotaltlsauth}{0}
\DefineRemark{30113011httptotaltlssuccess}{0}
\DefineRemark{30114911foxdate}{05-17}
\DefineRemark{30114911foxpctconnsuccess}{0.6}
\DefineRemark{30114911foxpcttlsvalidoftransport}{0.0}
\DefineRemark{30114911foxpctvalidtls}{0.5}
\DefineRemark{30114911foxtotalas}{0}
\DefineRemark{30114911foxtotalcertcns}{0}
\DefineRemark{30114911foxtotalcerts}{0}
\DefineRemark{30114911foxtotalconnerr}{4352764}
\DefineRemark{30114911foxtotalconnsuccess}{27598}
\DefineRemark{30114911foxtotalhosts}{4380362}
\DefineRemark{30114911foxtotalsuccess}{0}
\DefineRemark{30114911foxtotalsystemid}{0}
\DefineRemark{30114911foxtotaltls}{0}
\DefineRemark{30114911foxtotaltlsauth}{0}
\DefineRemark{30114911foxtotaltlssuccess}{0}
\DefineRemark{30114911tlsfoxdate}{05-17}
\DefineRemark{30114911tlsfoxpctconnsuccess}{0.6}
\DefineRemark{30114911tlsfoxpcttlsvalidoftransport}{43}
\DefineRemark{30114911tlsfoxpctvalidtls}{0.5}
\DefineRemark{30114911tlsfoxtotalas}{18}
\DefineRemark{30114911tlsfoxtotalcertcns}{20}
\DefineRemark{30114911tlsfoxtotalcerts}{22}
\DefineRemark{30114911tlsfoxtotalconnerr}{4352764}
\DefineRemark{30114911tlsfoxtotalconnsuccess}{27598}
\DefineRemark{30114911tlsfoxtotalhosts}{4380362}
\DefineRemark{30114911tlsfoxtotalsuccess}{26}
\DefineRemark{30114911tlsfoxtotalsystemid}{7}
\DefineRemark{30114911tlsfoxtotaltls}{11972}
\DefineRemark{30114911tlsfoxtotaltlsauth}{11971}
\DefineRemark{30114911tlsfoxtotaltlssuccess}{5453}
\DefineRemark{30115011httpdate}{05-17}
\DefineRemark{30115011httppctconnsuccess}{3.7}
\DefineRemark{30115011httppcttlsvalidoftransport}{0.0}
\DefineRemark{30115011httppctvalidtls}{0.3}
\DefineRemark{30115011httptotalas}{0}
\DefineRemark{30115011httptotalcertcns}{0}
\DefineRemark{30115011httptotalcerts}{0}
\DefineRemark{30115011httptotalconnerr}{4216542}
\DefineRemark{30115011httptotalconnsuccess}{163820}
\DefineRemark{30115011httptotalhosts}{4380362}
\DefineRemark{30115011httptotalsuccess}{0}
\DefineRemark{30115011httptotalsystemid}{0}
\DefineRemark{30115011httptotaltls}{0}
\DefineRemark{30115011httptotaltlsauth}{0}
\DefineRemark{30115011httptotaltlssuccess}{0}
\DefineRemark{30115011tlshttpdate}{05-17}
\DefineRemark{30115011tlshttppctconnsuccess}{3.7}
\DefineRemark{30115011tlshttppcttlsvalidoftransport}{87}
\DefineRemark{30115011tlshttppctvalidtls}{0.3}
\DefineRemark{30115011tlshttptotalas}{12}
\DefineRemark{30115011tlshttptotalcertcns}{18}
\DefineRemark{30115011tlshttptotalcerts}{21}
\DefineRemark{30115011tlshttptotalconnerr}{4216542}
\DefineRemark{30115011tlshttptotalconnsuccess}{163820}
\DefineRemark{30115011tlshttptotalhosts}{4380362}
\DefineRemark{30115011tlshttptotalsuccess}{22}
\DefineRemark{30115011tlshttptotalsystemid}{18}
\DefineRemark{30115011tlshttptotaltls}{142271}
\DefineRemark{30115011tlshttptotaltlsauth}{142269}
\DefineRemark{30115011tlshttptotaltlssuccess}{92529}
\DefineRemark{37823782siemensdate}{05-10}
\DefineRemark{37823782siemenspctconnsuccess}{1.0}
\DefineRemark{37823782siemenspcttlsvalidoftransport}{0.0}
\DefineRemark{37823782siemenspctvalidtls}{0.0}
\DefineRemark{37823782siemenstotalas}{0}
\DefineRemark{37823782siemenstotalcertcns}{0}
\DefineRemark{37823782siemenstotalcerts}{0}
\DefineRemark{37823782siemenstotalconnerr}{3348688}
\DefineRemark{37823782siemenstotalconnsuccess}{34958}
\DefineRemark{37823782siemenstotalhosts}{3383646}
\DefineRemark{37823782siemenstotalsuccess}{0}
\DefineRemark{37823782siemenstotalsystemid}{0}
\DefineRemark{37823782siemenstotaltls}{0}
\DefineRemark{37823782siemenstotaltlsauth}{0}
\DefineRemark{37823782siemenstotaltlssuccess}{0}
\DefineRemark{37823782tlssiemensdate}{05-10}
\DefineRemark{37823782tlssiemenspctconnsuccess}{1.0}
\DefineRemark{37823782tlssiemenspcttlsvalidoftransport}{16}
\DefineRemark{37823782tlssiemenspctvalidtls}{0.0}
\DefineRemark{37823782tlssiemenstotalas}{1}
\DefineRemark{37823782tlssiemenstotalcertcns}{1}
\DefineRemark{37823782tlssiemenstotalcerts}{1}
\DefineRemark{37823782tlssiemenstotalconnerr}{3348688}
\DefineRemark{37823782tlssiemenstotalconnsuccess}{34958}
\DefineRemark{37823782tlssiemenstotalhosts}{3383646}
\DefineRemark{37823782tlssiemenstotalsuccess}{1}
\DefineRemark{37823782tlssiemenstotalsystemid}{0}
\DefineRemark{37823782tlssiemenstotaltls}{5555}
\DefineRemark{37823782tlssiemenstotaltlsauth}{5555}
\DefineRemark{37823782tlssiemenstotaltlssuccess}{5497}
\DefineRemark{4481844818enipdate}{04-27}
\DefineRemark{4481844818enippctconnsuccess}{0.1}
\DefineRemark{4481844818enippcttlsvalidoftransport}{0.0}
\DefineRemark{4481844818enippctvalidtls}{6.4}
\DefineRemark{4481844818eniptotalas}{282}
\DefineRemark{4481844818eniptotalcertcns}{0}
\DefineRemark{4481844818eniptotalcerts}{0}
\DefineRemark{4481844818eniptotalconnerr}{3430937}
\DefineRemark{4481844818eniptotalconnsuccess}{5120}
\DefineRemark{4481844818eniptotalhosts}{3436057}
\DefineRemark{4481844818eniptotalsuccess}{1613}
\DefineRemark{4481844818eniptotalsystemid}{1487}
\DefineRemark{4481844818eniptotaltls}{0}
\DefineRemark{4481844818eniptotaltlsauth}{0}
\DefineRemark{4481844818eniptotaltlssuccess}{0}
\DefineRemark{4481844818tlsenipdate}{04-27}
\DefineRemark{4481844818tlsenippctconnsuccess}{0.1}
\DefineRemark{4481844818tlsenippcttlsvalidoftransport}{35}
\DefineRemark{4481844818tlsenippctvalidtls}{6.4}
\DefineRemark{4481844818tlseniptotalas}{0}
\DefineRemark{4481844818tlseniptotalcertcns}{0}
\DefineRemark{4481844818tlseniptotalcerts}{0}
\DefineRemark{4481844818tlseniptotalconnerr}{3430937}
\DefineRemark{4481844818tlseniptotalconnsuccess}{5120}
\DefineRemark{4481844818tlseniptotalhosts}{3436057}
\DefineRemark{4481844818tlseniptotalsuccess}{0}
\DefineRemark{4481844818tlseniptotalsystemid}{0}
\DefineRemark{4481844818tlseniptotaltls}{1816}
\DefineRemark{4481844818tlseniptotaltlsauth}{1816}
\DefineRemark{4481844818tlseniptotaltlssuccess}{1816}
\DefineRemark{48404840opcuadate}{07-26}
\DefineRemark{48404840opcuapctconnsuccess}{0.8}
\DefineRemark{48404840opcuapcttlsvalidoftransport}{0.0}
\DefineRemark{48404840opcuapctvalidtls}{0}
\DefineRemark{48404840opcuatotalas}{481}
\DefineRemark{48404840opcuatotalcertcns}{0}
\DefineRemark{48404840opcuatotalcerts}{0}
\DefineRemark{48404840opcuatotalconnerr}{2759232}
\DefineRemark{48404840opcuatotalconnsuccess}{21991}
\DefineRemark{48404840opcuatotalhosts}{2781223}
\DefineRemark{48404840opcuatotalsuccess}{2193}
\DefineRemark{48404840opcuatotalsystemid}{0}
\DefineRemark{48404840opcuatotaltls}{0}
\DefineRemark{48404840opcuatotaltlsauth}{0}
\DefineRemark{48404840opcuatotaltlssuccess}{0}
\DefineRemark{48404840tlsopcuadate}{07-26}
\DefineRemark{48404840tlsopcuapctconnsuccess}{1.1}
\DefineRemark{48404840tlsopcuapcttlsvalidoftransport}{19}
\DefineRemark{48404840tlsopcuapctvalidtls}{0.3}
\DefineRemark{48404840tlsopcuatotalas}{0}
\DefineRemark{48404840tlsopcuatotalcertcns}{0}
\DefineRemark{48404840tlsopcuatotalcerts}{0}
\DefineRemark{48404840tlsopcuatotalconnerr}{2748355}
\DefineRemark{48404840tlsopcuatotalconnsuccess}{30675}
\DefineRemark{48404840tlsopcuatotalhosts}{2779030}
\DefineRemark{48404840tlsopcuatotalsuccess}{0}
\DefineRemark{48404840tlsopcuatotalsystemid}{0}
\DefineRemark{48404840tlsopcuatotaltls}{5927}
\DefineRemark{48404840tlsopcuatotaltlsauth}{5921}
\DefineRemark{48404840tlsopcuatotaltlssuccess}{4681}
\DefineRemark{48434843httpdate}{07-24}
\DefineRemark{48434843httppctconnsuccess}{1.2}
\DefineRemark{48434843httppcttlsvalidoftransport}{0.0}
\DefineRemark{48434843httppctvalidtls}{0.3}
\DefineRemark{48434843httptotalas}{0}
\DefineRemark{48434843httptotalcertcns}{0}
\DefineRemark{48434843httptotalcerts}{0}
\DefineRemark{48434843httptotalconnerr}{2674012}
\DefineRemark{48434843httptotalconnsuccess}{31956}
\DefineRemark{48434843httptotalhosts}{2705968}
\DefineRemark{48434843httptotalsuccess}{0}
\DefineRemark{48434843httptotalsystemid}{0}
\DefineRemark{48434843httptotaltls}{0}
\DefineRemark{48434843httptotaltlsauth}{0}
\DefineRemark{48434843httptotaltlssuccess}{0}
\DefineRemark{48434843tlshttpdate}{07-24}
\DefineRemark{48434843tlshttppctconnsuccess}{1.2}
\DefineRemark{48434843tlshttppcttlsvalidoftransport}{19}
\DefineRemark{48434843tlshttppctvalidtls}{0.3}
\DefineRemark{48434843tlshttptotalas}{0}
\DefineRemark{48434843tlshttptotalcertcns}{0}
\DefineRemark{48434843tlshttptotalcerts}{0}
\DefineRemark{48434843tlshttptotalconnerr}{2674012}
\DefineRemark{48434843tlshttptotalconnsuccess}{31956}
\DefineRemark{48434843tlshttptotalhosts}{2705968}
\DefineRemark{48434843tlshttptotalsuccess}{0}
\DefineRemark{48434843tlshttptotalsystemid}{0}
\DefineRemark{48434843tlshttptotaltls}{6193}
\DefineRemark{48434843tlshttptotaltlsauth}{6185}
\DefineRemark{48434843tlshttptotaltlssuccess}{4898}
\DefineRemark{49111911foxdate}{04-18}
\DefineRemark{49111911foxpctconnsuccess}{0.3}
\DefineRemark{49111911foxpcttlsvalidoftransport}{0.0}
\DefineRemark{49111911foxpctvalidtls}{0.5}
\DefineRemark{49111911foxtotalas}{596}
\DefineRemark{49111911foxtotalcertcns}{0}
\DefineRemark{49111911foxtotalcerts}{0}
\DefineRemark{49111911foxtotalconnerr}{3668472}
\DefineRemark{49111911foxtotalconnsuccess}{12257}
\DefineRemark{49111911foxtotalhosts}{3680729}
\DefineRemark{49111911foxtotalsuccess}{4143}
\DefineRemark{49111911foxtotalsystemid}{564}
\DefineRemark{49111911foxtotaltls}{0}
\DefineRemark{49111911foxtotaltlsauth}{0}
\DefineRemark{49111911foxtotaltlssuccess}{0}
\DefineRemark{49111911tlsfoxdate}{04-18}
\DefineRemark{49111911tlsfoxpctconnsuccess}{0.3}
\DefineRemark{49111911tlsfoxpcttlsvalidoftransport}{19}
\DefineRemark{49111911tlsfoxpctvalidtls}{0.5}
\DefineRemark{49111911tlsfoxtotalas}{0}
\DefineRemark{49111911tlsfoxtotalcertcns}{0}
\DefineRemark{49111911tlsfoxtotalcerts}{0}
\DefineRemark{49111911tlsfoxtotalconnerr}{3668472}
\DefineRemark{49111911tlsfoxtotalconnsuccess}{12257}
\DefineRemark{49111911tlsfoxtotalhosts}{3680729}
\DefineRemark{49111911tlsfoxtotalsuccess}{0}
\DefineRemark{49111911tlsfoxtotalsystemid}{0}
\DefineRemark{49111911tlsfoxtotaltls}{2357}
\DefineRemark{49111911tlsfoxtotaltlsauth}{2342}
\DefineRemark{49111911tlsfoxtotaltlssuccess}{2348}
\DefineRemark{49113011httpdate}{04-18}
\DefineRemark{49113011httppctconnsuccess}{0.5}
\DefineRemark{49113011httppcttlsvalidoftransport}{0.0}
\DefineRemark{49113011httppctvalidtls}{0.3}
\DefineRemark{49113011httptotalas}{525}
\DefineRemark{49113011httptotalcertcns}{0}
\DefineRemark{49113011httptotalcerts}{0}
\DefineRemark{49113011httptotalconnerr}{3661653}
\DefineRemark{49113011httptotalconnsuccess}{19076}
\DefineRemark{49113011httptotalhosts}{3680729}
\DefineRemark{49113011httptotalsuccess}{3791}
\DefineRemark{49113011httptotalsystemid}{3656}
\DefineRemark{49113011httptotaltls}{0}
\DefineRemark{49113011httptotaltlsauth}{0}
\DefineRemark{49113011httptotaltlssuccess}{0}
\DefineRemark{49114911foxdate}{04-18}
\DefineRemark{49114911foxpctconnsuccess}{0.9}
\DefineRemark{49114911foxpcttlsvalidoftransport}{0.0}
\DefineRemark{49114911foxpctvalidtls}{0.5}
\DefineRemark{49114911foxtotalas}{0}
\DefineRemark{49114911foxtotalcertcns}{0}
\DefineRemark{49114911foxtotalcerts}{0}
\DefineRemark{49114911foxtotalconnerr}{3646941}
\DefineRemark{49114911foxtotalconnsuccess}{33788}
\DefineRemark{49114911foxtotalhosts}{3680729}
\DefineRemark{49114911foxtotalsuccess}{0}
\DefineRemark{49114911foxtotalsystemid}{0}
\DefineRemark{49114911foxtotaltls}{0}
\DefineRemark{49114911foxtotaltlsauth}{0}
\DefineRemark{49114911foxtotaltlssuccess}{0}
\DefineRemark{49114911tlsfoxdate}{04-18}
\DefineRemark{49114911tlsfoxpctconnsuccess}{0.9}
\DefineRemark{49114911tlsfoxpcttlsvalidoftransport}{38}
\DefineRemark{49114911tlsfoxpctvalidtls}{0.5}
\DefineRemark{49114911tlsfoxtotalas}{49}
\DefineRemark{49114911tlsfoxtotalcertcns}{51}
\DefineRemark{49114911tlsfoxtotalcerts}{54}
\DefineRemark{49114911tlsfoxtotalconnerr}{3646941}
\DefineRemark{49114911tlsfoxtotalconnsuccess}{33788}
\DefineRemark{49114911tlsfoxtotalhosts}{3680729}
\DefineRemark{49114911tlsfoxtotalsuccess}{115}
\DefineRemark{49114911tlsfoxtotalsystemid}{15}
\DefineRemark{49114911tlsfoxtotaltls}{12885}
\DefineRemark{49114911tlsfoxtotaltlsauth}{12882}
\DefineRemark{49114911tlsfoxtotaltlssuccess}{4293}
\DefineRemark{49115011httpdate}{04-18}
\DefineRemark{49115011httppctconnsuccess}{0.8}
\DefineRemark{49115011httppcttlsvalidoftransport}{0.0}
\DefineRemark{49115011httppctvalidtls}{0.3}
\DefineRemark{49115011httptotalas}{0}
\DefineRemark{49115011httptotalcertcns}{0}
\DefineRemark{49115011httptotalcerts}{0}
\DefineRemark{49115011httptotalconnerr}{3649653}
\DefineRemark{49115011httptotalconnsuccess}{31076}
\DefineRemark{49115011httptotalhosts}{3680729}
\DefineRemark{49115011httptotalsuccess}{0}
\DefineRemark{49115011httptotalsystemid}{0}
\DefineRemark{49115011httptotaltls}{0}
\DefineRemark{49115011httptotaltlsauth}{0}
\DefineRemark{49115011httptotaltlssuccess}{0}
\DefineRemark{49115011tlshttpdate}{04-18}
\DefineRemark{49115011tlshttppctconnsuccess}{0.8}
\DefineRemark{49115011tlshttppcttlsvalidoftransport}{35}
\DefineRemark{49115011tlshttppctvalidtls}{0.3}
\DefineRemark{49115011tlshttptotalas}{28}
\DefineRemark{49115011tlshttptotalcertcns}{27}
\DefineRemark{49115011tlshttptotalcerts}{28}
\DefineRemark{49115011tlshttptotalconnerr}{3649653}
\DefineRemark{49115011tlshttptotalconnsuccess}{31076}
\DefineRemark{49115011tlshttptotalhosts}{3680729}
\DefineRemark{49115011tlshttptotalsuccess}{70}
\DefineRemark{49115011tlshttptotalsystemid}{69}
\DefineRemark{49115011tlshttptotaltls}{11023}
\DefineRemark{49115011tlshttptotaltlsauth}{11021}
\DefineRemark{49115011tlshttptotaltlssuccess}{3187}
\DefineRemark{502502modbusdate}{06-28}
\DefineRemark{502502modbuspctconnsuccess}{1.8}
\DefineRemark{502502modbuspcttlsvalidoftransport}{0.0}
\DefineRemark{502502modbuspctvalidtls}{0.0}
\DefineRemark{502502modbustotalas}{2285}
\DefineRemark{502502modbustotalcertcns}{0}
\DefineRemark{502502modbustotalcerts}{0}
\DefineRemark{502502modbustotalconnerr}{2806481}
\DefineRemark{502502modbustotalconnsuccess}{50367}
\DefineRemark{502502modbustotalhosts}{2856848}
\DefineRemark{502502modbustotalsuccess}{30083}
\DefineRemark{502502modbustotalsystemid}{410}
\DefineRemark{502502modbustotaltls}{0}
\DefineRemark{502502modbustotaltlsauth}{0}
\DefineRemark{502502modbustotaltlssuccess}{0}
\DefineRemark{502502tlsmodbusdate}{06-28}
\DefineRemark{502502tlsmodbuspctconnsuccess}{1.8}
\DefineRemark{502502tlsmodbuspcttlsvalidoftransport}{12}
\DefineRemark{502502tlsmodbuspctvalidtls}{0.0}
\DefineRemark{502502tlsmodbustotalas}{0}
\DefineRemark{502502tlsmodbustotalcertcns}{0}
\DefineRemark{502502tlsmodbustotalcerts}{0}
\DefineRemark{502502tlsmodbustotalconnerr}{2806481}
\DefineRemark{502502tlsmodbustotalconnsuccess}{50367}
\DefineRemark{502502tlsmodbustotalhosts}{2856848}
\DefineRemark{502502tlsmodbustotalsuccess}{0}
\DefineRemark{502502tlsmodbustotalsystemid}{0}
\DefineRemark{502502tlsmodbustotaltls}{6153}
\DefineRemark{502502tlsmodbustotaltlsauth}{6153}
\DefineRemark{502502tlsmodbustotaltlssuccess}{6083}
\DefineRemark{56715671amqpdate}{06-13}
\DefineRemark{56715671amqppctconnsuccess}{3.7}
\DefineRemark{56715671amqppcttlsvalidoftransport}{0.0}
\DefineRemark{56715671amqppctvalidtls}{16}
\DefineRemark{56715671amqptotalas}{0}
\DefineRemark{56715671amqptotalcertcns}{0}
\DefineRemark{56715671amqptotalcerts}{0}
\DefineRemark{56715671amqptotalconnerr}{5340743}
\DefineRemark{56715671amqptotalconnsuccess}{203859}
\DefineRemark{56715671amqptotalhosts}{5544602}
\DefineRemark{56715671amqptotalsuccess}{0}
\DefineRemark{56715671amqptotalsystemid}{0}
\DefineRemark{56715671amqptotaltls}{0}
\DefineRemark{56715671amqptotaltlsauth}{0}
\DefineRemark{56715671amqptotaltlssuccess}{0}
\DefineRemark{56715671tlsamqpdate}{06-13}
\DefineRemark{56715671tlsamqppctconnsuccess}{3.7}
\DefineRemark{56715671tlsamqppcttlsvalidoftransport}{85}
\DefineRemark{56715671tlsamqppctvalidtls}{16}
\DefineRemark{56715671tlsamqptotalas}{709}
\DefineRemark{56715671tlsamqptotalcertcns}{2161}
\DefineRemark{56715671tlsamqptotalcerts}{3057}
\DefineRemark{56715671tlsamqptotalconnerr}{5340743}
\DefineRemark{56715671tlsamqptotalconnsuccess}{203859}
\DefineRemark{56715671tlsamqptotalhosts}{5544602}
\DefineRemark{56715671tlsamqptotalsuccess}{19466}
\DefineRemark{56715671tlsamqptotalsystemid}{6702}
\DefineRemark{56715671tlsamqptotaltls}{173531}
\DefineRemark{56715671tlsamqptotaltlsauth}{173464}
\DefineRemark{56715671tlsamqptotaltlssuccess}{130281}
\DefineRemark{56725672amqpdate}{06-14}
\DefineRemark{56725672amqppctconnsuccess}{5.5}
\DefineRemark{56725672amqppcttlsvalidoftransport}{0.0}
\DefineRemark{56725672amqppctvalidtls}{16}
\DefineRemark{56725672amqptotalas}{5194}
\DefineRemark{56725672amqptotalcertcns}{0}
\DefineRemark{56725672amqptotalcerts}{0}
\DefineRemark{56725672amqptotalconnerr}{5176253}
\DefineRemark{56725672amqptotalconnsuccess}{302585}
\DefineRemark{56725672amqptotalhosts}{5478838}
\DefineRemark{56725672amqptotalsuccess}{130485}
\DefineRemark{56725672amqptotalsystemid}{80753}
\DefineRemark{56725672amqptotaltls}{0}
\DefineRemark{56725672amqptotaltlsauth}{0}
\DefineRemark{56725672amqptotaltlssuccess}{0}
\DefineRemark{56725672tlsamqpdate}{06-14}
\DefineRemark{56725672tlsamqppctconnsuccess}{5.5}
\DefineRemark{56725672tlsamqppcttlsvalidoftransport}{49}
\DefineRemark{56725672tlsamqppctvalidtls}{16}
\DefineRemark{56725672tlsamqptotalas}{131}
\DefineRemark{56725672tlsamqptotalcertcns}{252}
\DefineRemark{56725672tlsamqptotalcerts}{326}
\DefineRemark{56725672tlsamqptotalconnerr}{5176253}
\DefineRemark{56725672tlsamqptotalconnsuccess}{302585}
\DefineRemark{56725672tlsamqptotalhosts}{5478838}
\DefineRemark{56725672tlsamqptotalsuccess}{6299}
\DefineRemark{56725672tlsamqptotalsystemid}{265}
\DefineRemark{56725672tlsamqptotaltls}{147280}
\DefineRemark{56725672tlsamqptotaltlsauth}{146901}
\DefineRemark{56725672tlsamqptotaltlssuccess}{106762}
\DefineRemark{56835683coapdate}{06-22}
\DefineRemark{56835683coappctconnsuccess}{89}
\DefineRemark{56835683coappcttlsvalidoftransport}{0.0}
\DefineRemark{56835683coappctvalidtls}{0.0}
\DefineRemark{56835683coaptotalas}{2821}
\DefineRemark{56835683coaptotalcertcns}{0}
\DefineRemark{56835683coaptotalcerts}{0}
\DefineRemark{56835683coaptotalconnerr}{54474}
\DefineRemark{56835683coaptotalconnsuccess}{425387}
\DefineRemark{56835683coaptotalhosts}{479861}
\DefineRemark{56835683coaptotalsuccess}{420780}
\DefineRemark{56835683coaptotalsystemid}{297}
\DefineRemark{56835683coaptotaltls}{0}
\DefineRemark{56835683coaptotaltlsauth}{0}
\DefineRemark{56835683coaptotaltlssuccess}{0}
\DefineRemark{56835683tlscoapdate}{06-22}
\DefineRemark{56835683tlscoappctconnsuccess}{89}
\DefineRemark{56835683tlscoappcttlsvalidoftransport}{0.0}
\DefineRemark{56835683tlscoappctvalidtls}{0.0}
\DefineRemark{56835683tlscoaptotalas}{0}
\DefineRemark{56835683tlscoaptotalcertcns}{0}
\DefineRemark{56835683tlscoaptotalcerts}{0}
\DefineRemark{56835683tlscoaptotalconnerr}{54474}
\DefineRemark{56835683tlscoaptotalconnsuccess}{425387}
\DefineRemark{56835683tlscoaptotalhosts}{479861}
\DefineRemark{56835683tlscoaptotalsuccess}{0}
\DefineRemark{56835683tlscoaptotalsystemid}{0}
\DefineRemark{56835683tlscoaptotaltls}{3}
\DefineRemark{56835683tlscoaptotaltlsauth}{3}
\DefineRemark{56835683tlscoaptotaltlssuccess}{1}
\DefineRemark{56845684coapdate}{06-21}
\DefineRemark{56845684coappctconnsuccess}{8.7}
\DefineRemark{56845684coappcttlsvalidoftransport}{0.0}
\DefineRemark{56845684coappctvalidtls}{0.0}
\DefineRemark{56845684coaptotalas}{0}
\DefineRemark{56845684coaptotalcertcns}{0}
\DefineRemark{56845684coaptotalcerts}{0}
\DefineRemark{56845684coaptotalconnerr}{48133}
\DefineRemark{56845684coaptotalconnsuccess}{4563}
\DefineRemark{56845684coaptotalhosts}{52696}
\DefineRemark{56845684coaptotalsuccess}{0}
\DefineRemark{56845684coaptotalsystemid}{0}
\DefineRemark{56845684coaptotaltls}{0}
\DefineRemark{56845684coaptotaltlsauth}{0}
\DefineRemark{56845684coaptotaltlssuccess}{0}
\DefineRemark{56845684tlscoapdate}{06-21}
\DefineRemark{56845684tlscoappctconnsuccess}{8.7}
\DefineRemark{56845684tlscoappcttlsvalidoftransport}{15}
\DefineRemark{56845684tlscoappctvalidtls}{0.0}
\DefineRemark{56845684tlscoaptotalas}{19}
\DefineRemark{56845684tlscoaptotalcertcns}{23}
\DefineRemark{56845684tlscoaptotalcerts}{30}
\DefineRemark{56845684tlscoaptotalconnerr}{48133}
\DefineRemark{56845684tlscoaptotalconnsuccess}{4563}
\DefineRemark{56845684tlscoaptotalhosts}{52696}
\DefineRemark{56845684tlscoaptotalsuccess}{76}
\DefineRemark{56845684tlscoaptotalsystemid}{25}
\DefineRemark{56845684tlscoaptotaltls}{702}
\DefineRemark{56845684tlscoaptotaltlsauth}{702}
\DefineRemark{56845684tlscoaptotaltlssuccess}{583}
\DefineRemark{802802modbusdate}{06-27}
\DefineRemark{802802modbuspctconnsuccess}{1.2}
\DefineRemark{802802modbuspcttlsvalidoftransport}{0.0}
\DefineRemark{802802modbuspctvalidtls}{0.0}
\DefineRemark{802802modbustotalas}{0}
\DefineRemark{802802modbustotalcertcns}{0}
\DefineRemark{802802modbustotalcerts}{0}
\DefineRemark{802802modbustotalconnerr}{3088196}
\DefineRemark{802802modbustotalconnsuccess}{36660}
\DefineRemark{802802modbustotalhosts}{3124856}
\DefineRemark{802802modbustotalsuccess}{0}
\DefineRemark{802802modbustotalsystemid}{0}
\DefineRemark{802802modbustotaltls}{0}
\DefineRemark{802802modbustotaltlsauth}{0}
\DefineRemark{802802modbustotaltlssuccess}{0}
\DefineRemark{802802tlsmodbusdate}{06-27}
\DefineRemark{802802tlsmodbuspctconnsuccess}{1.2}
\DefineRemark{802802tlsmodbuspcttlsvalidoftransport}{23}
\DefineRemark{802802tlsmodbuspctvalidtls}{0.0}
\DefineRemark{802802tlsmodbustotalas}{0}
\DefineRemark{802802tlsmodbustotalcertcns}{0}
\DefineRemark{802802tlsmodbustotalcerts}{0}
\DefineRemark{802802tlsmodbustotalconnerr}{3088196}
\DefineRemark{802802tlsmodbustotalconnsuccess}{36660}
\DefineRemark{802802tlsmodbustotalhosts}{3124856}
\DefineRemark{802802tlsmodbustotalsuccess}{0}
\DefineRemark{802802tlsmodbustotalsystemid}{0}
\DefineRemark{802802tlsmodbustotaltls}{8279}
\DefineRemark{802802tlsmodbustotaltlsauth}{8279}
\DefineRemark{802802tlsmodbustotaltlssuccess}{8211}
\DefineRemark{88838883mqttdate}{07-05}
\DefineRemark{88838883mqttpctconnsuccess}{15}
\DefineRemark{88838883mqttpcttlsvalidoftransport}{0.0}
\DefineRemark{88838883mqttpctvalidtls}{12}
\DefineRemark{88838883mqtttotalas}{0}
\DefineRemark{88838883mqtttotalcertcns}{0}
\DefineRemark{88838883mqtttotalcerts}{0}
\DefineRemark{88838883mqtttotalconnerr}{3620718}
\DefineRemark{88838883mqtttotalconnsuccess}{632221}
\DefineRemark{88838883mqtttotalhosts}{4252939}
\DefineRemark{88838883mqtttotalsuccess}{0}
\DefineRemark{88838883mqtttotalsystemid}{0}
\DefineRemark{88838883mqtttotaltls}{0}
\DefineRemark{88838883mqtttotaltlsauth}{0}
\DefineRemark{88838883mqtttotaltlssuccess}{0}
\DefineRemark{88838883tlsmqttdate}{07-05}
\DefineRemark{88838883tlsmqttpctconnsuccess}{15}
\DefineRemark{88838883tlsmqttpcttlsvalidoftransport}{85}
\DefineRemark{88838883tlsmqttpctvalidtls}{12}
\DefineRemark{88838883tlsmqtttotalas}{2243}
\DefineRemark{88838883tlsmqtttotalcertcns}{16543}
\DefineRemark{88838883tlsmqtttotalcerts}{17159}
\DefineRemark{88838883tlsmqtttotalconnerr}{3620718}
\DefineRemark{88838883tlsmqtttotalconnsuccess}{632221}
\DefineRemark{88838883tlsmqtttotalhosts}{4252939}
\DefineRemark{88838883tlsmqtttotalsuccess}{35652}
\DefineRemark{88838883tlsmqtttotalsystemid}{0}
\DefineRemark{88838883tlsmqtttotaltls}{540079}
\DefineRemark{88838883tlsmqtttotaltlsauth}{540055}
\DefineRemark{88838883tlsmqtttotaltlssuccess}{168617}
\DefineRemark{allamqppctvalidtls}{16} %
\DefineRemark{allcoappctvalidtls}{0.0} %
\DefineRemark{alldnp3pctvalidtls}{1.0} %
\DefineRemark{allenippctvalidtls}{6.4} %
\DefineRemark{allfoxpctvalidtls}{0.5} %
\DefineRemark{allfoxtotalas}{1565} %
\DefineRemark{allfoxtotalcertcns}{0} %
\DefineRemark{allfoxtotalsuccess}{23297}
\DefineRemark{allhttppctvalidtls}{0.3} %
\DefineRemark{allhttptotalas}{1564} %
\DefineRemark{allhttptotalcertcns}{0} %
\DefineRemark{allhttptotalsuccess}{23447}
\DefineRemark{alliec104pctvalidtls}{0.0} %
\DefineRemark{allmodbuspctvalidtls}{0.0} %
\DefineRemark{allmqttpctvalidtls}{12} %
\DefineRemark{allopcuapctvalidtls}{0.0} %
\DefineRemark{allsiemenspctvalidtls}{0.0} %
\DefineRemark{alltlsamqptotalas}{781} %
\DefineRemark{alltlsamqptotalcertcns}{2357} %
\DefineRemark{alltlsamqptotalsuccess}{23908} %
\DefineRemark{alltlscoaptotalas}{19} %
\DefineRemark{alltlscoaptotalcertcns}{23} %
\DefineRemark{alltlscoaptotalsuccess}{76} %
\DefineRemark{alltlsdnp3totalas}{3} %
\DefineRemark{alltlsdnp3totalcertcns}{3} %
\DefineRemark{alltlsdnp3totalsuccess}{5} %
\DefineRemark{alltlseniptotalas}{56} %
\DefineRemark{alltlseniptotalcertcns}{1} %
\DefineRemark{alltlseniptotalsuccess}{111} %
\DefineRemark{alltlsfoxtotalas}{52} %
\DefineRemark{alltlsfoxtotalcertcns}{55} %
\DefineRemark{alltlsfoxtotalsuccess}{123} %
\DefineRemark{alltlshttptotalas}{31} %
\DefineRemark{alltlshttptotalcertcns}{36} %
\DefineRemark{alltlshttptotalsuccess}{82} %
\DefineRemark{alltlsiec104totalas}{1} %
\DefineRemark{alltlsiec104totalcertcns}{1} %
\DefineRemark{alltlsiec104totalsuccess}{1} %
\DefineRemark{alltlsmodbustotalas}{0} %
\DefineRemark{alltlsmodbustotalcertcns}{0} %
\DefineRemark{alltlsmodbustotalsuccess}{0} %
\DefineRemark{alltlsmqtttotalas}{2318} %
\DefineRemark{alltlsmqtttotalcertcns}{19068} %
\DefineRemark{alltlsmqtttotalsuccess}{39061} %
\DefineRemark{alltlsopcuatotalas}{0} %
\DefineRemark{alltlsopcuatotalcertcns}{0} %
\DefineRemark{alltlsopcuatotalsuccess}{0} %
\DefineRemark{alltlssiemenstotalas}{1} %
\DefineRemark{alltlssiemenstotalcertcns}{1} %
\DefineRemark{alltlssiemenstotalsuccess}{1} %
\DefineRemark{optionaltlsamqptotalsuccess}{12347} %
\DefineRemark{optionaltlscoaptotalsuccess}{48} %
\DefineRemark{optionaltlsdnp3totalsuccess}{1} %
\DefineRemark{optionaltlseniptotalsuccess}{16} %
\DefineRemark{optionaltlsfoxtotalsuccess}{37} %
\DefineRemark{optionaltlshttptotalsuccess}{21} %
\DefineRemark{optionaltlsiec104totalsuccess}{1} %
\DefineRemark{optionaltlsmodbustotalsuccess}{0} %
\DefineRemark{optionaltlsmqtttotalsuccess}{18082} %
\DefineRemark{optionaltlsopcuatotalsuccess}{0} %
\DefineRemark{optionaltlssiemenstotalsuccess}{0} %

\DefineRemark{amqpcnttls10}{55}
\DefineRemark{amqpcnttls11}{0}
\DefineRemark{amqpcnttls12}{23853}
\DefineRemark{amqpcnttls13capable}{2960}
\DefineRemark{coapcnttls10}{0}
\DefineRemark{coapcnttls11}{0}
\DefineRemark{coapcnttls12}{76}
\DefineRemark{coapcnttls13capable}{0}
\DefineRemark{dnp3cnttls10}{2}
\DefineRemark{dnp3cnttls11}{0}
\DefineRemark{dnp3cnttls12}{3}
\DefineRemark{dnp3cnttls13capable}{2}
\DefineRemark{enipcnttls10}{0}
\DefineRemark{enipcnttls11}{0}
\DefineRemark{enipcnttls12}{111}
\DefineRemark{enipcnttls13capable}{0}
\DefineRemark{foxcnttls10}{5}
\DefineRemark{foxcnttls11}{0}
\DefineRemark{foxcnttls12}{118}
\DefineRemark{foxcnttls13capable}{0}
\DefineRemark{httpcnttls10}{2}
\DefineRemark{httpcnttls11}{0}
\DefineRemark{httpcnttls12}{80}
\DefineRemark{httpcnttls13capable}{7}
\DefineRemark{iec104cnttls10}{0}
\DefineRemark{iec104cnttls11}{0}
\DefineRemark{iec104cnttls12}{1}
\DefineRemark{iec104cnttls13capable}{0}
\DefineRemark{modbuscnttls10}{0}
\DefineRemark{modbuscnttls11}{0}
\DefineRemark{modbuscnttls12}{0}
\DefineRemark{modbuscnttls13capable}{0}
\DefineRemark{mqttcnttls10}{252}
\DefineRemark{mqttcnttls11}{61}
\DefineRemark{mqttcnttls12}{38748}
\DefineRemark{mqttcnttls13capable}{13087}
\DefineRemark{siemenscnttls10}{0}
\DefineRemark{siemenscnttls11}{0}
\DefineRemark{siemenscnttls12}{1}
\DefineRemark{siemenscnttls13capable}{0}

\DefineRemark{hostpctamqpdefaultassessmentFalse}{0.4} %
\DefineRemark{hostpctamqpdefaultassessmentinsecureboth}{0.3} %
\DefineRemark{hostpctamqpdefaultassessmentinsecurecipher}{0.2} %
\DefineRemark{hostpctamqpdefaultassessmentinsecuremac}{0.7} %
\DefineRemark{hostpctamqpdefaultassessmentsecure}{98} %
\DefineRemark{hostpctamqpinsecuresuccessFalse}{99} %
\DefineRemark{hostpctamqpinsecuresuccessTrue}{1.4} %
\DefineRemark{hostpctamqpnopfssuccessFalse}{100} %
\DefineRemark{hostpctamqpnopfssuccessTrue}{0.1} %
\DefineRemark{hostpctamqprecommendedsuccessFalse}{4.0} %
\DefineRemark{hostpctamqprecommendedsuccessTrue}{96} %
\DefineRemark{hostpctcoapdefaultassessmentFalse}{0.0} %
\DefineRemark{hostpctcoapdefaultassessmentinsecureboth}{0.0} %
\DefineRemark{hostpctcoapdefaultassessmentinsecurecipher}{0.0} %
\DefineRemark{hostpctcoapdefaultassessmentinsecuremac}{9.2} %
\DefineRemark{hostpctcoapdefaultassessmentsecure}{91} %
\DefineRemark{hostpctcoapinsecuresuccessFalse}{100} %
\DefineRemark{hostpctcoapinsecuresuccessTrue}{0.0} %
\DefineRemark{hostpctcoapnopfssuccessFalse}{99} %
\DefineRemark{hostpctcoapnopfssuccessTrue}{1.3} %
\DefineRemark{hostpctcoaprecommendedsuccessFalse}{9.2} %
\DefineRemark{hostpctcoaprecommendedsuccessTrue}{91} %
\DefineRemark{hostpctdnp3defaultassessmentFalse}{0.0} %
\DefineRemark{hostpctdnp3defaultassessmentinsecureboth}{40} %
\DefineRemark{hostpctdnp3defaultassessmentinsecurecipher}{0.0} %
\DefineRemark{hostpctdnp3defaultassessmentinsecuremac}{0.0} %
\DefineRemark{hostpctdnp3defaultassessmentsecure}{60} %
\DefineRemark{hostpctdnp3insecuresuccessFalse}{100} %
\DefineRemark{hostpctdnp3insecuresuccessTrue}{0.0} %
\DefineRemark{hostpctdnp3nopfssuccessFalse}{100} %
\DefineRemark{hostpctdnp3nopfssuccessTrue}{0.0} %
\DefineRemark{hostpctdnp3recommendedsuccessFalse}{60} %
\DefineRemark{hostpctdnp3recommendedsuccessTrue}{40} %
\DefineRemark{hostpctenipdefaultassessmentFalse}{100} %
\DefineRemark{hostpctenipdefaultassessmentinsecureboth}{0.0} %
\DefineRemark{hostpctenipdefaultassessmentinsecurecipher}{0.0} %
\DefineRemark{hostpctenipdefaultassessmentinsecuremac}{0.0} %
\DefineRemark{hostpctenipdefaultassessmentsecure}{0.0} %
\DefineRemark{hostpctenipinsecuresuccessFalse}{100} %
\DefineRemark{hostpctenipinsecuresuccessTrue}{0.0} %
\DefineRemark{hostpctenipnopfssuccessFalse}{100} %
\DefineRemark{hostpctenipnopfssuccessTrue}{0.0} %
\DefineRemark{hostpcteniprecommendedsuccessFalse}{0.0} %
\DefineRemark{hostpcteniprecommendedsuccessTrue}{100} %
\DefineRemark{hostpctfoxdefaultassessmentFalse}{0.0} %
\DefineRemark{hostpctfoxdefaultassessmentinsecureboth}{2.4} %
\DefineRemark{hostpctfoxdefaultassessmentinsecurecipher}{0.0} %
\DefineRemark{hostpctfoxdefaultassessmentinsecuremac}{4.9} %
\DefineRemark{hostpctfoxdefaultassessmentsecure}{93} %
\DefineRemark{hostpctfoxinsecuresuccessFalse}{98} %
\DefineRemark{hostpctfoxinsecuresuccessTrue}{2.4} %
\DefineRemark{hostpctfoxnopfssuccessFalse}{100} %
\DefineRemark{hostpctfoxnopfssuccessTrue}{0.0} %
\DefineRemark{hostpctfoxrecommendedsuccessFalse}{14} %
\DefineRemark{hostpctfoxrecommendedsuccessTrue}{86} %
\DefineRemark{hostpcthttpdefaultassessmentFalse}{1.2} %
\DefineRemark{hostpcthttpdefaultassessmentinsecureboth}{1.2} %
\DefineRemark{hostpcthttpdefaultassessmentinsecurecipher}{0.0} %
\DefineRemark{hostpcthttpdefaultassessmentinsecuremac}{3.7} %
\DefineRemark{hostpcthttpdefaultassessmentsecure}{94} %
\DefineRemark{hostpcthttpinsecuresuccessFalse}{99} %
\DefineRemark{hostpcthttpinsecuresuccessTrue}{1.2} %
\DefineRemark{hostpcthttpnopfssuccessFalse}{100} %
\DefineRemark{hostpcthttpnopfssuccessTrue}{0.0} %
\DefineRemark{hostpcthttprecommendedsuccessFalse}{11} %
\DefineRemark{hostpcthttprecommendedsuccessTrue}{89} %
\DefineRemark{hostpctiec104defaultassessmentFalse}{0.0} %
\DefineRemark{hostpctiec104defaultassessmentinsecureboth}{0.0} %
\DefineRemark{hostpctiec104defaultassessmentinsecurecipher}{0.0} %
\DefineRemark{hostpctiec104defaultassessmentinsecuremac}{0.0} %
\DefineRemark{hostpctiec104defaultassessmentsecure}{100} %
\DefineRemark{hostpctiec104insecuresuccessFalse}{100} %
\DefineRemark{hostpctiec104insecuresuccessTrue}{0.0} %
\DefineRemark{hostpctiec104nopfssuccessFalse}{100} %
\DefineRemark{hostpctiec104nopfssuccessTrue}{0.0} %
\DefineRemark{hostpctiec104recommendedsuccessFalse}{0.0} %
\DefineRemark{hostpctiec104recommendedsuccessTrue}{100} %
\DefineRemark{hostpctmqttdefaultassessmentFalse}{0.4} %
\DefineRemark{hostpctmqttdefaultassessmentinsecureboth}{0.1} %
\DefineRemark{hostpctmqttdefaultassessmentinsecurecipher}{0.1} %
\DefineRemark{hostpctmqttdefaultassessmentinsecuremac}{12} %
\DefineRemark{hostpctmqttdefaultassessmentsecure}{88} %
\DefineRemark{hostpctmqttinsecuresuccessFalse}{90} %
\DefineRemark{hostpctmqttinsecuresuccessTrue}{9.8} %
\DefineRemark{hostpctmqttnopfssuccessFalse}{100} %
\DefineRemark{hostpctmqttnopfssuccessTrue}{0.2} %
\DefineRemark{hostpctmqttrecommendedsuccessFalse}{10} %
\DefineRemark{hostpctmqttrecommendedsuccessTrue}{90} %
\DefineRemark{hostpctsiemensdefaultassessmentFalse}{0.0} %
\DefineRemark{hostpctsiemensdefaultassessmentinsecureboth}{100} %
\DefineRemark{hostpctsiemensdefaultassessmentinsecurecipher}{0.0} %
\DefineRemark{hostpctsiemensdefaultassessmentinsecuremac}{0.0} %
\DefineRemark{hostpctsiemensdefaultassessmentsecure}{0.0} %
\DefineRemark{hostpctsiemensinsecuresuccessFalse}{0.0} %
\DefineRemark{hostpctsiemensinsecuresuccessTrue}{100} %
\DefineRemark{hostpctsiemensnopfssuccessFalse}{100} %
\DefineRemark{hostpctsiemensnopfssuccessTrue}{0.0} %
\DefineRemark{hostpctsiemensrecommendedsuccessFalse}{100} %
\DefineRemark{hostpctsiemensrecommendedsuccessTrue}{0.0} %

\DefineRemark{connpcthoneypotmaxreduction}{98} %
\DefineRemark{connpcthoneypotminreduction}{17} %
\DefineRemark{hostscnt}{967551} %
\DefineRemark{hostscntacceptclientcert}{7445} %
\DefineRemark{hostscntamqp}{142046} %
\DefineRemark{hostscntamqpaccessible}{18991} %
\DefineRemark{hostscntamqpcertappservergeneratedsbcatls10}{45} %
\DefineRemark{hostscntamqprejectedcerthint}{15121} %
\DefineRemark{hostscntcoap}{420808} %
\DefineRemark{hostscntdefault}{934736} %
\DefineRemark{hostscntdefaultonly}{904183} %
\DefineRemark{hostscntdepreactedtlswithversioninfo}{98} %
\DefineRemark{hostscntdeprecatedtlsversion}{377} %
\DefineRemark{hostscntdnp3}{507} %
\DefineRemark{hostscntdnp3orangelivebox}{1334} %
\DefineRemark{hostscntenip}{1708} %
\DefineRemark{hostscntfox}{23383} %
\DefineRemark{hostscntfoxplatform}{23508} %
\DefineRemark{hostscntiec104}{4484} %
\DefineRemark{hostscntmodbus}{30083} %
\DefineRemark{hostscntmqtt}{312323} %
\DefineRemark{hostscntmqttaccessible}{255857} %
\DefineRemark{hostscntmqttmesdiff}{160645}
\DefineRemark{hostscntmqttrejectedcerthint}{22749} %
\DefineRemark{hostscntopcua}{2193} %
\DefineRemark{hostscntrejectclientcert}{19263} %
\DefineRemark{hostscntrequestclientcert}{29646} %
\DefineRemark{hostscntrequestecho}{98} %
\DefineRemark{hostscntsiemens}{6508} %
\DefineRemark{hostscnttls}{63368} %
\DefineRemark{hostscnttls10withrecommended}{1} %
\DefineRemark{hostscnttlsallowinsecure}{4292} %
\DefineRemark{hostscnttlsamqp}{23908} %
\DefineRemark{hostscnttlscoap}{76} %
\DefineRemark{hostscnttlsdnp3}{5} %
\DefineRemark{hostscnttlsenip}{111} %
\DefineRemark{hostscnttlsenipsharedserial}{19} %
\DefineRemark{hostscnttlsfox}{123} %
\DefineRemark{hostscnttlsfoxplatform}{82} %
\DefineRemark{hostscnttlsfoxsharedserial}{5} %
\DefineRemark{hostscnttlsgeneralerror}{722825} %
\DefineRemark{hostscnttlsiec104}{1} %
\DefineRemark{hostscnttlsmodbus}{0} %
\DefineRemark{hostscnttlsmqtt}{39061} %
\DefineRemark{hostscnttlsonly}{32815} %
\DefineRemark{hostscnttlsopcua}{0} %
\DefineRemark{hostscnttlsoptional}{30553} %
\DefineRemark{hostscnttlssiemens}{1} %
\DefineRemark{hostscnttlssiemensmodbusiec104dnp3}{7} %
\DefineRemark{hostscnttlsweak}{26665} %
\DefineRemark{hostspcacceptclientcertofrequested}{0} %
\DefineRemark{hostspctacceptclientcert}{25} %
\DefineRemark{hostspctamqpaccessible}{13} %
\DefineRemark{hostspctamqpaccessibledefault}{14} %
\DefineRemark{hostspctamqpaccessibletls}{1.4} %
\DefineRemark{hostspctdeprecatedtlsversion}{0.6} %
\DefineRemark{hostspctmqttaccessible}{82} %
\DefineRemark{hostspctmqttaccessibledefault}{84} %
\DefineRemark{hostspctmqttaccessibletls}{28} %
\DefineRemark{hostspctmqttmsgrecv}{9.0} %
\DefineRemark{hostspctrequestclientcert}{11} %
\DefineRemark{hostspcttls}{6.5} %
\DefineRemark{hostspcttlsaccessibleofaccesscontrol}{18} %
\DefineRemark{hostspcttlsallowinsecure}{6.0} %
\DefineRemark{hostspcttlsrejectnopfs}{100} %
\DefineRemark{hostspcttlsretrofitprotocol}{0.4} %
\DefineRemark{hostspcttlssecurebydesignprotocol}{7.2} %
\DefineRemark{hostspcttlsweak}{42} %
\DefineRemark{hostspcttlsweakduetoauth}{18} %
\DefineRemark{hostspcttlsweakduetocipher}{6.1} %
\DefineRemark{hostspcttlsweakduetodeprecatedcert}{2.2} %
\DefineRemark{hostspcttlsweakduetoreuse}{30} %
\DefineRemark{hostspcttlsweakduetoversion}{0.6} %
\DefineRemark{hostspcttlsworecommended}{8.4} %
\DefineRemark{protohoneypotmaxreduction}{Siemens S7} %
\DefineRemark{protohoneypotminreduction}{Fox Platform} %
\DefineRemark{pvaltlsage}{0} %
\DefineRemark{uvaltlsage}{66690676} %

\DefineRemark{ascnttop3reusedcerts}{348} %
\DefineRemark{certcnt}{23595} %
\DefineRemark{certcntclusteramazon}{151} %
\DefineRemark{certcntclustermean}{6} %
\DefineRemark{certcntclustermedian}{4} %
\DefineRemark{certcntclusterowntracks}{207} %
\DefineRemark{certcntinteras}{528} %
\DefineRemark{certcntinterasloadbalancing}{12} %
\DefineRemark{certcntintraas}{1260} %
\DefineRemark{certcntintraasloadbalancing}{67} %
\DefineRemark{certcntlargestcluster}{207} %
\DefineRemark{certcntmd5}{36} %
\DefineRemark{certcntmd5templated}{28} %
\DefineRemark{certcntnoreuse}{21807} %
\DefineRemark{certcntsha1}{473} %
\DefineRemark{certcntsha1templated}{22} %
\DefineRemark{certpctmd5templated}{78} %
\DefineRemark{certpctnoreuse}{92} %
\DefineRemark{certpctsha1templated}{4.7} %
\DefineRemark{clustercnt}{547} %
\DefineRemark{clustercntsha1templated}{3} %
\DefineRemark{hostcnteniptopreuse}{14} %
\DefineRemark{hostscntamqpcertmd5}{31} %
\DefineRemark{hostscntamqpdeprecatedhash}{322} %
\DefineRemark{hostscntamqpshortkey}{119} %
\DefineRemark{hostscntcasignedexpired}{3062}
\DefineRemark{hostscntinterasloadbalancing}{8893} %
\DefineRemark{hostscntintraasloadbalancing}{10119} %
\DefineRemark{hostscntmqttcasignedweak}{0} %
\DefineRemark{hostscntmqttcertmd5}{20} %
\DefineRemark{hostscntmqttdeprecatedhash}{603} %
\DefineRemark{hostscntmqttshortkey}{344} %
\DefineRemark{hostscntprivatecaexpired}{3375}
\DefineRemark{hostscnttop3reusedcerts}{6858} %
\DefineRemark{hostspctamqpacmeofcasigned}{8.7}
\DefineRemark{hostspctamqpcasignedexpired}{1.8} %
\DefineRemark{hostspctamqpcertmd5privateca}{2.2} %
\DefineRemark{hostspctamqpdeprecatedhashprivateca}{17} %
\DefineRemark{hostspctamqpexpired}{3.0} %
\DefineRemark{hostspctamqpnoncasignedweakcertissued2021}{4.0}
\DefineRemark{hostspctamqpprivatecaexpired}{20} %
\DefineRemark{hostspctamqpselfsignedexpired}{5.9} %
\DefineRemark{hostspctamqpshortkeyprivateca}{3.3} %
\DefineRemark{hostspctamqpshortkeyselfsigned}{11} %
\DefineRemark{hostspctcasignedfollowcab}{100} %
\DefineRemark{hostspctcoapcasignedexpired}{38} %
\DefineRemark{hostspctcoapexpired}{21} %
\DefineRemark{hostspctcoapprivatecaexpired}{10} %
\DefineRemark{hostspctcoapselfsigned3yexpired}{100} %
\DefineRemark{hostspctcoapselfsignedexpired}{50} %
\DefineRemark{hostspctdnp3casignedexpired}{100} %
\DefineRemark{hostspctdnp3privatecaexpired}{100} %
\DefineRemark{hostspctenipselfsigned3yexpired}{13} %
\DefineRemark{hostspctenipselfsignedexpired}{13} %
\DefineRemark{hostspctfoxcasignedexpired}{29} %
\DefineRemark{hostspctfoxprivatecaexpired}{1.6} %
\DefineRemark{hostspcthttpcasignedexpired}{29} %
\DefineRemark{hostspcthttpprivatecaexpired}{0.0} %
\DefineRemark{hostspcthttpselfsignedexpired}{0.0} %
\DefineRemark{hostspctiec104casignedexpired}{100} %
\DefineRemark{hostspctmaxcasignedexpired}{38}
\DefineRemark{hostspctmaxprivatecaexpired}{21}
\DefineRemark{hostspctmincasignedexpired}{1.8}
\DefineRemark{hostspctminprivatecaexpired}{0.0}
\DefineRemark{hostspctmqttacmeofcasigned}{45}
\DefineRemark{hostspctmqttcasignedexpired}{13} %
\DefineRemark{hostspctmqttcertmd5privateca}{0.0} %
\DefineRemark{hostspctmqttdeprecatedhashprivateca}{2.9} %
\DefineRemark{hostspctmqttnoncasignedweakcertissued2021}{11}
\DefineRemark{hostspctmqttprivatecaexpired}{21} %
\DefineRemark{hostspctmqttselfsigned3yexpired}{11} %
\DefineRemark{hostspctmqttselfsignedexpired}{8.9} %
\DefineRemark{hostspctmqttshortkeyprivateca}{1.4} %
\DefineRemark{hostspctmqttshortkeyselfsigned}{3.6} %
\DefineRemark{hostspctnoncasignedweak}{5.2}
\DefineRemark{hostspctprivatecadisregardcab}{60} %
\DefineRemark{hostspctretrofitweakcertissued2021}{25}
\DefineRemark{hostspctsiemensselfsignedexpired}{0.0} %
\DefineRemark{maxascntinterasloadbalancing}{5} %
\DefineRemark{protomaxcasignedexpired}{CoAP}
\DefineRemark{protomaxprivatecaexpired}{MQTT}
\DefineRemark{protomincasignedexpired}{AMQP}
\DefineRemark{protominprivatecaexpired}{Fox Platform}

\DefineRemark{emailcntmqttpayload}{5092} %
\DefineRemark{hostscntemailpayload}{372} %

\begin{document}

\title[Measuring the Untapped TLS Support in the Industrial Internet of Things]{Missed Opportunities: Measuring the Untapped TLS Support\\in the Industrial Internet of Things}

\author{Markus Dahlmanns\(^*\), Johannes Lohmöller\(^*\), Jan Pennekamp\(^*\),\\ Jörn Bodenhausen\(^*\), Klaus Wehrle\(^*\), Martin Henze\(^{\S,\ddagger}\)}
\def\cleanauthors{Markus Dahlmanns, Johannes Lohmöller, Jan Pennekamp, Jörn Bodenhausen, Klaus Wehrle, Martin Henze}
\affiliation{
\(^*\)\textit{Communication and Distributed Systems}, RWTH Aachen University \city{Aachen} \country{Germany} \\
\(^\S\)\textit{Security and Privacy in Industrial Cooperation}, RWTH Aachen University \city{Aachen} \country{Germany} \\
\(^\ddagger\)\textit{Cyber Analysis \& Defense}, Fraunhofer FKIE \city{Wachtberg} \country{Germany}\\
\{dahlmanns, lohmoeller, pennekamp, bodenhausen, wehrle\}@comsys.rwth-aachen.de \(\cdot\)
henze@cs.rwth-aachen.de
}

\renewcommand{\shortauthors}{Dahlmanns, et al.}

\begin{abstract}
The ongoing trend to move industrial appliances from previously isolated networks to the Internet requires fundamental changes in security to uphold secure and safe operation.
Consequently, to ensure end-to-end secure communication and authentication,
(i)~traditional industrial protocols, e.g., Modbus, are retrofitted with TLS support, and
(ii)~modern protocols, e.g., MQTT, are directly designed to use TLS.
To understand whether these changes indeed lead to secure Industrial Internet of Things deployments, i.e., using TLS-based protocols, which are configured according to security best practices, we perform an Internet-wide security assessment of ten industrial protocols covering the complete IPv4 address space.

Our results show that both, retrofitted existing protocols and newly developed secure alternatives, are barely noticeable in the wild.
While we find that new protocols have a higher TLS adoption rate than traditional protocols~(\SI{\Remark{hostspcttlssecurebydesignprotocol}}{\percent} vs.\ \SI{\Remark{hostspcttlsretrofitprotocol}}{\percent}), the overall adoption of TLS is comparably low (\SI{\Remark{hostspcttls}}{\percent} of hosts).
Thus, most industrial deployments (\num{\Remark{hostscntdefault}} hosts) are insecurely connected to the Internet. %
Furthermore, we identify that \SI{\Remark{hostspcttlsweak}}{\percent} of hosts with TLS support (\num{\Remark{hostscnttlsweak}} hosts) show security deficits, e.g., missing access control.
Finally, we show that support in configuring systems securely, e.g., via configuration templates, is promising to strengthen security.
\end{abstract}



\maketitle

\section{Introduction}

Traditionally, industrial networks, i.e., networks used for process and factory automation, were explicitly designed to be isolated from other computer networks~\cite{galloway2013,cheminod-industrialsecurityissues-2013,mirian-icsmes-2016}.
Consequently, industrial protocols, e.g., Modbus, were specified without any security measures, e.g., encryption, authentication, or integrity protection~\cite{dnp3-spec,ethernetip-spec,MODBUSMe20:online}.
However, similar to paradigms like the Internet of Things~(IoT) which increasingly interconnects consumer hardware, also industrial networks converge with other networks resulting in an Industrial Internet of Things~(IIoT)~\cite{serror2021iiot, sadeghi-secprivchallengesiiot-2015, BOYES20181}.
While this convergence enables a plethora of novel functionality~\cite{KHAN2020106522, lade2017, serror2021iiot,pennekamp-towardsiop-2019}, it also significantly increases the number of attack vectors and thus demands tighter security requirements~\cite{sadeghi-secprivchallengesiiot-2015}, especially end-to-end secure communication and authentication.
Most importantly, traditional (insecure) industrial protocols without any security measures must not be used anymore, especially for communication via the Internet.

Indeed, related work confirms the availability of such traditional and insecure protocols on the Internet:
While more than \num{60000}~publicly accessible systems use traditional protocols~\cite{mirian-icsmes-2016} to provide their services, the protocols are also heavily used by Industrial Control Systems for communication via the Internet~\cite{nawrocki-passics-20}.
Communication between such unsecured industrial endpoints is not protected against attacks such as eavesdropping or message alternation, allowing attackers to retrieve confidential business information or even control production lines~\cite{hemsley-icsattacks-2018,quarta2017experimental,stellios2018survey,humayed-survey-2017,cheminod-industrialsecurityissues-2013}.
Such attacks can have a significant impact on the environment, machinery, or even human life, as evidenced by NotPetya and other attacks~\cite{hemsley-icsattacks-2018,stellios2018survey}, as well as are expected to be more frequent in the future~\cite{dodson2020plcmirai}.

As a countermeasure, traditional protocols have been retrofitted~(as of 2013) to account for these new security requirements:
They now rely on TLS to end-to-end protect the communication and allow for client authentication as well as access control~\cite{MBTCPSec70:online, ieee1815-2010-spec, iec60870-5-104:2000, iec62351-3, cipsecurity-spec}.
Complementing these efforts, modern protocols such as AMQP, MQTT, and CoAP, specifically targeting IIoT communication but also commonly used in the IoT, have been explicitly designed to provide security via TLS~\cite{mqtt-spec, RFC7252}.
However, to capitalize on their promised security benefits, these protocols need to be deployed on Internet-facing appliances and configured securely, e.g., w.r.t.\ used cryptographic ciphers and authentication methods~\cite{rfc7525}.

While a first Internet-wide analysis of the secure-by-design industrial protocol OPC~UA identified thousands of \emph{theoretically} secure Internet-facing deployments, a surprisingly high number of \SI{92}{\percent} exhibited \emph{practical} security issues, mostly due to configuration deficits~\cite{2020-dahlmanns-imc-opcua}.
Given that TLS is the de-facto standard for secure communication, e.g., on the Web, the question arises whether \mbox{similar} issues prevail for TLS-based (I)IoT protocols.
As little is known whether these retrofitted and modern (I)IoT protocols relying on TLS for security are currently deployed in the wild and, if so, configured securely, we set out to shed light on these issues.

In this paper, we thus study to which extent (I)IoT deployments rely on TLS-based protocol variants to end-to-end secure their communication and whether they are configured securely, i.e., we look into today's attack surface.
To this end, we actively scan the entire IPv4 address space~($\sim$\num{3.7}~billion IP addresses) on \num{20}~ports to identify TLS-enabled (I)IoT deployments, assess their configuration relying on well-known best practices~\cite{rfc7525,cabforum,bsi-ics-security,nistsp800-tls,tr2102-2}, and derive guidelines to increase the security of future deployments.

\noindent\textbf{Contributions:}
Our main contributions are as follows.
\begin{itemize}[noitemsep,topsep=0pt,leftmargin=9pt]
    \item We identify ten TLS-based (I)IoT protocols and scan the entire IPv4 address space to quantify TLS adoption among these protocols.
    While indicating a higher TLS adoption rate of \SI{\Remark{hostspcttlssecurebydesignprotocol}}{\percent} on hosts relying on secure-by-design protocols, our results show that TLS adoption on hosts employing protocols offering a retrofitted variant is comparatively low (only \SI{\Remark{hostspcttlsretrofitprotocol}}{\percent} offer TLS support).
    Thus, the majority of \num{\Remark{hostscnt}} found (I)IoT deployments remain insecure and do not benefit from the security offered by TLS.
    \item For TLS-enabled systems, we assess whether their configuration in practice indeed results in a sufficient security level.
    To this end, we analyze whether the found deployments adhere to common TLS configuration guidelines~\cite{rfc7525,nistsp800-tls,tr2102-2}.
    Here, \SI{\Remark{hostspcttlsweak}}{\percent} of TLS-enabled deployments suffer from configuration issues that impact security, ranging from deprecated TLS versions over selection of insecure ciphers to massive certificate reuse.
    \item To foster open science and allow for reproducibility of our results, we release our implementations of TLS-based industrial protocols for \texttt{zgrab2}~\cite{COMSYSzg99:online} and our anonymized dataset~\cite{dahlmanns-tlsdataset-2021}.
\end{itemize}

\section{Secure Industrial Communication}
\label{sec:background}

Automated industrial production relies on communication protocols enabling plant components to exchange control commands and sensor values~\cite{galloway2013,MODBUSMe20:online}.
In this section, we discuss the evolution of industrial protocols for isolated networks to secure (I)IoT protocols.
While these protocols, in theory, are able to provide confidential, integrity-protected, and authenticated communication, we deliberate that the actual security level depends on their configuration.

\textbf{Evolution of Industrial Communication:}
Relying on bus communication, e.g., via RS-485, first industrial networks were designed for isolated communication where, e.g., SCADA terminals communicated with local Programmable Logic Controllers~(PLCs) to control production processes and visualize their status~\cite{galloway2013}.
To communicate via these buses, vendors of industrial devices each specified their own protocol, e.g., Siemens developed S7, Allen-Bradley~(Rockwell Automation) formalized EtherNet/IP~\cite{ethernetip-spec}, and Modicon~(Schneider Electric) established Modbus~\cite{modbusrtu-spec}.
We provide a concise overview of relevant industrial communication protocols and their respective evolution over time in Table~\ref{tab:protocol-evolution}.

Even though these protocols differ among other things in terms of packet layout, they share the same core features: reading and writing values of plant components~\cite{MODBUSMe20:online, iec60870-5-104:2000, ethernetip-spec, wolsing2021ipal}, e.g., to get sensor readings or control processes.
To simplify the communication with office networks managing productions, vendors ported their protocols for usage via Ethernet, Internet Protocol~(IP), and in most cases TCP~(cf.\ Table~\ref{tab:protocol-evolution}).
Hence, PLCs nowadays act as servers allowing clients to read and write values or commands.
We refer to these \emph{client/server~(C/S)}-based protocols as \emph{traditional protocols}.

\begin{table}[!t]
\footnotesize
\raggedleft

\begin{tabular}{rl@{\hspace{2\tabcolsep}}c@{\hspace{2\tabcolsep}}ccc}
& \textbf{Protocol} & \textbf{Type} & \textbf{Bus} & \textbf{IP} & \textbf{TLS}\\

\addlinespace[-\aboverulesep]
\cmidrule[\heavyrulewidth]{2-6}

\tikzmark{m}\multirow{6}{2.5pt}{\rotatebox[origin=c]{90}{\emph{Traditional}}} &
Modbus &
C/S & 1979~\cite{MODBUSMe20:online} &
2006~\cite{MODBUSMe20:online} &
2018~\cite{MBTCPSec70:online}\\

&
\cellcolor[HTML]{C0C0C0}DNP3 &
\cellcolor[HTML]{C0C0C0}C/S &
\cellcolor[HTML]{C0C0C0}1993~\cite{dnp3-spec} &
\multicolumn{2}{c}{\cellcolor[HTML]{C0C0C0}2014~\cite{ieee1815-2010-spec}} \\

&
IEC 104 &
C/S &
1995\(^*\)~\cite{iec60870-5-101:1995} &
2000~\cite{iec60870-5-104:2000} &
2014~\cite{iec62351-3} \\

&
\cellcolor[HTML]{C0C0C0}EtherNet/IP &
\cellcolor[HTML]{C0C0C0}C/S &
\cellcolor[HTML]{C0C0C0}-- &
\cellcolor[HTML]{C0C0C0}2001~\cite{ethernetip-spec} &
\cellcolor[HTML]{C0C0C0}2015~\cite{cipsecurity-spec}\\

&
Siemens S7 &
C/S &
\(\leq\) 2004\(^+\) &
\(\leq\) 2004\(^{+,\dagger}\)~\cite{RFC1006} &
2016\(^+\)\\

&
\cellcolor[HTML]{C0C0C0}Tridium Fox &
\cellcolor[HTML]{C0C0C0}C/S &
\cellcolor[HTML]{C0C0C0}-- &
\cellcolor[HTML]{C0C0C0}\(\leq\) 2004\(^+\) &
\cellcolor[HTML]{C0C0C0}\(\leq\) 2013\(^+\)\\

\cmidrule[\lightrulewidth]{2-6}
\addlinespace[-2pt]

\tikzmark{n}\multirow{4}{2.5pt}{\rotatebox[origin=c]{90}{\emph{Modern}}} &
AMQP &
PubSub &
-- &
2003\(^+\) &
2007\(^+\)\\

&
\cellcolor[HTML]{C0C0C0}OPC UA &
\cellcolor[HTML]{C0C0C0}C/S \&
\cellcolor[HTML]{C0C0C0}PubSub &
\cellcolor[HTML]{C0C0C0}-- &
\multicolumn{2}{c}{\cellcolor[HTML]{C0C0C0}2006~\cite{opcua-concepts-2006,opcua-mappings-2006}}\\

&
MQTT &
PubSub &
-- &
\multicolumn{2}{c}{2013~\cite{mqtt-spec}}\\

&
\cellcolor[HTML]{C0C0C0}CoAP\(^\ddagger\) &
\cellcolor[HTML]{C0C0C0}C/S &
\cellcolor[HTML]{C0C0C0}-- &
\multicolumn{2}{c}{\cellcolor[HTML]{C0C0C0}2014~\cite{RFC7252}}\\

\cmidrule[\heavyrulewidth]{2-6}
\addlinespace[-\belowrulesep]
\end{tabular}

\qquad
\scriptsize
\(^*\) IEC 101. \,
\(^+\) No official specification available. \,
\(^\dagger\) ISO-on-TCP. \,
\(^\ddagger\) UDP/DTLS-based.
\begin{tikzpicture}[remember picture,overlay]
   \draw[arrows=->,line width=0.75pt]  ($(pic cs:m) + (-2pt,3pt)$) -- ($(pic cs:n) + (-2pt,-25pt)$) node [rotate=180, midway, above, sloped] (TextNode) {\emph{Time}};
\end{tikzpicture}
\caption{%
  Types (Client/Server (C/S) \& Publish/Subscribe (PubSub)) and specification years of industrial protocol variants.
}
\vspace{-3.7em}
\label{tab:protocol-evolution}
\end{table}

Recently, a string of \emph{modern protocols} surfaced focusing on a small resource footprint~(CoAP)~\cite{RFC7252} or implementing the modern \emph{Publish/Subscribe~(PubSub)} communication pattern~(AMQP, MQTT)~\cite{amqp-spec,mqtt-spec}.
In the latter, sending and receiving entities connect to a central broker instance, facilitating many-to-many communication by distributing received data to all clients that previously subscribed to a specific topic.
Implementing both of these communication concepts, OPC~UA~\cite{opcua-concepts-2006, opcua-pubsub-2017} accounts for diverse needs.

More importantly, to address the need for security, especially in Internet-facing deployments, OPC~UA provides attested~\cite{bsi-opcua-analysis} security measures by design~\cite{opcua-profiles-2017}, using specialized security paradigms but also with a TLS-based variant.
While modern protocols were designed with security in mind to account for security and safety needs of modern (I)IoT deployments, traditional protocols underwent a retrofitting process.
Since 2013, the specifications of traditional protocols have been updated to enable TLS communication.

As secure communication is an inherent building block for the safe operation of (I)IoT appliances to prevent attacks leading to production outages or harm to humans, the question arises whether these variants arrived in practice at Internet-reachable deployments.

\textbf{Pitfalls in Secure Communication:}
TLS, the predominantly used security protocol on the Internet, and DTLS provide confidential, integrity-protected, and authentic communication for TCP and UDP connections~\cite{rfc5246, rfc6347}, respectively (we use (D)TLS as an abbreviation for TLS or DTLS in the remainder of this paper).
However, the achieved level of communication security depends on the (D)TLS \emph{version} and \emph{cipher suite} negotiated during the handshake.

Since specific (D)TLS versions and cryptographic primitives included in cipher suites lose their security promises from time to time, %
servers should not rely on deprecated versions and should not offer cipher suites with problematic security primitives~\cite{rfc7525}.
Hence, operators are responsible to keep the configuration of their servers up to date, which is especially challenging with long-life industrial hardware, but essential to uphold security and safety guarantees.

Apart from confidentiality and integrity protection, authentication is an inherent part of secure communication.
The exchange of certificates allows both server and client to deny connections for unwanted identities ensuring a reliable access control~\cite{MODBUSMe20:online, iec60870-5-104:2000, ethernetip-spec}.
While self-signed certificates force operators to manually safelist certificates, %
CA-signed certificates simplify the key management and include a public entity in the issuing process~(whenever a public CA is used).
The use of CAs is common practice on the Web.

However, cryptographic primitives within issued certificates must be kept updated to retain their security promises~\cite{tr2102-2}.
Hence, to ensure secure deployments and prevent impersonation attacks, operators need to replace certificates relying on insecure primitives.

\textit{\textbf{Takeaway:}
With the evolving communication needs in (I)IoT deployments, protocols increasingly adopt security and authentication best practices from the Web.
Protocols are either retrofitted with TLS support or secured communication by design.
However, available configuration options influence the security of ((I)IoT) appliances.
Thus, investigating whether deployments are really secure and conform to today's security recommendations is important.
}

\section{Related Work}
\label{sec:relatedwork}
Our Internet-wide assessment of the security configuration of TLS-secured industrial deployments is motivated by related work on the security of industrial deployments, assessments of the security of Internet-reachable web services, and the combination of both.

\textbf{Security of Industrial Deployments:}
Although the security issues in industrial deployments are well understood~\cite{hemsley-icsattacks-2018,quarta2017experimental,stellios2018survey,humayed-survey-2017,cheminod-industrialsecurityissues-2013, sadeghi-secprivchallengesiiot-2015}, and security incidents can be catastrophic~\cite{nawrocki-passics-20}, to date, they occur comparably seldom~\cite{miller2012survey}.
Still, deployments are often insecure:
Traditional industrial protocols are heavily used for unprotected communication via the Internet~\cite{nawrocki-passics-20}.
Additionally, more than \num{60000}~protocol deployments are connected to the Internet~\cite{mirian-icsmes-2016}, identified as real PLCs~\cite{feng-characterizing-2016} and robots~\cite{demarinis-rosscanning-2019} not restricting access, enabling everyone on the Internet to control them.
Furthermore, the number of exposed deployments increases continuously~\cite{xu2018increase}.

Different Internet scan services, e.g., Shodan~\cite{shodan} or Censys~\cite{durumeric2015search}, collect and share meta-information on Internet-facing services, including industrial deployments~\cite{leverett-shodanclassification-2011,hansson-analysisshodan-2018}, often listed quickly after deployment~\cite{bodenheim-shodaneval-2014}, but by far do not see all deployments communicating via the Internet~\cite{barbieri2021assessing}.
Still, such meta-information serves as the basis for security assessments~\cite{roepert-opcuaassessment-2020} of traditional industrial services, finding many affected by known vulnerabilities~\cite{ceron2020online, kiravuo-vulnerabilitiesfinland-2015,genge-shovat-2016}.

Other researchers~\cite{fachkha-cpsprobing-2017,husak-cpscyberawareness-2018,cabana-detectics-attacks-2019} collect their data via active Internet measurement tools, e.g., ZMap~\cite{durumeric-zmap-2013}.
While such tools typically do not interfere with the operation of industrial equipment~\cite{coffey-scanninganalysis-2018}, malicious attacks can.
As researchers expect that the number of attacks against Internet-connected deployments increases~\cite{dodson2020plcmirai}, %
and exploits allow enslaving PLCs as network proxy to internal networks using their native programming language~\cite{klick2015plcbackdoor}, securing these Internet-facing deployments is of utmost importance.
Thus, the need to rely on secure (I)IoT protocols is unquestionable.
However, until now, an analysis on deployments relying on these secure (I)IoT protocols and whether they are configured securely is still missing.

\textbf{Security-related Internet Measurements:}
The configuration of security protocols was subject to research basing on active and passive Internet measurements.
Especially TLS and PKI usage were in the scope of these works:
The TLS and certificate configuration of communication services~\cite{holz-tlscomm-2016} and Internet-facing embedded devices~\cite{cui-insecureembedded-2010} showed deficits.
Furthermore, related work showed that security flaws in key generation~\cite{heninger-mining-2012}, fails in Diffie--Hellman~\cite{adrian15dhfails} and TLS implementations~\cite{springall-tlsshortcuts-2016} are widespread, as well as analyzed the shift of HTTP servers to TLS~1.3~\cite{2020-holz-tls13}.
Focusing on certificates of HTTP hosts, different works investigate their configuration~\cite{holz-509pki-2011}, validity~\cite{chung-certificates-2016}, wrong issuance~\cite{kumar-certificates-2018}, prevalence in certificate \mbox{transparency} logs~\cite{gasser-ctpractices-2018}, and insecurities induced by cross-signing~\cite{hiller20crosssigning}.

With a slightly different focus, Internet measurements also cover SSH configurations~\cite{gasser-ssh-2014}, cloud usage, and communication security of IoT~devices~\cite{ren-informationexposure-2019}, as well as compromised IoT~devices~\cite{shaikh-correlating-2018,neshenko-survey-2019,mangino-insecurity-2020}.

\textbf{Internet-wide Industrial Security Assessments:}
While the security of deployments in other areas is often widely assessed, the configuration of industrial deployments relying on security features was rarely covered by research.
Focusing on the secure-by-design OPC~UA protocol, related work showed that operators frequently fail to configure deployments securely~\cite{2020-dahlmanns-imc-opcua}, possibly due to constrained protocol implementations~\cite{erba2021practical}, and presented mechanisms for a more secure device provisioning~\cite{kohnhauser21opcuaprovisioning}.

\textit{\textbf{Takeaway:}
While security on the Web increases, Internet measurements have shown that many industrial deployments are connected to the Internet insecurely due to usage of insecure protocols or misconfiguration.
However, works assessing industrial appliances left out TLS-based deployments, leaving open whether they improve security.
}

\section{Methodology and Dataset}
\label{sec:methodology}

To identify potential attack vectors of (I)IoT deployments relying on TLS to theoretically secure their communication, we select active Internet measurements---a suitable tool to assess such deployments---to study the security of services relying on retrofitted and modern TLS-based (I)IoT protocols.
We describe our measurement methodology %
in Section~\ref{subsec:measurements} and our resulting dataset in Section~\ref{subsec:dataset}.

\subsection{Actively Finding Industrial Deployments}
\label{subsec:measurements}

Prior to our Internet scans, we first surveyed relevant protocols.

\begin{table*}[!ht]
\footnotesize
\centering
\vspace{0.5em}
\setlength{\tabcolsep}{3.5pt}

\begin{tabular}{cccccccccclccccc}
\cline{1-3} \cline{5-10} \cline{12-16}
\multicolumn{3}{c}{\textit{Section~\ref{subsec:protocolselection} (Selected Protocols)}} &
  \textit{} &
  \multicolumn{6}{c}{\textit{Section~\ref{subsec:validating} (Validation Process)}} &
   &
  \multicolumn{5}{c}{\textit{Section~\ref{subsec:deployments} (Deployments)}} \\ \cline{1-10} \cline{12-16} 
\multirow{2}{*}{\textbf{Protocol}} &
  \multirow{2}{*}{\textbf{Port}} &
  \multirow{2}{*}{\textbf{\begin{tabular}[c]{@{}c@{}}Variant\\ (Port iff diff.)\end{tabular}}} &
  \multirow{2}{*}{\textbf{\begin{tabular}[c]{@{}c@{}}Date\\ (2021)\end{tabular}}} &
  \multirow{2}{*}{\textbf{Hosts}} &
  \multirow{2}{*}{\textbf{Transport}} &
  \multicolumn{3}{c}{\textbf{(D)TLS}} &
  \multirow{2}{*}{\textbf{Valid}} &
   &
  \multirow{2}{*}{\textbf{\begin{tabular}[c]{@{}c@{}}Total\\ (opt. TLS)\end{tabular}}} &
  \multirow{2}{*}{\textbf{\begin{tabular}[c]{@{}c@{}}\%\\ (D)TLS\end{tabular}}} &
   &
  \multicolumn{2}{c}{\textbf{Distinct}} \\
 &
   &
   &
   &
   &
   &
  \textbf{Valid} &
  \textbf{Auth. OK} &
  \textbf{Succ.} &
   &
   &
   &
   &
   &
  \textbf{ASes} &
  \textbf{Cert. CNs} \\ \cline{1-10} \cline{12-13} \cline{15-16}
\multirow{3}{*}{Modbus} &
  \multirow{2}{*}{502\(^\dagger\)} &
  Standard &
  \multirow{2}{*}{\Remark{502502modbusdate}} &
  \multirow{2}{*}{\SI{\Remark{502502modbustotalhosts}}{}} &
  \multirow{2}{*}{\SI{\Remark{502502modbustotalconnsuccess}}{} (\SI{\Remark{502502modbuspctconnsuccess}}{\percent})} &
  \multicolumn{3}{c}{---} &
  \SI{\Remark{502502modbustotalsuccess}}{} &
   &
  \SI{\Remark{502502modbustotalsuccess}}{} &
  $\rceil$ &
   &
  \SI{\Remark{502502modbustotalas}}{} &
  --- \\ \cline{12-12} \cline{15-16}
 &
   &
  TLS &
   &
   &
   &
  \SI{\Remark{502502tlsmodbustotaltls}}{} (\SI{\Remark{502502tlsmodbuspcttlsvalidoftransport}}{\percent}) &
  \SI{\Remark{502502tlsmodbustotaltlsauth}}{} &
  \SI{\Remark{502502tlsmodbustotaltlssuccess}}{} &
  \SI{\Remark{502502tlsmodbustotalsuccess}}{} &
  \multirow{2}{*}{} &
  \multirow{2}{*}{\begin{tabular}[c]{@{}c@{}}\SI{\Remark{alltlsmodbustotalsuccess}}{}\\ (\SI{\Remark{optionaltlsmodbustotalsuccess}}{})\end{tabular}} &
  \SI{\Remark{allmodbuspctvalidtls}}{\percent} &
   &
  \multirow{2}{*}{\SI{\Remark{alltlsmodbustotalas}}{}} &
  \multirow{2}{*}{\SI{\Remark{alltlsmodbustotalcertcns}}{}} \\ \cline{2-10} 
 &
  802\(^\ddagger\) &
  TLS &
  \Remark{802802tlsmodbusdate} &
  \SI{\Remark{802802tlsmodbustotalhosts}}{} &
  \SI{\Remark{802802tlsmodbustotalconnsuccess}}{} (\SI{\Remark{802802tlsmodbuspctconnsuccess}}{\percent}) &
  \SI{\Remark{802802tlsmodbustotaltls}}{} (\SI{\Remark{802802tlsmodbuspcttlsvalidoftransport}}{\percent}) &
  \SI{\Remark{802802tlsmodbustotaltlsauth}}{} &
  \SI{\Remark{802802tlsmodbustotaltlssuccess}}{} &
  \SI{\Remark{802802tlsmodbustotalsuccess}}{} &
   &
   &
  $\rfloor$ &
   &
   &
  \\ \cline{1-10} \cline{12-13} \cline{15-16} 
\multirow{3}{*}{DNP3} &
  \multirow{2}{*}{20000\(^\dagger\)} &
  Standard &
  \multirow{2}{*}{\Remark{2000020000dnp3date}} &
  \multirow{2}{*}{\SI{\Remark{2000020000dnp3totalhosts}}{}} &
  \multirow{2}{*}{\SI{\Remark{2000020000dnp3totalconnsuccess}}{} (\SI{\Remark{2000020000dnp3pctconnsuccess}}{\percent})} &
  \multicolumn{3}{c}{---} &
  \SI{\Remark{2000020000dnp3totalsuccess}}{} &
   &
  \SI{\Remark{2000020000dnp3totalsuccess}}{} &
  $\rceil$ &
   &
  \SI{\Remark{2000020000dnp3totalas}}{} &
  --- \\ \cline{12-12} \cline{15-16}
 &
   &
  TLS &
   &
   &
   &
  \SI{\Remark{2000020000tlsdnp3totaltls}}{} (\SI{\Remark{2000020000tlsdnp3pcttlsvalidoftransport}}{\percent}) &
  \SI{\Remark{2000020000tlsdnp3totaltlsauth}}{} &
  \SI{\Remark{2000020000tlsdnp3totaltlssuccess}}{} &
  \SI{\Remark{2000020000tlsdnp3totalsuccess}}{} &
  \multirow{2}{*}{} &
  \multirow{2}{*}{\begin{tabular}[c]{@{}c@{}}\SI{\Remark{alltlsdnp3totalsuccess}}{}\\ (\SI{\Remark{optionaltlsdnp3totalsuccess}}{})\end{tabular}} &
  \SI{\Remark{alldnp3pctvalidtls}}{\percent} &
   &
  \multirow{2}{*}{\SI{\Remark{alltlsdnp3totalas}}{}} &
  \multirow{2}{*}{\SI{\Remark{alltlsdnp3totalcertcns}}{}} \\ \cline{2-10} 
 &
  19999\(^\ddagger\) &
  TLS &
  \Remark{1999919999tlsdnp3date} &
  \SI{\Remark{1999919999tlsdnp3totalhosts}}{} &
  \SI{\Remark{1999919999tlsdnp3totalconnsuccess}}{} (\SI{\Remark{1999919999tlsdnp3pctconnsuccess}}{\percent}) &
  \SI{\Remark{1999919999tlsdnp3totaltls}}{} (\SI{\Remark{1999919999tlsdnp3pcttlsvalidoftransport}}{\percent}) &
  \SI{\Remark{1999919999tlsdnp3totaltlsauth}}{} &
  \SI{\Remark{1999919999tlsdnp3totaltlssuccess}}{} &
  \SI{\Remark{1999919999tlsdnp3totalsuccess}}{} &
   &
   &
  $\rfloor$ &
   &
   &
  \\ \cline{1-10} \cline{12-13} \cline{15-16} 
\multirow{3}{*}{IEC 104} &
  \multirow{2}{*}{2404\(^\dagger\)} &
  Standard &
  \multirow{2}{*}{\Remark{24042404iec104date}} &
  \multirow{2}{*}{\SI{\Remark{24042404iec104totalhosts}}{}} &
  \multirow{2}{*}{\SI{\Remark{24042404iec104totalconnsuccess}}{} (\SI{\Remark{24042404iec104pctconnsuccess}}{\percent})} &
  \multicolumn{3}{c}{---} &
  \SI{\Remark{24042404iec104totalsuccess}}{} &
   &
  \SI{\Remark{24042404iec104totalsuccess}}{} &
  $\rceil$ &
   &
  \SI{\Remark{24042404iec104totalas}}{} &
  --- \\ \cline{12-12} \cline{15-16}
 &
   &
  TLS &
   &
   &
   &
  \SI{\Remark{24042404tlsiec104totaltls}}{} (\SI{\Remark{24042404tlsiec104pcttlsvalidoftransport}}{\percent}) &
  \SI{\Remark{24042404tlsiec104totaltlsauth}}{} &
  \SI{\Remark{24042404tlsiec104totaltlssuccess}}{} &
  \SI{\Remark{24042404tlsiec104totalsuccess}}{} &
  \multirow{2}{*}{} &
  \multirow{2}{*}{\begin{tabular}[c]{@{}c@{}}\SI{\Remark{alltlsiec104totalsuccess}}{}\\ (\SI{\Remark{optionaltlsiec104totalsuccess}}{})\end{tabular}} &
  \SI{\Remark{alliec104pctvalidtls}}{\percent} &
   &
  \multirow{2}{*}{\SI{\Remark{alltlsiec104totalas}}{}} &
  \multirow{2}{*}{\SI{\Remark{alltlsiec104totalcertcns}}{}} \\ \cline{2-10}
 &
  19998\(^\ddagger\) &
  TLS &
  \Remark{1999819998tlsiec104date} &
  \SI{\Remark{1999819998tlsiec104totalhosts}}{} &
  \SI{\Remark{1999819998tlsiec104totalconnsuccess}}{} (\SI{\Remark{1999819998tlsiec104pctconnsuccess}}{\percent}) &
  \SI{\Remark{1999819998tlsiec104totaltls}}{} (\SI{\Remark{1999819998tlsiec104pcttlsvalidoftransport}}{\percent}) &
  \SI{\Remark{1999819998tlsiec104totaltlsauth}}{} &
  \SI{\Remark{1999819998tlsiec104totaltlssuccess}}{} &
  \SI{\Remark{1999819998tlsiec104totalsuccess}}{} &
   &
   &
  $\rfloor$ &
   &
   &
  \\ \cline{1-10} \cline{12-13} \cline{15-16} 
\multirow{3}{*}{\begin{tabular}[c]{@{}c@{}}EtherNet/IP\\ (EnIP)\end{tabular}} &
  \multirow{2}{*}{44818\(^\dagger\)} &
  Standard &
  \multirow{2}{*}{\Remark{4481844818enipdate}} &
  \multirow{2}{*}{\SI{\Remark{4481844818eniptotalhosts}}{}} &
  \multirow{2}{*}{\SI{\Remark{4481844818eniptotalconnsuccess}}{} (\SI{\Remark{4481844818enippctconnsuccess}}{\percent})} &
  \multicolumn{3}{c}{---} &
  \SI{\Remark{4481844818eniptotalsuccess}}{} &
   &
  \SI{\Remark{4481844818eniptotalsuccess}}{} &
  $\rceil$ &
   &
  \SI{\Remark{4481844818eniptotalas}}{} &
  --- \\ \cline{12-12} \cline{15-16}
 &
   &
  TLS &
   &
   &
   &
  \SI{\Remark{4481844818tlseniptotaltls}}{} (\SI{\Remark{4481844818tlsenippcttlsvalidoftransport}}{\percent}) &
  \SI{\Remark{4481844818tlseniptotaltlsauth}}{} &
  \SI{\Remark{4481844818tlseniptotaltlssuccess}}{} &
  \SI{\Remark{4481844818tlseniptotalsuccess}}{} &
  \multirow{2}{*}{} &
  \multirow{2}{*}{\begin{tabular}[c]{@{}c@{}}\SI{\Remark{alltlseniptotalsuccess}}{}\\ (\SI{\Remark{optionaltlseniptotalsuccess}}{})\end{tabular}} &
  \SI{\Remark{allenippctvalidtls}}{\percent} &
   &
  \multirow{2}{*}{\SI{\Remark{alltlseniptotalas}}{}} &
  \multirow{2}{*}{\SI{\Remark{alltlseniptotalcertcns}}{}} \\ \cline{2-10} 
 &
  2221\(^\ddagger\) &
  TLS &
  \Remark{22212221tlsenipdate} &
  \SI{\Remark{22212221tlseniptotalhosts}}{} &
  \SI{\Remark{22212221tlseniptotalconnsuccess}}{} (\SI{\Remark{22212221tlsenippctconnsuccess}}{\percent}) &
  \SI{\Remark{22212221tlseniptotaltls}}{} (\SI{\Remark{22212221tlsenippcttlsvalidoftransport}}{\percent}) &
  \SI{\Remark{22212221tlseniptotaltlsauth}}{} &
  \SI{\Remark{22212221tlseniptotaltlssuccess}}{} &
  \SI{\Remark{22212221tlseniptotalsuccess}}{} &
   &
   &
  $\rfloor$ &
   &
   &
  \\ \cline{1-10} \cline{12-13} \cline{15-16} 
\multirow{3}{*}{Siemens S7} &
  \multirow{2}{*}{102\(^\dagger\)} &
  Standard &
  \multirow{2}{*}{\Remark{102102siemensdate}} &
  \multirow{2}{*}{\SI{\Remark{102102siemenstotalhosts}}{}} &
  \multirow{2}{*}{\SI{\Remark{102102siemenstotalconnsuccess}}{} (\SI{\Remark{102102siemenspctconnsuccess}}{\percent})} &
  \multicolumn{3}{c}{---} &
  \SI{\Remark{102102siemenstotalsuccess}}{} &
   &
  \SI{\Remark{102102siemenstotalsuccess}}{} &
  $\rceil$ &
   &
  \SI{\Remark{102102siemenstotalas}}{} &
  --- \\ \cline{12-12} \cline{15-16}
 &
   &
  TLS &
   &
   &
   &
  \SI{\Remark{102102tlssiemenstotaltls}}{} (\SI{\Remark{102102tlssiemenspcttlsvalidoftransport}}{\percent}) &
  \SI{\Remark{102102tlssiemenstotaltlsauth}}{} &
  \SI{\Remark{102102tlssiemenstotaltlssuccess}}{} &
  \SI{\Remark{102102tlssiemenstotalsuccess}}{} &
  \multirow{2}{*}{} &
  \multirow{2}{*}{\begin{tabular}[c]{@{}c@{}}\SI{\Remark{alltlssiemenstotalsuccess}}{}\\ (\SI{\Remark{optionaltlssiemenstotalsuccess}}{})\end{tabular}} &
  \SI{\Remark{allsiemenspctvalidtls}}{\percent} &
   &
  \multirow{2}{*}{\SI{\Remark{alltlssiemenstotalas}}{}} &
  \multirow{2}{*}{\SI{\Remark{alltlssiemenstotalcertcns}}{}} \\ \cline{2-10} 
 &
  3782\(^\ddagger\) &
  TLS &
  \Remark{37823782tlssiemensdate} &
  \SI{\Remark{37823782tlssiemenstotalhosts}}{} &
  \SI{\Remark{37823782tlssiemenstotalconnsuccess}}{} (\SI{\Remark{37823782tlssiemenspctconnsuccess}}{\percent}) &
  \SI{\Remark{37823782tlssiemenstotaltls}}{} (\SI{\Remark{37823782tlssiemenspcttlsvalidoftransport}}{\percent}) &
  \SI{\Remark{37823782tlssiemenstotaltlsauth}}{} &
  \SI{\Remark{37823782tlssiemenstotaltlssuccess}}{} &
  \SI{\Remark{37823782tlssiemenstotalsuccess}}{} &
   &
   &
  $\rfloor$ &
   &
   &
  \\ \cline{1-10} \cline{12-13} \cline{15-16} 
\multicolumn{1}{l}{\multirow{8}{*}{\begin{tabular}[c]{@{}c@{}}Tridium Fox\\ (TF)\end{tabular}}} &
  \multirow{4}{*}{3011\(^\dagger\)} &
  Standard (1911) &
  \multirow{4}{*}{\Remark{30113011httpdate}} &
  \multirow{4}{*}{\SI{\Remark{30113011httptotalhosts}}{}} &
  \SI{\Remark{30111911foxtotalconnsuccess}}{} (\SI{\Remark{30111911foxpctconnsuccess}}{\percent}) &
  \multicolumn{3}{c}{---} &
  \SI{\Remark{30111911foxtotalsuccess}}{} &
  \multirow{8}{*}{} &
  Standard &
  $\rceil$ &
   &
   &
  \\
\multicolumn{1}{l}{} &
   &
  Platform &
   &
   &
  \SI{\Remark{30113011httptotalconnsuccess}}{} (\SI{\Remark{30113011httppctconnsuccess}}{\percent}) &
  \multicolumn{3}{c}{---} &
  \SI{\Remark{30113011httptotalsuccess}}{} &
   &
  \SI{\Remark{allfoxtotalsuccess}}{} &
  \multirow{2}{*}{\SI{\Remark{allfoxpctvalidtls}}{\percent}} &
   &
  \SI{\Remark{allfoxtotalas}}{} &
  --- \\ \cline{12-12} \cline{15-16}
\multicolumn{1}{l}{} &
   &
  TLS (4911) &
   &
   &
  \SI{\Remark{30114911tlsfoxtotalconnsuccess}}{} (\SI{\Remark{30114911tlsfoxpctconnsuccess}}{\percent}) &
  \SI{\Remark{30114911tlsfoxtotaltls}}{} (\SI{\Remark{30114911tlsfoxpcttlsvalidoftransport}}{\percent}) &
  \SI{\Remark{30114911tlsfoxtotaltlsauth}}{} &
  \SI{\Remark{30114911tlsfoxtotaltlssuccess}}{} &
  \SI{\Remark{30114911tlsfoxtotalsuccess}}{} &
   &
  TLS &
   &
   &
   &
  \\
\multicolumn{1}{l}{} &
   &
  Plat.-TLS (5011) &
   &
   &
  \SI{\Remark{30115011tlshttptotalconnsuccess}}{} (\SI{\Remark{30115011tlshttppctconnsuccess}}{\percent}) &
  \SI{\Remark{30115011tlshttptotaltls}}{} (\SI{\Remark{30115011tlshttppcttlsvalidoftransport}}{\percent}) &
  \SI{\Remark{30115011tlshttptotaltlsauth}}{} &
  \SI{\Remark{30115011tlshttptotaltlssuccess}}{} &
  \SI{\Remark{30115011tlshttptotalsuccess}}{} &
   &
  \SI{\Remark{alltlsfoxtotalsuccess}}{} (\SI{\Remark{optionaltlsfoxtotalsuccess}}{}) &
  $\rfloor$ &
   &
   \SI{\Remark{alltlsfoxtotalas}}{} &
   \SI{\Remark{alltlsfoxtotalcertcns}}{} \\ \cline{2-10} \cline{12-13} \cline{15-16} 
\multicolumn{1}{l}{} &
  \multirow{4}{*}{4911\(^\ddagger\)} &
  Standard (1911) &
  \multirow{4}{*}{\Remark{49114911tlsfoxdate}} &
  \multirow{4}{*}{\SI{\Remark{49114911tlsfoxtotalhosts}}{}} &
  \SI{\Remark{49111911foxtotalconnsuccess}}{} (\SI{\Remark{49111911foxpctconnsuccess}}{\percent}) &
  \multicolumn{3}{c}{---} &
  \SI{\Remark{49111911foxtotalsuccess}}{} &
   &
  Platform &
  $\rceil$ &
   &
   &
  \\
\multicolumn{1}{l}{} &
   &
  Platform (3011) &
   &
   &
  \SI{\Remark{49113011httptotalconnsuccess}}{} (\SI{\Remark{49113011httppctconnsuccess}}{\percent}) &
  \multicolumn{3}{c}{---} &
  \SI{\Remark{49113011httptotalsuccess}}{} &
   &
  \SI{\Remark{allhttptotalsuccess}}{} &
  \multirow{2}{*}{\SI{\Remark{allhttppctvalidtls}}{\percent}} &
   &
  \SI{\Remark{allhttptotalas}}{} &
  --- \\ \cline{12-12} \cline{15-16}
\multicolumn{1}{l}{} &
   &
  TLS &
   &
   &
  \SI{\Remark{49114911tlsfoxtotalconnsuccess}}{} (\SI{\Remark{49114911tlsfoxpctconnsuccess}}{\percent}) &
  \SI{\Remark{49114911tlsfoxtotaltls}}{} (\SI{\Remark{49114911tlsfoxpcttlsvalidoftransport}}{\percent}) &
  \SI{\Remark{49114911tlsfoxtotaltlsauth}}{} &
  \SI{\Remark{49114911tlsfoxtotaltlssuccess}}{} &
  \SI{\Remark{49114911tlsfoxtotalsuccess}}{} &
   &
  Plat.-TLS &
   &
   &
   &
  \\
\multicolumn{1}{l}{} &
   &
  Plat.-TLS (5011) &
   &
   &
  \SI{\Remark{49115011tlshttptotalconnsuccess}}{} (\SI{\Remark{49115011tlshttppctconnsuccess}}{\percent}) &
  \SI{\Remark{49115011tlshttptotaltls}}{} (\SI{\Remark{49115011tlshttppcttlsvalidoftransport}}{\percent}) &
  \SI{\Remark{49115011tlshttptotaltlsauth}}{} &
  \SI{\Remark{49115011tlshttptotaltlssuccess}}{} &
  \SI{\Remark{49115011tlshttptotalsuccess}}{} &
   &
  \SI{\Remark{alltlshttptotalsuccess}}{} (\SI{\Remark{optionaltlshttptotalsuccess}}{}) &
  $\rfloor$ &
   &
  \SI{\Remark{alltlshttptotalas}}{} &
  \SI{\Remark{alltlshttptotalcertcns}}{} \\ \cline{1-10} \cline{12-13} \cline{15-16} 
\multirow{3}{*}{AMQP} &
  \multirow{2}{*}{5672\(^\dagger\)} &
  Standard &
  \multirow{2}{*}{\Remark{56725672amqpdate}} &
  \multirow{2}{*}{\SI{\Remark{56725672amqptotalhosts}}{}} &
  \multirow{2}{*}{\SI{\Remark{56725672amqptotalconnsuccess}}{} (\SI{\Remark{56725672amqppctconnsuccess}}{\percent})} &
  \multicolumn{3}{c}{---} &
  \SI{\Remark{56725672amqptotalsuccess}}{} &
   &
  \SI{\Remark{56725672amqptotalsuccess}}{} &
  $\rceil$ &
   &
  \SI{\Remark{56725672amqptotalas}}{} &
  --- \\ \cline{12-12} \cline{15-16}
 &
   &
  TLS &
   &
   &
   &
  \SI{\Remark{56725672tlsamqptotaltls}}{} (\SI{\Remark{56725672tlsamqppcttlsvalidoftransport}}{\percent}) &
  \SI{\Remark{56725672tlsamqptotaltlsauth}}{} &
  \SI{\Remark{56725672tlsamqptotaltlssuccess}}{} &
  \SI{\Remark{56725672tlsamqptotalsuccess}}{} &
  \multirow{2}{*}{} &
  \multirow{2}{*}{\begin{tabular}[c]{@{}c@{}}\SI{\Remark{alltlsamqptotalsuccess}}{}\\ (\SI{\Remark{optionaltlsamqptotalsuccess}}{})\end{tabular}} &
  \SI{\Remark{allamqppctvalidtls}}{\percent} &
   &
  \multirow{2}{*}{\SI{\Remark{alltlsamqptotalas}}{}} &
  \multirow{2}{*}{\SI{\Remark{alltlsamqptotalcertcns}}{}} \\ \cline{2-10} 
 &
  5671\(^\ddagger\) &
  TLS &
  \Remark{56715671tlsamqpdate} &
  \SI{\Remark{56715671tlsamqptotalhosts}}{} &
  \SI{\Remark{56715671tlsamqptotalconnsuccess}}{} (\SI{\Remark{56715671tlsamqppctconnsuccess}}{\percent}) &
  \SI{\Remark{56715671tlsamqptotaltls}}{} (\SI{\Remark{56715671tlsamqppcttlsvalidoftransport}}{\percent}) &
  \SI{\Remark{56715671tlsamqptotaltlsauth}}{} &
  \SI{\Remark{56715671tlsamqptotaltlssuccess}}{} &
  \SI{\Remark{56715671tlsamqptotalsuccess}}{} &
   &
   &
  $\rfloor$ &
   &
   &
  \\ \cline{1-10} \cline{12-13} \cline{15-16} 
\multirow{3}{*}{OPC UA} &
  \multirow{2}{*}{4840\(^\dagger\)} &
  Binary &
  \multirow{2}{*}{\Remark{48404840opcuadate}} &
  \multirow{2}{*}{\SI{\Remark{48404840opcuatotalhosts}}{}} &
  \multirow{2}{*}{\SI{\Remark{48404840opcuatotalconnsuccess}}{} (\SI{\Remark{48404840opcuapctconnsuccess}}{\percent})} &
  \multicolumn{3}{c}{---} &
  \SI{\Remark{48404840opcuatotalsuccess}}{} &
   &
  \SI{\Remark{48404840opcuatotalsuccess}}{} &
  $\rceil$ &
   &
  \SI{\Remark{48404840opcuatotalas}}{} &
  --- \\ \cline{12-12} \cline{15-16}
 &
   &
  TLS &
   &
   &
   &
  \SI{\Remark{48404840tlsopcuatotaltls}}{} (\SI{\Remark{48404840tlsopcuapcttlsvalidoftransport}}{\percent}) &
  \SI{\Remark{48404840tlsopcuatotaltlsauth}}{} &
  \SI{\Remark{48404840tlsopcuatotaltlssuccess}}{} &
  \SI{\Remark{48404840tlsopcuatotalsuccess}}{} &
  \multirow{2}{*}{} &
  \multirow{2}{*}{\begin{tabular}[c]{@{}c@{}}\SI{\Remark{alltlsopcuatotalsuccess}}{}\\ (\SI{\Remark{optionaltlsopcuatotalsuccess}}{})\end{tabular}} &
  \SI{\Remark{allopcuapctvalidtls}}{\percent} &
   &
  \multirow{2}{*}{\SI{\Remark{alltlsopcuatotalas}}{}} &
  \multirow{2}{*}{\SI{\Remark{alltlsopcuatotalcertcns}}{}} \\ \cline{2-10} 
 &
  4843\(^\ddagger\) &
  TLS &
  \Remark{48434843tlshttpdate} &
  \SI{\Remark{48434843tlshttptotalhosts}}{} &
  \SI{\Remark{48434843tlshttptotalconnsuccess}}{} (\SI{\Remark{48434843tlshttppctconnsuccess}}{\percent}) &
  \SI{\Remark{48434843tlshttptotaltls}}{} (\SI{\Remark{48434843tlshttppcttlsvalidoftransport}}{\percent}) &
  \SI{\Remark{48434843tlshttptotaltlsauth}}{} &
  \SI{\Remark{48434843tlshttptotaltlssuccess}}{} &
  \SI{\Remark{48434843tlshttptotalsuccess}}{} &
   &
   &
  $\rfloor$ &
   &
   &
  \\ \cline{1-10} \cline{12-13} \cline{15-16} 
\multirow{3}{*}{MQTT} &
  \multirow{2}{*}{1883\(^\dagger\)} &
  Standard &
  \multirow{2}{*}{\Remark{18831883mqttdate}} &
  \multirow{2}{*}{\SI{\Remark{18831883mqtttotalhosts}}{}} &
  \multirow{2}{*}{\SI{\Remark{18831883mqtttotalconnsuccess}}{} (\SI{\Remark{18831883mqttpctconnsuccess}}{\percent})} &
  \multicolumn{3}{c}{---} &
  \SI{\Remark{18831883mqtttotalsuccess}}{} &
   &
  \SI{\Remark{18831883mqtttotalsuccess}}{} &
  $\rceil$ &
   &
  \SI{\Remark{18831883mqtttotalas}}{} &
  --- \\ \cline{12-12} \cline{15-16}
 &
   &
  TLS &
   &
   &
   &
  \SI{\Remark{18831883tlsmqtttotaltls}}{} (\SI{\Remark{18831883tlsmqttpcttlsvalidoftransport}}{\percent}) &
  \SI{\Remark{18831883tlsmqtttotaltlsauth}}{} &
  \SI{\Remark{18831883tlsmqtttotaltlssuccess}}{} &
  \SI{\Remark{18831883tlsmqtttotalsuccess}}{} &
  \multirow{2}{*}{} &
  \multirow{2}{*}{\begin{tabular}[c]{@{}c@{}}\SI{\Remark{alltlsmqtttotalsuccess}}{}\\ (\SI{\Remark{optionaltlsmqtttotalsuccess}}{})\end{tabular}} &
  \SI{\Remark{allmqttpctvalidtls}}{\percent} &
   &
  \multirow{2}{*}{\SI{\Remark{alltlsmqtttotalas}}{}} &
  \multirow{2}{*}{\SI{\Remark{alltlsmqtttotalcertcns}}{}} \\ \cline{2-10} 
 &
  8883\(^\ddagger\) &
  TLS &
  \Remark{88838883tlsmqttdate} &
  \SI{\Remark{88838883tlsmqtttotalhosts}}{} &
  \SI{\Remark{88838883tlsmqtttotalconnsuccess}}{} (\SI{\Remark{88838883tlsmqttpctconnsuccess}}{\percent}) &
  \SI{\Remark{88838883tlsmqtttotaltls}}{} (\SI{\Remark{88838883tlsmqttpcttlsvalidoftransport}}{\percent}) &
  \SI{\Remark{88838883tlsmqtttotaltlsauth}}{} &
  \SI{\Remark{88838883tlsmqtttotaltlssuccess}}{} &
  \SI{\Remark{88838883tlsmqtttotalsuccess}}{} &
   &
   &
  $\rfloor$ &
   &
   &
  \\ \cline{1-10} \cline{12-13} \cline{15-16} 
\multirow{3}{*}{CoAP} &
  \multirow{2}{*}{5683\(^\dagger\)} &
  Standard &
  \multirow{2}{*}{\Remark{56835683coapdate}} &
  \multirow{2}{*}{\SI{\Remark{56835683coaptotalhosts}}{}} &
  \multirow{2}{*}{\SI{\Remark{56835683coaptotalconnsuccess}}{} (\SI{\Remark{56835683coappctconnsuccess}}{\percent})} &
  \multicolumn{3}{c}{---} &
  \SI{\Remark{56835683coaptotalsuccess}}{} &
   &
  \SI{\Remark{56835683coaptotalsuccess}}{} &
  $\rceil$ &
   &
  \SI{\Remark{56835683coaptotalas}}{} &
  --- \\ \cline{12-12} \cline{15-16}
 &
   &
  DTLS &
   &
   &
   &
  \SI{\Remark{56835683tlscoaptotaltls}}{} (\SI{\Remark{56835683coappcttlsvalidoftransport}}{\percent}) &
  \SI{\Remark{56835683tlscoaptotaltlsauth}}{} &
  \SI{\Remark{56835683tlscoaptotaltlssuccess}}{} &
  \SI{\Remark{56835683tlscoaptotalsuccess}}{} &
  \multirow{2}{*}{} &
  \multirow{2}{*}{\begin{tabular}[c]{@{}c@{}}\SI{\Remark{alltlscoaptotalsuccess}}{}\\ (\SI{\Remark{optionaltlscoaptotalsuccess}}{})\end{tabular}} &
  \SI{\Remark{allcoappctvalidtls}}{\percent} &
   &
  \multirow{2}{*}{\SI{\Remark{alltlscoaptotalas}}{}} &
  \multirow{2}{*}{\SI{\Remark{alltlscoaptotalcertcns}}{}} \\ \cline{2-10} 
 &
  5684\(^\ddagger\) &
  DTLS &
  \Remark{56845684tlscoapdate} &
  \SI{\Remark{56845684tlscoaptotalhosts}}{} &
  \SI{\Remark{56845684tlscoaptotalconnsuccess}}{} (\SI{\Remark{56845684tlscoappctconnsuccess}}{\percent}) &
  \SI{\Remark{56845684tlscoaptotaltls}}{} (\SI{\Remark{56845684tlscoappcttlsvalidoftransport}}{\percent}) &
  \SI{\Remark{56845684tlscoaptotaltlsauth}}{} &
  \SI{\Remark{56845684tlscoaptotaltlssuccess}}{} &
  \SI{\Remark{56845684tlscoaptotalsuccess}}{} &
   &
   &
  $\rfloor$ &
   &
   &
  \\ \cline{1-10} \cline{12-13} \cline{15-16} 
\end{tabular}
\caption{%
\emph{Left}:
Selected protocols and their variants (\(^\dagger\)standard and \(^\ddagger\)secure port), measurement dates~(in 2021), and the number of Internet-facing hosts subject to our step-by-step protocol validation process.
\emph{Right}:
Number and spread of hosts implementing a protocol with and w/o (optional) TLS-support.
PubSub protocols have the highest TLS adoption.
}%
\vspace{-3.8em}
\label{tab:dataset}
\end{table*}

\subsubsection{Selection of Industrial Protocols}
\label{subsec:protocolselection}
The collection of our dataset compiled using active Internet measurements requires the selection of retrofitted industrial and (I)IoT protocols to scan.
Our selection process is a three-step approach~(which we elaborate in more detail in Appendix~\ref{sec:protocolselection}).
\textbf{\textit{First}}, we thoroughly analyze several related work in the area of Internet measurements with industrial background~\cite{mirian-icsmes-2016,feng-characterizing-2016,barbieri2021assessing,xu2018increase,2020-dahlmanns-imc-opcua} and compose a list of \num{30}~traditional industrial and (I)IoT protocols incl.\ specified standard ports subject in their analyses.
\textbf{\textit{Second}}, we employed the IANA Port Number Registry to find corresponding (retrofitted) secure variants.
From the \num{30}~detected protocols, we find \num{18} having an entry in the port number registry and \num{9}~entries indicating a (D)TLS-secured variant: Modbus, DNP3, IEC~104, Siemens~S7, EtherNet/IP, AMQP, OPC~UA, MQTT, and CoAP.
\textbf{\textit{Third}}, we survey the protocol specifications and technical guidelines looking for (D)TLS support and find EtherNet/IP~(EnIP) and Tridium Fox~(TF; always offered together with Fox Platform~(FP)).
Overall, we compile a set of ten TLS-securable industrial protocols and their standard port numbers~(cf.\ Table~\ref{tab:dataset}~(left)), which we target in our active Internet-wide scans.

\subsubsection{Internet-wide Measurements}
To find (I)IoT deployments, %
we scan the complete IPv4 address space for services running on the identified ports.
Apart from the TLS-secured protocol variant, we also measure the standard variant (for comparisons). %
Furthermore, to eliminate short-term influences on our measurements, we \mbox{repeated} them twice and ensured that the results do not differ significantly\footnote{Only for MQTT (port~1883), we noticed \SI{\Remark{hostscntmqttmesdiff}}{}~newly reachable brokers in South Korea without TLS support in comparison to our measurement on \textbf{2020}-08-31.}.
Overall, the design and execution of our measurements were subject to comprehensive ethical considerations and best practices for Internet-wide active measurements~(cf.\ Appendix~\ref{sec:ethics}).

Our Internet measurements consist of three steps:
(i)~identifying systems having a specific port open,
(ii)~when measuring (D)TLS-enabled protocols variants, verifying that the server indeed responds with (D)TLS messages, and
(iii)~validating the application layer protocol.
Technically, our measurements rely on \texttt{zmap}~\cite{durumeric-zmap-2013} to detect Internet-facing systems behind each of the ports, i.e., to find open TCP or UDP ports with replying services.
To subsequently retrieve (configuration) data from active systems, we use \texttt{zgrab2}~\cite{zgrab2}, which we extended with DTLS support~(via~\cite{piondtls75:online}).
Using this setup, we collect detailed information on the (D)TLS configuration of hosts, e.g., their certificates or selected cipher suites.
Therefore, we perform four (D)TLS handshakes with each server.

To validate protocol support on the application layer, we extended ZGrab2 with implementations for AMQP~(basing on~\cite{streadwa3:online, Azuregoa73:online}), CoAP, EtherNet/IP, IEC~104, MQTT~(\cite{eclipsep63:online}), and OPC~UA~(\cite{COMSYSzg99:online}).
Basing on these implementations, we perform protocol-compliant handshakes whenever a (D)TLS connection could be established.

As enabling (D)TLS often is independently configurable from the port~\cite{mosquitt59:online,Networki25:online}, we assume to find (D)TLS deployments on the port reserved for the standard variant.
Hence, when scanning ports for standard variants and cannot verify the expected protocol, we also try to perform (D)TLS handshakes analog to the secure port.

\subsection{Composing Our Analysis Data}
\label{subsec:dataset}

Since servers can provide different services on ports originally reserved for (I)IoT protocols, we must validate all findings.
Thus, we next report on how we identified them and labeled their type. %

\subsubsection{Validating Responses}
\label{subsec:validating}

Table~\ref{tab:dataset}~(center) guides through our five-step approach identifying valid protocol deployments, i.e., proving (D)TLS-support and verifying (I)IoT protocol answers.
\textbf{\textit{First}}, out of all hosts signalizing an open port during our ZMap scan (column \textit{Hosts}), we select systems that do not close the TCP connection immediately and actually respond~(\textit{Transport}).
Here, we filter out a significant number of IP addresses, which was expected due to background noise that related work reported~\cite{holz-tlscomm-2016}, %
i.e., up to \SI{1}{\permille} of the IPv4 address space reply to \texttt{SYN} packets but do not complete a handshake.
During our UDP measurements, a large share of hosts responded with invalid packets, e.g., wrong length fields, which we do not consider. %
Notably, we detect that ports for the modern protocols AMQP, MQTT, and CoAP have a comparably large fraction of hosts completing a handshake and responding with data, already indicating a larger number of protocol deployments.
However, it also comprises port numbers easy to remember~(19999~and~20000).

The next three steps are only relevant for protocol variants relying on (D)TLS:
\textbf{\textit{Second}}, when performing a (D)TLS handshake, we only select hosts that reply with a valid \texttt{Server} \texttt{Hello}~(column \textit{(D)TLS Valid}).
Behind the ports where we expected to find (D)TLS deployments, we identified between \SI{100}{\percent} of the \mbox{responding} hosts~(EtherNet/IP) and only \SI{6.5}{\percent}~(IEC~104) as (D)TLS \mbox{deployments}.
Interestingly, we also find significant numbers~(up to \SI{49}{\percent} for AMQP) of (D)TLS hosts behind the protocol's standard ports.
These numbers show that operators indeed use ports originally reserved for standard protocol variants for their (D)TLS-enabled deployments.

\textbf{\textit{Third}}, we focus on hosts not denying our connection due to failed client authentication~(column \textit{(D)TLS Auth.\ OK}).
\mbox{Specifically}, we select hosts that do not refuse our connection after receiving our self-signed client certificate, as we would not be able to validate the protocols running on these hosts.
Instead, we manually revised the server certificates to get an intuition for the type of running services and protocols.
Here, we do not find any subjects inside the certificates indicating industrial use, e.g., `scada' or well-known industrial manufacturers.
However, for the PubSub protocols, we find \SI{\Remark{hostscntmqttrejectedcerthint}}{}~hosts~(MQTT) and \SI{\Remark{hostscntamqprejectedcerthint}}{}~hosts~(AMQP) delivering a certificate with accompanying product names, e.g., `rabbitmq'~(AMQP broker implementation), usage description, e.g., `messaging', or protocol names.
On both ports reserved for DNP3 we recognize certificates indicating Orange Livebox router devices in \SI{\Remark{hostscntdnp3orangelivebox}}{} cases.
Since Internet routers typically do not offer a DNP3 interface, they are not pertinent for our analysis.
This observation indicates that operators relying on secure-by-design protocols more often make use of security features, although specifications of retrofitted industrial protocols explicitly mention client authentication~\cite{MBTCPSec70:online,iec62351-3}.

While an effective method to prevent unwanted clients, e.g., attackers, to connect---especially in the area of (I)IoT deployments---we find the usage of client authentication comparably low:
Only \SI{\Remark{hostspctrequestclientcert}}{\percent} of all TLS deployments (\SI{\Remark{hostscntrequestclientcert}}{}~hosts) request a client certificate during the handshake.
Notably, \SI{\Remark{hostscntacceptclientcert}}{}~of these hosts~(\SI{\Remark{hostspctacceptclientcert}}{\percent}) still accept our connection and hence do not validate client certificates.

\textbf{\textit{Fourth}}, we only select hosts completing the (D)TLS handshake without any error (column \textit{(D)TLS Succ.}).
Next to denying a connection due to rejection of our client certificate, we recognized \SI{\Remark{hostscnttlsgeneralerror}}{}~hosts aborting the handshake for other reasons without clearly stating why (generic handshake error), e.g., due to an incomplete implementation or an extravagant form of access control.

\textbf{\textit{Finally}}, and relevant for measurements with and without established (D)TLS channel, we send protocol-conform messages and validate whether the response message of the server is also protocol compliant~(column \textit{Valid}), i.e., parsable and having protocol fields, e.g., the length field, valid.
In total, we find \SI{\Remark{hostscnt}}{}~(I)IoT services, of which only \SI{\Remark{hostscnttls}}{} (\SI{\Remark{hostspcttls}}{\percent}) are (D)TLS-secured~(with a strong focus on modern PubSub protocols).
While our measurements of traditional insecure variants show an increase from 2016~\cite{mirian-icsmes-2016}, they also show that the number of (D)TLS-secured deployments is sparse.

\subsubsection{Identifying Single Deployments and Characteristics}
\label{subsec:deployments}
For our subsequent security assessment and an indication of its impact, we revise our dataset and describe the characteristics of deployments.

As hosts offering TLS-endpoints with the same configuration on both the standard and TLS port would distort our assessment, we consider such services as one.
Furthermore, several operators~(especially for PubSub) decide to provide only optional TLS support by offering an insecure and secure endpoint on the respective ports simultaneously~(counted as TLS-adopting).
Table~\ref{tab:dataset}~(right) denotes such behavior~(column \textit{Total (opt. (D)TLS)}) and shows the share of deployments relying on (D)TLS~(\textit{\%~(D)TLS}).
Here, we identify three groups:
(i)~protocols with \emph{small TLS-adoption}, i.e., no hosts rely on the TLS variant of Modbus and only fewer than ten deployments on DNP3, IEC~104, Siemens~S7, and OPC~UA.
(ii)~We find protocols with a \emph{medium TLS-adoption}, i.e., EtherNet/IP, Tridium Fox, and CoAP, and
(iii)~\emph{large TLS-adoption} for PubSub~(AMQP and MQTT).

Although leaving the choice for secure communication to the client, modern PubSub deployments rely on TLS more often than deployments with traditional protocols and CoAP.
Hence, such variants are likely not in use by operators due to deficit configuration, the long lifetime of devices not implementing such new variants, and keeping the resource footprint on constrained devices low.

To analyze the spread of found deployments over the Internet and supervision of different operators, we also report on the distinct number of ASes and Certificate Common Names on TLS-based services in Table~\ref{tab:dataset}~(right).
While PubSub protocol deployments are generally widely distributed over the Internet and present certificates with many distinct common names, MQTT brokers are more widely distributed than AMQP brokers, where a significant number is hosted by cloud providers, e.g., Microsoft and CloudAMQP.
Deployments of protocols with medium TLS-adoption are mainly located in distinct ASes and, for Tridium Fox, Fox Platform, and CoAP, presenting different common names.
The single common name of all EtherNet/IP devices backtracks to Rockwell Automation FactoryTalk Linx deployments, indicating that the manufacturer specifies the common name.
Lastly, all minor used protocols are test deployments~(except for two DNP3 deployments indicating an affiliation with a South Korean company in their certificate).

\textit{\textbf{Takeaway:}
Only \SI{\Remark{hostscnttls}}{} (I)IoT deployments (\SI{\Remark{hostspcttls}}{\percent}) secure their communication, with a focus on employing modern PubSub instead of retrofitted protocols~(\SI{\Remark{hostspcttlssecurebydesignprotocol}}{\percent} vs.\ \SI{\Remark{hostspcttlsretrofitprotocol}}{\percent}).
Hence, the majority of deployments is open for attacks, jeopardizing a secure and safe operation.
}

\begin{figure*}[!t]
\centering
\begin{subfigure}{.166\linewidth}
  \centering
  \includegraphics[width=\linewidth]{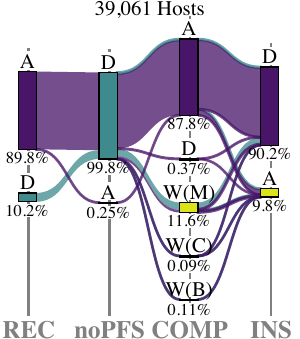}
  \caption{MQTT}
\end{subfigure}%
\begin{subfigure}{.166\linewidth}
  \centering
  \includegraphics[width=\linewidth]{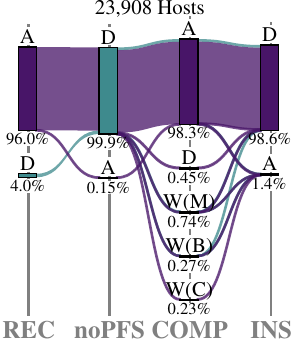}
  \caption{AMQP}
\end{subfigure}%
\begin{subfigure}{.166\linewidth}
  \centering
  \includegraphics[width=\linewidth]{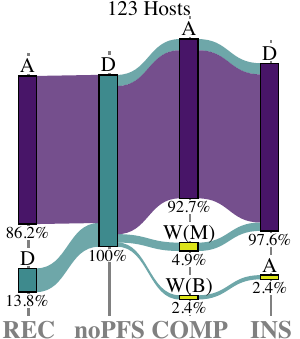}
  \caption{Tridium Fox}
\end{subfigure}%
\begin{subfigure}{.166\linewidth}
  \centering
  \includegraphics[width=\linewidth]{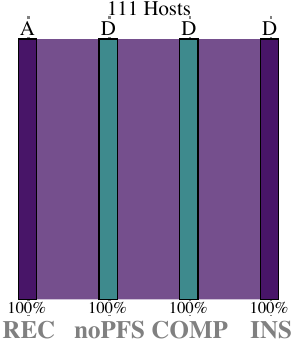}
  \caption{EtherNet/IP}
\end{subfigure}%
\begin{subfigure}{.166\linewidth}
  \centering
  \includegraphics[width=\linewidth]{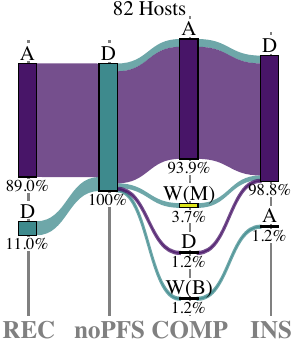}
  \caption{Fox Platform}
\end{subfigure}%
\begin{subfigure}{.166\linewidth}
  \centering
  \includegraphics[width=\linewidth]{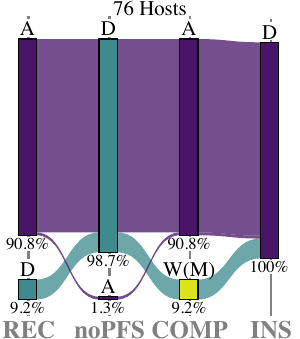}
  \caption{CoAP}
\end{subfigure}%
\vspace{-1em}
\caption{
Most hosts act securely~(\textcolor{perfecthost}{violet}) independently of the protocol and cipher suite sets~(REC, noPFS, COMP, INS), i.e., accept connections~(A) under a secure suite and deny~(D) otherwise.
Some hosts deny connections of clients not impacting their security (\textcolor{sensiblehost}{green}), and only a few hosts act insecurely~(yellow) accepting connections under weak cipher suites (W(\(\bullet\))).
}
\vspace{-1.75em}
\label{fig:tls-cipher}
\end{figure*}

\section{Breaking Down Configurations}
\label{sec:analysis}

Although TLS-adoption is generally improvable in the (I)IoT, examining the security configurations of today's deployments is crucial. %
To this end, we perform an Internet-wide security assessment.
For a benign analysis, we focus on protocols with more than 50~deployments, i.e., we focus on MQTT, AMQP, Tridium Fox and Fox Platform, EtherNet/IP, as well as CoAP.
For the detected deployments, we analyze foundational parameters for (D)TLS in Section~\ref{subsec:tlsfoundations}, advanced configuration of certificates and PKI in Section~\ref{subsec:certificates}, and security configuration of the protocol~(if available) in Section~\ref{subsec:applicationsecurity}.

\subsection{TLS Foundations}
\label{subsec:tlsfoundations}

Since (D)TLS is only secure when configured correctly, we assess whether operators set parameters of the (D)TLS deployments sufficiently for secure communication.
To this end, we use official guidelines listing adequate TLS~versions~\cite{bsi-ics-security, rfc8996} and cipher suites~\cite{rfc7525}.

\subsubsection{Negotiated Protocol Version}

TLS deployments should not use any TLS version prior to 1.2~\cite{bsi-ics-security}, as all of them are nowadays deprecated and insecure due to their reliance on SHA1~\cite{rfc8996,rfc7457}.
While the majority of industrial deployments use TLS~1.2, we also find deployments falling back to older TLS versions and see deployments already indicating support for the future-proof TLS~1.3~\cite{rfc8446}.

\begin{table}[!t]
\vspace{0.25em}
\centering
\begin{tabular}{ll@{\hspace{4\tabcolsep}}cccc}

\textbf{Protocol}    &  & \textbf{TLS~1.0}                  & \textbf{TLS~1.1}                  & \textbf{TLS~1.2}                  & \textbf{TLS~1.3\(^*\)}  \\

\addlinespace[-\aboverulesep]
\cmidrule[\heavyrulewidth]{1-6}

MQTT        &  & \num{\Remark{mqttcnttls10}}    & \num{\Remark{mqttcnttls11}}    & \num{\Remark{mqttcnttls12}}    & \num{\Remark{mqttcnttls13capable}} \\
AMQP        &  & \num{\Remark{amqpcnttls10}}    & \num{\Remark{amqpcnttls11}}    & \num{\Remark{amqpcnttls12}}    & \num{\Remark{amqpcnttls13capable}} \\
DNP3        &  & \num{\Remark{dnp3cnttls10}}    & \num{\Remark{dnp3cnttls11}}    & \num{\Remark{dnp3cnttls12}}    & \num{\Remark{dnp3cnttls13capable}} \\
Tridium Fox &  & \num{\Remark{foxcnttls10}}     & \num{\Remark{foxcnttls11}}     & \num{\Remark{foxcnttls12}}     & \num{\Remark{foxcnttls13capable}} \\
Fox Platform&  & \num{\Remark{httpcnttls10}}    & \num{\Remark{httpcnttls11}}    & \num{\Remark{httpcnttls12}}    & \num{\Remark{httpcnttls13capable}} \\

\cmidrule[\heavyrulewidth]{1-6}
\addlinespace[-\belowrulesep]

\end{tabular}
\qquad
\scriptsize
\mbox{\(^*\) indicated via known byte sequence in the server random number when falling back to TLS~1.2~\cite{rfc8446}}
\caption{Most deployments rely on (D)TLS~1.2 (Protocols where all deployments negotiate (D)TLS~1.2 are omitted).}
\vspace{-5.25em}
\label{tab:tls-versions}
\vspace{0.25em}
\end{table}

Table~\ref{tab:tls-versions} lists the number of protocol deployments relying on a specific TLS version.
Altogether, \num{\Remark{hostscntdeprecatedtlsversion}}~deployments fall back to insecure TLS versions~(\SI{\Remark{hostspctdeprecatedtlsversion}}{\percent}), inherently weakening their communication security by allowing for impersonation attacks.
Looking into the certificates' \texttt{NotBefore} dates, we notice that hosts relying on older TLS versions are operated this way for long periods~(Mann-Whitney-U test: \(U{=}\num{\Remark{uvaltlsage}}; p{\approx}\num{1.05e-180}\)).
Hence, they were not updated in light of changes in the security landscape, i.e., no secure TLS version is enabled.
\SI{\Remark{hostscntamqpcertappservergeneratedsbcatls10}}{}~AMQP hosts relying on TLS~1.0, i.e., allowing for impersonation or downgrading attacks, deliver a certificate issued by \texttt{AppServerGeneratedSBCA}~(as per certificate's \texttt{Issuer CA}).
While we are not entirely sure which software generates these certificates, we assume an association with Microsoft's Windows Service Bus after reviewing websites mentioning the CA's name.
As software providing this bus is deprecated, these deployments likely are old and not updated to recent alternatives, which now impacts their security.
Contrarily, we note that server software revealing their version~(\SI{\Remark{hostscntdepreactedtlswithversioninfo}}{\percent} of deployments falling back to TLS version <1.2 provide version information) in all but one case were published after major TLS libraries added TLS~1.2 support.
Thus, operators must have set the maximum TLS version manually to a deprecated TLS version, e.g., to reduce compatibility problems with ancient clients.
This procedure inherently weakens the communication security with clients that otherwise would choose TLS~1.2.

Deployments indicating support for TLS~1.3 are predominantly located at cloud hosters or in the case of Fox Platform and DNP3 proxied via their infrastructure, covering the findings of related work where the rise of TLS~1.3 was attributed to such providers~\cite{2020-holz-tls13}.
These hosts already profit from significant security and performance advantages, e.g., entirely encrypted and 0-RTT handshakes.

\subsubsection{Security of Offered Cipher Suites}

While a recent TLS version is an important foundation, selecting a secure cipher suite is equally crucial to ensure secure communication. %
Servers should not select cipher suites with deprecated cryptographic primitives, especially not when clients indicate support for completely secure suites.
To analyze this aspect, we performed four subsequent handshakes with servers indicating support for different cipher suites, i.e., \textbf{\textit{REC}} covering cipher suites recommended by official guidelines~\cite{nistsp800-tls,rfc7525,tr2102-2},
\textbf{\textit{noPFS}} containing recommended cipher suites to be used when no perfect forward secrecy~(PFS) can be implemented~(due to hardware limitations; thus enabling attackers to recover information from data encrypted with older key material)~\cite{rfc7525},
\textbf{\textit{COMP}} offering cipher suites the standard ZGrab2 implements for maximum compatibility, and
\textbf{\textit{INS}} consisting of insecure ciphers~(cf.\ Appendix~\ref{sec:cipherclasses}).

Figure~\ref{fig:tls-cipher} shows the share of hosts for every protocol accepting~(A) or denying~(D) a connection under the cipher suite sets indicated by boxes on respective vertical lines --- for MQTT, \SI{\Remark{hostpctmqttrecommendedsuccessTrue}}{\percent} accept a connection under REC.
For COMP, the figure denotes the reason for the insecurity, i.e., due to the included cipher~(W(C)), MAC~(W(M)), or both~(W(B))).
Shapes between the cipher suite sets allow to trace dependencies between host decisions, e.g., which share of hosts accept a connection under COMP but deny a connection under INS --- for MQTT, the largest share of hosts accepting a connection under COMP denies a connection under INS.
The color of these shapes indicates the host decision of the host fraction on the REC set --- for MQTT, most hosts accepting a connection under COMP, but denying under INS, already accepted a connection under REC.

Over all protocols, most hosts accept a connection when clients only offer recommended cipher suites~(REC), i.e., such clients can communicate securely.
Still, \SI{\Remark{hostspcttlsworecommended}}{\percent} of all servers deny a connection in this setting, i.e., they do not support recommended suites, weakening the communication security.
This negligence comprises all but one server using TLS~1.0 and all servers relying on TLS~1.1, showing that outdated configurations also have an impact on the cipher suite selection rendering the communication even more insecure. %

Besides, independently of the protocol, most servers deny a connection when clients only request recommended ciphers w/o perfect forward secrecy~(noPFS), most likely, as the respective cipher suites are typically excluded from the server's suite set by (OpenSSL's) default.
While not negatively influencing the security, resource-constraint (I)IoT clients cannot connect to these servers. %

When clients request one of a broad range of suites~(COMP), servers accept most connection attempts and predominantly choose secure cipher suites.
Still, a few hosts of each protocol, except for EtherNet/IP, choose weak cipher suites.
Most notably, this also comprises servers previously accepting a connection when only recommended suites are offered by the client, indicating that these servers
(i)~do not support all recommended cipher suites or prioritize weaker cipher suites over more secure ones, and
(ii)~are configured for backward compatibility, i.e., not denying connections when clients request no recommended suite.
While the selection of a suite relying on a weak MAC enables attackers to alter communication, potentially disturbing production processes, weak ciphers allow for eavesdropping of possibly sensitive data.

Hence, servers should not allow clients to connect via insecure ciphers~(INS), and indeed, over all protocols, most servers deny such attempts.
However, \num{\Remark{hostscnttlsallowinsecure}}~servers~(\SI{\Remark{hostspcttlsallowinsecure}}{\percent})~still accept insecure connections, significantly weakening the communication security.
Notably, for CoAP, Tridium Fox, and Fox Platform, servers accepting a connection with these insecure settings also accepted connections with weak cipher suites when clients offer a mix of recommended and insecure ciphers~(COMP).
However, CoAP servers also accept connections when clients only indicate support for recommended cipher suites, indicating that these servers do not have valid security policies or are intentionally configured for backward compatibility.

Notably, hosts relying on EtherNet/IP stand out as all of these hosts only accept a cipher from the recommended cipher suite set, i.e., have a very restrictive selection likely induced by their vendor.

\textit{\textbf{Takeaway:}
\SI{\Remark{hostspctdeprecatedtlsversion}}{\percent} of all deployments rely on deprecated TLS versions allowing for impersonation attacks, and \SI{\Remark{hostspcttlsallowinsecure}}{\percent} of servers allow clients to connect via insecure ciphers easing eavesdropping or alteration of sensitive (I)IoT data.
Both result from outdated configurations today impacting the security and safety of affected deployments.
}

\subsection{Authentication Based on Certificates}
\label{subsec:certificates}

Another important aspect for secure communication is server authentication relying on certificates and optionally a PKI to prevent attackers from performing Man-in-the-Middle~(MitM) attacks.
To understand whether operators protect their deployments against MitM attacks, we analyze whether operators make use of PKIs, inspect the age and lifetime of certificates to conclude for implemented strategies for certificate renewal, and whether the cryptographic primitives used to generate certificates are still secure.

\begin{figure}[!t]
\centering
\includegraphics[width=\linewidth]{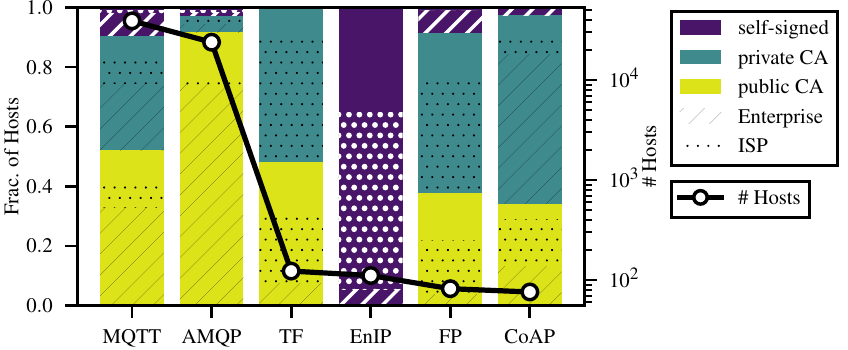}
\vspace{-2.5em}
\caption{\mbox{Most certificates are issued by (private) CAs. For} \mbox{PubSub, public CAs are highly frequented by cloud providers} \mbox{and large companies operating enterprise networks.}}
\label{fig:cert-trust}
\vspace{-1.75em}
\end{figure}

\subsubsection{Reducing Certificate Management Overhead}
To analyze the trust anchor operators rely on (either using self-signed certificates, or utilizing Certificate Authorities~(CAs)), we check whether \mbox{certificates} are validatable against root certificates in typical trust stores\footnote{Stores from iOS/MacOS, Windows, Android, OpenJDK, Mozilla NSS, and Oracle JDK.}.
Furthermore, we consider certificates where the issuer and common name are different but not validatable against one of the root certificates in the trust stores as signed by a private CA.
We further mark certificates where these fields are equal as self-signed.

Figure~\ref{fig:cert-trust} shows the trust anchor distribution of hosts split by protocol and indicates whether hosts are located in enterprise networks or networks of Internet Service Providers\footnote{\label{fnt:peeringdb}We rely on the AS's entry in PeeringDB~(\url{peeringdb.com}) for classification (We count content providers to enterprise and (educational) network services to ISP).}.
All EtherNet/IP servers use self-signed certificates and are mainly located in ISP networks indicating that
(i)~these hosts do not belong to large companies which would operate own enterprise ASes and
(ii)~that authentication in automation networks often relies on trust lists manually maintained on every client, increasing the management overhead.

Most Tridium Fox and Fox Platform instances are also located in ISP networks but rely on certificates issued by private CAs.
Hence, these small companies issue certificates themselves.
While still not involving external entities in the certificate issuance, operators must deploy their own CA root certificate on connecting clients easing the setup of new deployments as no trust lists must be updated.

The majority of MQTT and AMQP brokers as well as CoAP servers rely on certificates issued by public CAs.
Here, MQTT brokers are often hosted by cloud providers also taking care of valid certificates as part of their service.
In contrast, AMQP brokers relying on certificates issued by public CAs are mostly located in company networks, indicating that operating companies have strong security policies.
Notably, \SI{\Remark{hostspctmqttacmeofcasigned}}{\percent}~of MQTT and \SI{\Remark{hostspctamqpacmeofcasigned}}{\percent}~of AMQP deployments profit from certificates issued by CAs supporting automated certificate management, e.g., Let's Encrypt.
Hence, automated management allows to decrease the management overhead in the (I)IoT to likely increase authentication security similar to the Web~\cite{aas2019let}.

\subsubsection{The Less Overhead the Less Lifetime}
\label{subsec:certlifetime}

Limiting the validity period of certificates can help to improve the security as CA's reverify requesting identities, operators are forced to revisit their used cryptographic primitives, and contain the negative impact of compromised certificates.
As general guidelines significantly decreased the recommended lifetime of certificates~(generated after June~2016: lifetime $\leq$~39~months, after February~2018: $\leq$~825~days, after September~2020: $\leq$~398~days)~\cite{cabforum}, the question arises whether operators act accordingly or whether servers present expired certificates.

\begin{figure}[!t]
\centering
\includegraphics[width=\linewidth]{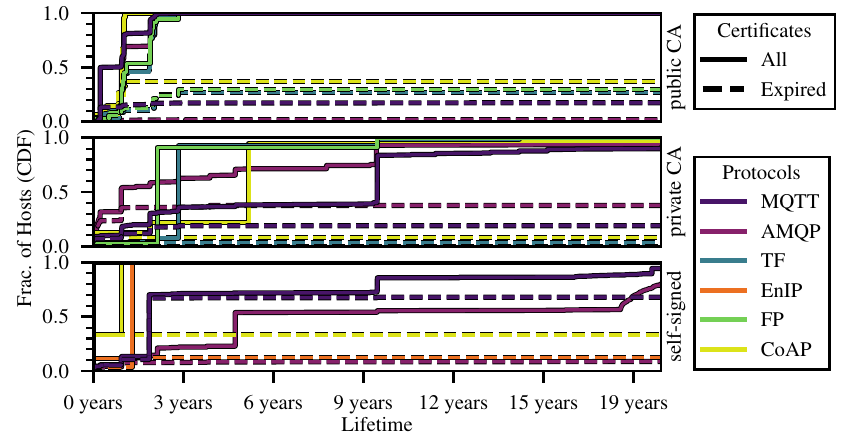}
\vspace{-2.5em}
\caption{The lifetime of certificates issued by public CAs is significantly shorter, forcing operators to update certificates frequently. Still, many servers deliver expired certificates.}
\label{fig:cert-lifetime}
\vspace{-1.5em}
\end{figure}

Figure~\ref{fig:cert-lifetime} shows the share of servers with a certain certificate lifetime for different trust anchors.
Moreover, it denotes the share of hosts distributing expired certificates. %
The majority of hosts relying on public CA-signed certificates, independently of the protocol, deliver certificates with a lifetime of two years or less.
Thereby, public CAs obey the aforementioned guidelines on certificate lifetimes.
However, \num{\Remark{hostscntcasignedexpired}}~hosts rely on expired certificates~(between \SI{\Remark{hostspctmincasignedexpired}}{\percent}~(\Remark{protomincasignedexpired}) and \SI{\Remark{hostspctmaxcasignedexpired}}{\percent}~(\Remark{protomaxcasignedexpired})), showing that
(i)~these servers do not benefit from regularly replaced certificates and
(ii)~clients of such servers do not validate certificate expiry accepting potentially compromised certificates, both allowing for impersonation attacks.

Private CAs do not obey the guidelines on certificate issuance, i.e., \SI{\Remark{hostspctprivatecadisregardcab}}{\percent} of hosts relying on private CAs send certificates with lifetimes beyond the recommended timespans.
Hence, operators do not plan to replace the certificates regularly, potentially affecting their security on deprecation of used primitives.
Furthermore, hosts use already expired private CA-signed certificates~(up to \SI{\Remark{hostspctmaxprivatecaexpired}}{\percent}~(\Remark{protomaxprivatecaexpired})), again showing that clients do not check on certificate expiry.

The lifetime of self-signed certificates is even longer, and certificates are more often expired~(EtherNet/IP:~\SI{\Remark{hostspctenipselfsigned3yexpired}}{\percent}, MQTT:~\SI{\Remark{hostspctmqttselfsigned3yexpired}}{\percent}, CoAP:~\SI{\Remark{hostspctcoapselfsigned3yexpired}}{\percent} of hosts deliver expired self-signed certificates having a lifetime longer than three years).
Hence, operators replace certificates less likely when not forced by external entities, which does not necessarily impact the security now, but maybe in the future.
Thus, public CAs enforcing update processes increase security.

\subsubsection{Most Certificates Are Secure --- Currently}

While the regular renewal of certificates helps to update the cryptographic primitives certificates rely on, the question is whether operators do so.
When operators still rely on certificates with too short public keys, the authentication is at stake independently of the trust anchor.
Moreover, deprecated hash functions open doors for impersonation attacks on certificates signed by public or private CAs.
Hence, guidelines recommend to not use certificates with an RSA key length lower than 2000~bit, alternatively to RSA, to rely on ECDSA~\cite{tr2102-2, nistsp800-tls}, and to not use SHA1 and MD5 for signatures~\cite{tr2102-1}.

\begin{figure}[!t]
\centering
\includegraphics[width=\linewidth]{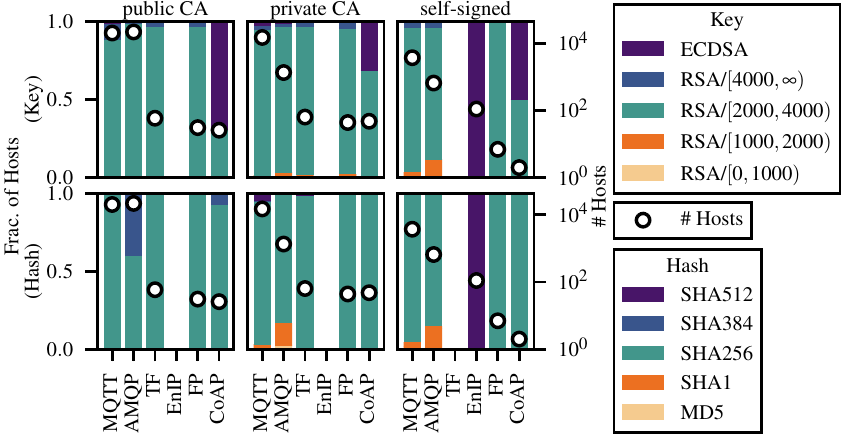}
\vspace{-2.5em}
\caption{While all certificates issued by public CAs are secure, some others are not. Insecure parameters are hatched.}
\label{fig:cert-key-types}
\vspace{-1.5em}
\end{figure}

Figure~\ref{fig:cert-key-types} visualizes the share of hosts delivering certificates with a certain key type and length in the case of RSA, as well as the used hash function separated by their certificate's trust anchor and used (I)IoT protocol.
All servers relying on public CA-signed certificates use cryptographic primitives considered as secure.
Contrarily, servers relying on certificates issued by private CAs or self-signed certificates use too short keys and deprecated hash functions more frequently.
Hence, these systems jeopardize their authenticity.

Altogether, \SI{\Remark{hostscntamqpshortkey}}{}~AMQP brokers (\SI{\Remark{hostspctamqpshortkeyprivateca}}{\percent} of private CA authenticated and \SI{\Remark{hostspctamqpshortkeyselfsigned}}{\percent} of self-authenticated brokers) rely on too short asymmetric keys, opening the door for attackers to eavesdrop and alter their communication.
Furthermore, the certificates of \SI{\Remark{hostscntamqpdeprecatedhash}}{}~brokers~(\SI{\Remark{hostspctamqpdeprecatedhashprivateca}}{\percent} of private CA authenticated brokers) rely on a deprecated hash function for signature generation easing impersonation attacks.
Thereby, \SI{\Remark{hostscntamqpcertmd5}}{}~brokers~(\SI{\Remark{hostspctamqpcertmd5privateca}}{\percent} of private CA authenticated brokers) use certificates still relying on MD5.
Similar results are visible for private CA and self-authenticated MQTT brokers, i.e., \SI{\Remark{hostscntmqttshortkey}}{}~MQTT brokers (\SI{\Remark{hostspctmqttshortkeyprivateca}}{\percent} of private CA authenticated and \SI{\Remark{hostspctmqttshortkeyselfsigned}}{\percent} of self-authenticated brokers) have a too short key included in their certificate.
\SI{\Remark{hostspctmqttnoncasignedweakcertissued2021}}{\percent} of these hosts use certificates generated in 2021, showing that operators still generate certificates with weak parameters.

The same holds for deployments using retrofitted protocols, i.e., a minority relies on certificates with weak parameters for authentication not significantly older than strong certificates~(\SI{\Remark{hostspctretrofitweakcertissued2021}}{\percent} of weak certificates were generated in 2021).
In contrast, all EtherNet/IP hosts use strongest cryptographic primitives, i.e., ECDSA and SHA512, indicating efficacy of well-applied security policies.

\textit{
\textbf{Takeaway:}
Public CAs force operators to update their certificates regularly, increasing the security of (I)IoT deployments by impeding impersonation.
Still, \SI{\Remark{hostspctnoncasignedweak}}{\percent} of deployments (not relying on public CAs) use certificates with weak parameters putting their security at stake.
}

\subsubsection{Large-Scale Leakage of Secrets}

While using strong cryptographic primitives is a fundamental requirement for secure authentication, the correct handling of cryptographic secrets is even more important.
Especially, the private key must not be leaked to any other entity or spread over numerous (I)IoT devices located on the field~\cite{2020-dahlmanns-imc-opcua}.
Thus, we analyze how many hosts use a single certificate and how widespread they are on the Internet, i.e., over how many autonomous systems~(ASes) these hosts are distributed.
On this basis, we conclude whether (compromised) certificates are in use.

To this end, we classify certificates into three categories according to how many hosts in how many ASes use them for authentication:
(i)~We consider certificates in a single AS behind one or two IP addresses~(accounting for changes of dynamic IP addresses) as not reused.
(ii)~Certificates on more than two hosts in a single AS are \emph{intra-AS} reused, and
(iii)~certificates on more than two hosts in different ASes are \emph{inter-AS} reused.
While intra-AS reuse can indicate load-balancing %
and their affiliation to a single operator, intra-AS reuse likely implies the reuse of key material across operators.

\begin{figure}[!t]
\centering
\includegraphics[width=\linewidth]{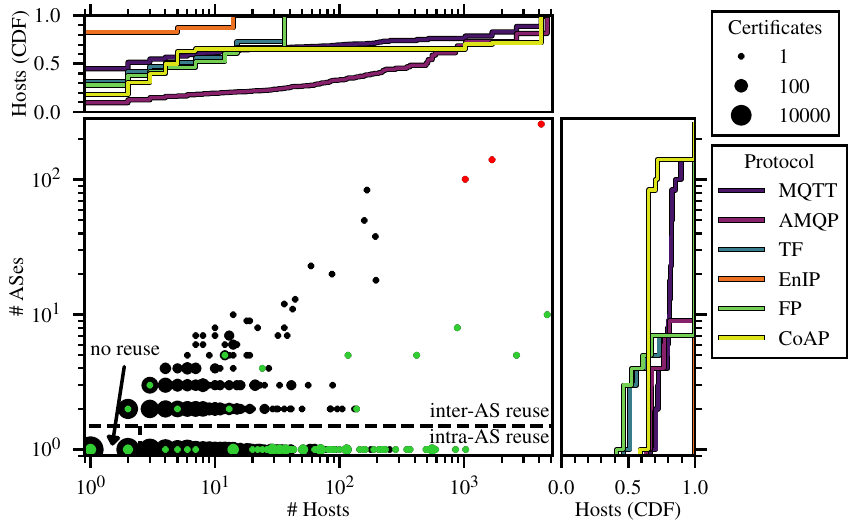}
\vspace{-2.5em}
\caption{
Certificates are reused over different IPs/ASes, indicating load balancing~(\textcolor{limegreen}{\(\bullet\)}) or compromised secrets~(\textcolor{red}{\(\bullet\)} or \(\bullet\)).
}
\label{fig:cert-reuse}
\vspace{-1.5em}
\end{figure}

Figure~\ref{fig:cert-reuse} assigns the count of certificates to a number of hosts which use these certificates for authentication and the number of ASes the hosts are located in (nodes increase their size in log-scale).
Furthermore, Figure~\ref{fig:cert-reuse} breaks down which share of hosts per protocol is affected by the reuse of certain certificates.
While we detected \SI{\Remark{certpctnoreuse}}{\percent} of the certificates (\SI{\Remark{certcntnoreuse}} of \SI{\Remark{certcnt}} certificates) used by hosts behind one or two IP addresses in a single AS, i.e., not reused, we discovered \SI{\Remark{certcntintraas}} intra-AS and \SI{\Remark{certcntinteras}} inter-AS reused.

In both classes of reuse, we detected certificates delivered by AMQP or MQTT brokers indicating load balancing, e.g., hosted by major cloud providers like Amazon.
We identified such certificates by their \texttt{Common} \texttt{Name} and \texttt{Organization} field and colored corresponding shares in Figure~\ref{fig:cert-reuse} green.
This practice affects \SI{\Remark{certcntintraasloadbalancing}} certificates reused on in total \SI{\Remark{hostscntintraasloadbalancing}} hosts in the same AS, and \SI{\Remark{certcntinterasloadbalancing}} certificates reused over \SI{\Remark{hostscntinterasloadbalancing}}{}~hosts in different ASes~(up to \SI{\Remark{maxascntinterasloadbalancing}} ASes for increased availability).
As such hosts are typically installed in secure data centers and in control of the same entity, e.g., the cloud provider, these cases of reuse do not negatively impact authenticity.

Contrarily, we found certificates by MQTT and CoAP hosts operated by different entities.
The three most reused certificates~(red in Figure~\ref{fig:cert-reuse}) are used by \SI{\Remark{hostscnttop3reusedcerts}}{}~MQTT and CoAP hosts distributed over in total \SI{\Remark{ascnttop3reusedcerts}} ASes indicating that several entities are in possession of the certificate's private key, massively impacting their authenticity.
All entities are able to perform impersonation attacks on each other to eavesdrop on confidential information or alter communication.
On further investigation, we found these certificates and the corresponding private keys as an example in a GitHub repository of an MQTT broker software with CoAP extension\footnote{\url{https://github.com/emqx/emqx}}.
Furthermore, the repository serves as the basis for a frequently used Docker image providing this functionality, which also includes the certificate and cryptographic secrets.
This aspect shows that delivering exemplary cryptographic material inherently instigates operators to weaken their security.
We contacted the broker developer via email and issue on GitHub suggesting to remove the example and achieved approval.
However, the newest version of the certificate and the private key is still online and still part of the Docker Image.

Regarding (I)IoT protocols offside of modern PubSub protocols, we also identify practices of reuse.
For EtherNet/IP, we found a single certificate used by \SI{\Remark{hostcnteniptopreuse}}{}~hosts at a smaller scale.
After manual inspection of the payload data available on these hosts, we assume that project files, including the certificate and cryptographic secrets, have been copied between devices operating in similar appliances.

\textit{\textbf{Takeaway:}
  \SI{\Remark{hostspcttlsweakduetoreuse}}{\percent} of hosts reuse certificates with (potentially) compromised secrets, e.g., included as examples in software.
  As partly available on the Internet, these secrets can be used by attackers to perform impersonation attacks to intercept communication.
}

\subsubsection{Templates to the Rescue?}
Reusing already available but compromised secrets is easy for operators but also puts authenticity at stake.
Hence, ideally, operators receive support in generating cryptographic secrets.
Guides providing instructions and predefined scripts can help to determine cryptographic parameters, e.g., a hash function.
However, until now, their practical use and impact are unclear. %
Thus, we shed light on this uncertainty to give directions for best practices in templating and analyze their current influences.

A manual revision of all received certificates shows widely distributed similar subject names, although subject names only have a few naming conventions.
As this detail indicates a first coherence, we cluster all received certificates by their subject name and subsequently analyze the similarity of other parameters in each cluster.
Our clustering relies on text mining approaches, i.e., (1,3)-grams, term frequency-inverse document frequency~(TF-IDF), DBSCAN, and a manual review of all calculated clusters~(cf.\ Appendix~\ref{sec:certclustering}).

On the basis of the subject name, we find \SI{\Remark{clustercnt}} certificate clusters of up to \SI{\Remark{certcntlargestcluster}}{}~certificates~(mean size: \SI{\Remark{certcntclustermean}}{}, median: \SI{\Remark{certcntclustermedian}}{}).
The largest cluster comprises certificates with the subject name leading to a bash script in the GitHub repository of OwnTracks, an open-source location tracking software often used in smart homes.
Hence, operators also often use templates from the Web to generate their certificates, adopting potentially predefined security preferences.
In contrast, the second-largest cluster includes~\SI{\Remark{certcntclusteramazon}}{}~certificates.
It can be assigned to a major cloud provider generating certificates for different MQTT broker instances showing that templating is also used in enterprise environments, most likely for fast and automated configuration.
Further clusters relate to certificates, including names of large companies active in the communication, virtualization, and IIoT sector.
However, we were not able to find any information on these templates on the Web, i.e., the templates used to generate these certificates are most likely company-internal.

\begin{figure}[!t]
  \centering
  \includegraphics[width=\linewidth]{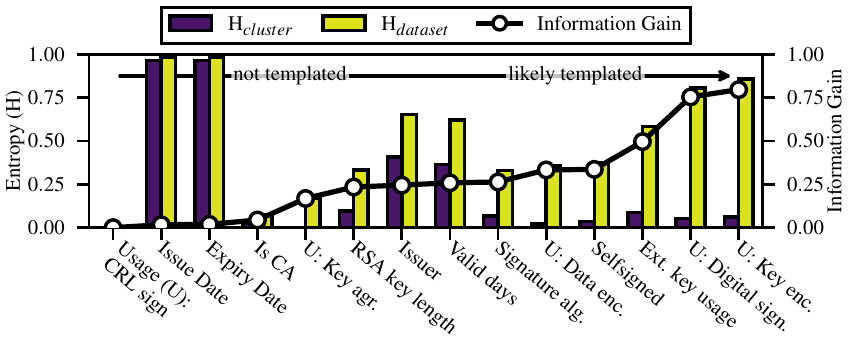}
  \vspace{-2.5em}
  \caption{Templates influence selected usage parameters and signature algorithm but not issue and expiry dates.}
  \label{fig:eval-cert-entropy}
  \vspace{-1.5em}
\end{figure}

Next, we analyze which certificate parameters other than the subject name are influenced by templates.
To this end, we compare the entropy of single certificate parameters over the complete dataset with the average normalized entropy weighted by cluster size of each cluster and assume that a lower cluster entropy signalizes that templates set a respective parameter.
Figure~\ref{fig:eval-cert-entropy} visualizes both entropies and the gain of similarity of a single parameter within all clusters.
Notably, templates specify used cryptographic primitives and their parameters, e.g., the RSA key length or the signature algorithm, more regularly than issuing or expiry dates, having a strong influence on the certificate's security.

To assess the impact of templating on the security of generated certificates, we analyze the occurrence of deprecated parameters in our clusters.
Here, we identify \SI{\Remark{certcntmd5templated}} of \SI{\Remark{certcntmd5}}{}~(\SI{\Remark{certpctmd5templated}}{\percent}) detected certificates relying on MD5 signatures in a single cluster, i.e., generated using a single template.
Notably, all of these certificates were generated since 2020, showing that old templates weaken the security of today's systems.
Moreover, we found three clusters with in total \SI{\Remark{certcntsha1templated}}{}~certificates relying on SHA1, implying that operators use three distinct templates still predefining outdated primitives. %

Templating is already widely used in practice and helps operators to configure their system fastly and also can have a positive impact on security.
However, our analysis shows that templates specifying deprecated parameters, e.g., SHA1 and MD5, are still used and were not updated.
While regularly updated templates can influence the security of systems positively, templates with deprecated parameters can put security at stake.

\textit{\textbf{Takeaway:}
  \mbox{Templates help operators to generate fresh key material}
  \mbox{and are widely used in practice. However, templates with deprecated}
  \mbox{settings weaken the security of systems at scale. Hence, operators must}
  \mbox{use regularly updated templates that take deprecation into account.}
}

\subsection{Open Door On PubSub Brokers}
\label{subsec:applicationsecurity}
For now, our analysis focused on server authentication via their certificates for hosts that do not deny our connection attempt during the TLS handshake.
Consequently, if these deployments do not enforce access control on the application layer, they are unprotected against malicious direct access.
Especially the \SI{\Remark{hostscnttlsenip}} hosts relying on the TLS-based variant of EtherNet/IP but also the other seven devices~(today mostly test deployments) communicating via the retrofitted variants of S7, Modbus, IEC104, and DNP3 allow espionage or changes of machine behavior. %
While espionage puts the security of business secrets at stake, changing machine behavior allows attackers to disturb production processes.

For protocols with support for access control, the question arises whether operators make use of these mechanisms or leave their systems open for attackers.
However, not all protocols support an analysis of found deployments for properly configured access control mechanisms under our ethical guidelines~(cf.\ Appendix~\ref{sec:ethics}).
As CoAP does not provide any access control mechanism~(to achieve a small resource footprint)~\cite{RFC7252} but leaves this task to downstream applications, no process is available on how to check for such mechanisms.
Hence, an analysis would require systematically checking various URLs, leading to a significant overhead on these systems.

Tridium Fox requires access control of remote clients in every case~\cite{niagara-2019}.
As we are not aware of any default password, and we do not guess various passwords, we refrain from further analysis.
Hence, we focus our analysis of activated access control \mbox{mechanisms} on AMQP, MQTT, and Fox Platform~(relying on HTTP).

Fox Platform hosts, independently of using TLS for secure communication or not, reply with either HTTP error codes indicating the servers actively denying the connection, sending back empty responses, or login forms requesting credentials.
Hence, we also consider all these devices as secure regarding access control.

Contrarily, several AMQP and MQTT brokers do not implement proper access control, allowing everyone on the Internet~(including attackers) to connect and publish/retrieve data via/from these brokers, potentially gaining access to confidential production information and disturbing productions.
The default configuration of RabbitMQ, a major AMQP broker implementation, already contains user credentials, initially limited to local network access.
To understand operator behavior, i.e., whether operators change predefined credentials when offering services to the Internet, we check whether brokers accept these credentials.
Indeed, \SI{\Remark{hostspctamqpaccessible}}{\percent} of all \num{\Remark{hostscntamqp}} AMQP brokers (\SI{\Remark{hostspctamqpaccessibledefault}}{\percent} of brokers w/o TLS, \SI{\Remark{hostspctamqpaccessibletls}}{\percent} with TLS) allow access using these credentials.
Thus, operators use default credentials for login via the Internet that are also known by potential attackers.
Hence, default configurations should not contain predefined secrets but must be generated for every deployment independently.

For MQTT, \SI{\Remark{hostscntmqttaccessible}}{} of \SI{\Remark{hostscntmqtt}}{} found brokers~(\SI{\Remark{hostspctmqttaccessible}}{\percent}) do not restrict access of any kind.
This finding affects only \SI{\Remark{hostspctmqttaccessibletls}}{\percent} of TLS-protected brokers but \SI{\Remark{hostspctmqttaccessibledefault}}{\percent} of non-TLS-protected brokers, indicating that operators more often have access control in mind when simultaneously configuring their communication security appropriately.

We connect to these systems to gain any intuition on the usage of affected systems and search contact information in payload data.
Other than for AMQP, where retrieving payload data would disturb ongoing communication~(as AMQP brokers do not duplicate messages), on MQTT, we subscribed to the root topic to receive all messages exchanged via the broker~(in agreement with our institution's data protection officer, cf.\ Appendix~\ref{sec:ethics}).
Similar to related work~\cite{maggi2018fragility}, we find exchanged payloads to often cover privacy sensitive types, including feeds of private video surveillance, location services, smart homes, hospitals, and further (I)IoT deployments, underpinning how problematic the availability of such systems is.
On these systems, we found \SI{\Remark{emailcntmqttpayload}}{}~email addresses, which we contacted to responsibly disclose our findings~(cf.\ Appendix~\ref{sec:ethics}).

\textit{\textbf{Takeaway:}
\SI{\Remark{hostspcttlsaccessibleofaccesscontrol}}{\percent} of deployments using TLS for secure communication do not implement access control, i.e., attackers can access the devices to espionage confidential information or alter the system's state.
Default configurations used by operators, including pre-configured credentials or disabling authentication, put the systems at risk.
}

\section{Discussion, Limitations \& Guidelines}
\label{sec:discussion}

The outcome of our work is two-fold.
First, we have seen that our Internet measurements are suitable to identify weaknesses in the configuration of Internet-facing (I)IoT deployments.
Future work could look into approach-induced \textit{limitations}, which we report in the following.
Second, based on our analysis, we are able to derive \textit{guidelines} to secure the configurations of (I)IoT deployments.

\textbf{No Influence By Honeypots:}
Honeypots act like real production systems attracting attackers to perform their exploits. %
Such systems could, in theory, influence our analysis as operators might configure honeypots less securely.
However, the lack of publicly-available honeypot implementations of retrofitted or modern (I)IoT protocols decreases the likelihood of such interference.
Still, we manually inspected found systems~(with a focus on sparse \mbox{retrofitted} protocol deployments) to find any abnormalities.
While we did not find any deployments traceable to TLS-enabled standard honeypots, e.g., Conpot~\cite{mushorgc91:online}, we identified~(on top of excluded testing deployments) \SI{\Remark{hostscnttlsenipsharedserial}} Tridium Fox devices and \SI{\Remark{hostscnttlsfoxsharedserial}} EtherNet/IP hosts sharing an---otherwise---unique serial number, respectively.
Based on the low number and lack of other anomalies, we conclude that, on a large scale, honeypots do not influence our analysis.

\textbf{Limitations of Internet Scans:}
Like any other active Internet scans, we are not able to (i)~detect (I)IoT protocols behind hosts denying our connection attempts, %
(ii)~analyze deployments that are only reachable internally or via VPNs, and
(iii)~find deployments running on non-scanned ports.
Thus, we might miss properly-secured hosts in our analysis.
Still, the absolute volume of insecure systems is alarming on its own.
However, long lifetimes of industrial equipment, generally slow protocol rollouts, the increasing number of deployments relying on traditional protocols, and our presented results indicate that retrofitted variants are not widely in use today.
To answer with certainty whether more properly-secured hosts exist (e.g., behind VPNs), future work should analyze factory, automation, and other (I)IoT networks on site and at large scale.

\textbf{Guidelines to Improve (I)IoT Security:}
Our analysis shows that external entities must regularly force operators to revisit their security configurations to increase the provided security level~(e.g., CAs mandating operators to re-create certificates).
To ease secure configurations, secure-by-design templates are a useful approach, e.g., scripts to automatically generate certificates.
Even more important, hardware and software vendors must not include any examples or defaults into their product as operators are otherwise tempted to use them in Internet-facing deployments.
Simultaneously, for a secure operation, rigorous maintenance of templates and all security configurations is crucial.
Finally, to reliably protect all connections and prevent downgrading attacks, operators may not provide an insecure endpoint simultaneously to a TLS-secured endpoint.

\textbf{The Calm Before the Storm?:}
To analyze whether attackers target TLS-based deployments, we deployed seven low interaction honeypots~(excluded from our own analysis).
We operated our honeypots over six months~(\mbox{2021-01-20} till \mbox{2021-07-25}) in five ASes providing services on all ports subject to our measurements and perform TLS handshakes on the respective TLS ports.
Across all deployments, ports reserved for the secure protocol variant are between \SI{\Remark{connpcthoneypotminreduction}}{\percent}~(\Remark{protohoneypotminreduction}) and \SI{\Remark{connpcthoneypotmaxreduction}}{\percent}~(\Remark{protohoneypotmaxreduction}) less often subject to connection attempts than the port for the insecure variant, indicating that secure variants are not yet heavily subject to attacks and measurements.
While most connection attempts concern modern protocols, operators should also urgently secure their retrofitted protocol deployments as they are still reachable and insecure.

\section{Conclusion}
\label{sec:conclusion}
The convergence of industrial appliances with the Internet requires rolling out end-to-end secure communication and access control to formerly isolated networks~\cite{sadeghi-secprivchallengesiiot-2015}.
To assess whether secure protocols are indeed adequately used and securely configured in the (I)IoT, we perform active Internet measurements to find deployments relying on ten (I)IoT protocols specified for usage via TLS.

Our results show that TLS adoption in the (I)IoT is low: Only~\SI{\Remark{hostspcttls}}{\percent} of \num{\Remark{hostscnt}}~deployments protect their communication.
Hence, a large share of deployments still communicates insecurely, enabling attackers to eavesdrop on confidential information or alter potentially safety-critical communication.
Furthermore, we find that TLS adoption tends to focus on hosts using modern (I)IoT protocols designed with security in mind (\SI{\Remark{hostspcttlssecurebydesignprotocol}}{\percent}) rather than deployments relying on secure variants of traditional industrial protocols (\SI{\Remark{hostspcttlsretrofitprotocol}}{\percent}).
Overall, the evolution towards secure industrial protocols is only barely noticeable, likely due to the long lifetimes of industrial devices.

To support operators in upgrading their systems for secure communication, we assess the security configuration of existing TLS-enabled systems.
We reveal that \SI{\Remark{hostspcttlsweak}}{\percent} of (seemingly secure) TLS-enabled deployments (\num{\Remark{hostscnttlsweak}} appliances) are configured insecurely, resulting from outdated protocol versions~(\SI{\Remark{hostspcttlsweakduetoversion}}{\percent}), ciphers~(\SI{\Remark{hostspcttlsweakduetocipher}}{\percent}), certificates relying on deprecated primitives~(\SI{\Remark{hostspcttlsweakduetodeprecatedcert}}{\percent}), reuse of compromised secrets~(\SI{\Remark{hostspcttlsweakduetoreuse}}{\percent}), or disabled access control~(\SI{\Remark{hostspcttlsweakduetoauth}}{\percent}).

Moreover, we indicate that configuration templates can ease the secure configuration of industrial appliances, e.g., by predefining cryptographic parameters when generating certificates.
While already in use and partly responsible for outdated configurations today, e.g., recently issued certificates relying on MD5, we see huge potential in regularly updated templates to assist operators in securely configuring industrial software and hardware.

To conclude, our work shows that the evolution of industrial protocols towards secure end-to-end communication is not widely reflected in real-world deployments.
Even worse, when used at all, these protocols are often configured insecurely.
To remedy this situation, operators need support in securely configuring (I)IoT deployments, thus requiring approaches to ease the secure configuration of industrial appliances for operators and practitioners.

\begin{acks}
Funded by the Deutsche Forschungsgemeinschaft (DFG, German Research Foundation) under Germany's Excellence Strategy -- EXC-2023 Internet of Production -- 390621612.
\end{acks}

\bibliographystyle{ACM-Reference-Format-limit}
\bibliography{paper}

\appendix

\section{Protocol Selection}
\label{sec:protocolselection}

\begin{table*}[!ht]
\footnotesize
\centering

\begin{tabular}{cccccccc}
  \hline
  \multirow{2}{*}{Protocol} & \multirow{2}{*}{Related Work}                                                           & \multirow{2}{*}{Standard Port} & \multicolumn{2}{c}{IANA}                & \multicolumn{2}{c}{(D)TLS mentioned}                                            & \multirow{2}{*}{Secure Port} \\
                            &                                                                                         &                                & Assigned               & Secure Variant & \multicolumn{1}{c}{Specification}    & \multicolumn{1}{c}{Technical Guidelines} &                              \\ \hline
  BACnet                    & \cite{mirian-icsmes-2016,feng-characterizing-2016,xu2018increase,barbieri2021assessing} & 47808                          & bacnet                 & ---            & \cite{bacnet-spec}                   & ---                                      & ---                          \\
  \rowcolor[HTML]{C0C0C0}
  DNP3                      & \cite{mirian-icsmes-2016,feng-characterizing-2016,xu2018increase,barbieri2021assessing} & 20000                          & dnp                    & dnp-sec        & \cite{ieee1815-2010-spec,iec62351-3} & ---                                      & 19999                        \\
  \rowcolor[HTML]{C0C0C0}
  Modbus                    & \cite{mirian-icsmes-2016,feng-characterizing-2016,xu2018increase,barbieri2021assessing} & 502                            & mbap                   & mbap-s         & \cite{MBTCPSec70:online}             & ---                                      & 802                          \\
  \rowcolor[HTML]{C0C0C0}
  Siemens S7                & \cite{mirian-icsmes-2016,feng-characterizing-2016,xu2018increase,barbieri2021assessing} & 102                            & iso-tsap               & iso-tp0s       & ---                                  & \cite{siemens-2019}                      & 3782                         \\
  \rowcolor[HTML]{C0C0C0}
  TridiumFox                & \cite{mirian-icsmes-2016,feng-characterizing-2016,xu2018increase,barbieri2021assessing} & 1911                           & mtp*                   & ---            & ---                                  & \cite{niagara-2019,niagaraax-2012}       & 4911                         \\
  \rowcolor[HTML]{C0C0C0}
  Ethernet/IP               & \cite{mirian-icsmes-2016,feng-characterizing-2016,barbieri2021assessing}                & 44818                          & EtherNet-IP-2          & ethernet-ip-s  & \cite{cipsecurity-spec}              & ---                                      & 2221                         \\
  HART-IP                   & \cite{mirian-icsmes-2016,feng-characterizing-2016,barbieri2021assessing}                & 5094                           & hart-ip                & ---            & ---                                  & ---                                      & ---                          \\
  \rowcolor[HTML]{C0C0C0}
  OPC~UA                    & \cite{2020-dahlmanns-imc-opcua,barbieri2021assessing}                                   & 4840                           & opcua-tcp              & opcua-tls      & \cite{opcua-profiles-2017}           & ---                                      & 4843                         \\
  Automatic Tank Gauge      & \cite{feng-characterizing-2016,barbieri2021assessing}                                   & 10001                          & scp-config*            & ---            & ---                                  & ---                                      & ---                          \\
  CodeSys                   & \cite{feng-characterizing-2016,barbieri2021assessing}                                   & 2455                           & wago-io-system         & ---            & ---                                  & ---                                      & ---                          \\
  General Electric SRTP     & \cite{feng-characterizing-2016,barbieri2021assessing}                                   & 18245--18246                   & ---                    & ---            & ---                                  & ---                                      & ---                          \\
  \rowcolor[HTML]{C0C0C0}
  IEC~104                   & \cite{feng-characterizing-2016,barbieri2021assessing}                                   & 2404                           & iec-104                & iec-104-sec    & \cite{iec62351-3}                    & ---                                      & 19998                        \\
  Melsec-Q                  & \cite{feng-characterizing-2016,barbieri2021assessing}                                   & 5006--5007                     & wsm-server*            & ---            & ---                                  & ---                                      & ---                          \\
  OMRON FINS                & \cite{feng-characterizing-2016,barbieri2021assessing}                                   & 9600                           & micromuse-ncpw*        & ---            & ---                                  & ---                                      & ---                          \\
  PC~Worx                   & \cite{feng-characterizing-2016,barbieri2021assessing}                                   & 1962                           & biap-mp*               & ---            & ---                                  & ---                                      & ---                          \\
  ProConOS                  & \cite{feng-characterizing-2016,barbieri2021assessing}                                   & 20547                          & ---                    & ---            & ---                                  & ---                                      & ---                          \\
  Red Lion Crimson V3       & \cite{feng-characterizing-2016,barbieri2021assessing}                                   & 789                            & ---                    & ---            & ---                                  & ---                                      & ---                          \\
  ANSI C12.22               & \cite{mirian-icsmes-2016,barbieri2021assessing}                                         & 1153                           & c1222-acse             & ---            & ---                                  & ---                                      & ---                          \\
  ICCP                      & \cite{mirian-icsmes-2016,barbieri2021assessing}                                         & 102                            & iso-tsap*              & ---            & ---                                  & ---                                      & ---                          \\
  IEC~61850                 & \cite{mirian-icsmes-2016,barbieri2021assessing}                                         & 102                            & iso-tsap*              & ---            & ---                                  & ---                                      & ---                          \\
  \rowcolor[HTML]{C0C0C0}
  AMQP                      & \cite{barbieri2021assessing}                                                            & 5672                           & amqp                   & amqps          & \cite{amqp-spec}                     & ---                                      & 5671                         \\
  ATG                       & \cite{barbieri2021assessing}                                                            & 10001                          & scp-config*            & ---            & ---                                  & ---                                      & ---                          \\
  \rowcolor[HTML]{C0C0C0}
  CoAP                      & \cite{barbieri2021assessing}                                                            & 5683                           & coap                   & coaps          & \cite{RFC7252}                       & ---                                      & 5684                         \\
  EtherCAT                  & \cite{barbieri2021assessing}                                                            & 34980                          & ethercat               & ---            & ---                                  & ---                                      & ---                          \\
  FF HSE                    & \cite{barbieri2021assessing}                                                            & 1089--1091                     & ff-\{annunc,fms,sm\}   & ---            & ---                                  & ---                                      & ---                          \\
  FL-net                    & \cite{barbieri2021assessing}                                                            & 55000--55003                   & ---                    & ---            & ---                                  & ---                                      & ---                          \\
  \rowcolor[HTML]{C0C0C0}
  MQTT                      & \cite{barbieri2021assessing}                                                            & 1883                           & mqtt                   & secure-mqtt    & \cite{mqtt-spec}                     & ---                                      & 8883                         \\
  PROFINET                  & \cite{barbieri2021assessing}                                                            & 34962--34964                   & profinet-\{rt,rtm,cm\} & ---            & ---                                  & ---                                      & ---                          \\
  Zigbee IP                 & \cite{barbieri2021assessing}                                                            & 11754--11755                   & zep, zigbee-ip         & zigbee-ips     & ---                                  & ---                                      & 11756                        \\
  CSPV4                     & \cite{feng-characterizing-2016}                                                         & 2222                           & EtherNet-IP-1          & ethernet-ip-s  & ---                                  & ---                                      & 2221                         \\ \hline
  \end{tabular}

\qquad
\scriptsize
\(^*\) IANA registration unrelated to (I)IoT protcol. \,
Protocols marked in grey are subject to our analysis.
\caption{(I)IoT protocols subject in related work undergoing our selection process.}
\vspace{-2.5em}
\label{tab:protocol-selection}
\end{table*}

Without any prior knowledge on the deployment of (D)TLS-based (I)IoT protocols, we need to systematically summarize (I)IoT protocols subject to our study.
To select TLS-enabled (I)IoT protocols for which we assess deployments during our analysis, we follow a three-step process guided through in Table~\ref{tab:protocol-selection}.
First, we compiled a list of (I)IoT protocols~(without TLS-support) subject in related work focusing on Internet measurements with an industrial background~\cite{mirian-icsmes-2016,feng-characterizing-2016,barbieri2021assessing,xu2018increase,2020-dahlmanns-imc-opcua}.
Here, we found 30~protocols subject to active and passive Internet-wide measurements indicating numerous deployments for many of the protocols.

Second, we inspected IANA port registrations for listed protocols and searched for counterparts indicating security within their name.
To this end, we performed an automated full-text search on protocol names registered for standard ports of identified protocols and manually checked for results indicating a secure variant in their name, e.g., \texttt{TLS}, \texttt{-sec}, or \texttt{secure}.
From 30~protocols, we found 18 with an entry in the IANA registrations associated with the standard (insecure) port and the industrial protocol.
Out of these 18 entries, we identified 11 registrations with an associated secure protocol variant.
These registrations also state a secure port to scan during our Internet-wide measurements to find and assess the security of deployments implementing these protocols.

Third, we investigate whether protocol specifications or technical guidelines and manuals of devices implementing these protocols indicate TLS support.
While we find eight protocols specified via TLS~\cite{ieee1815-2010-spec,iec62351-3,MBTCPSec70:online,cipsecurity-spec,opcua-profiles-2017,iec62351-3,amqp-spec,RFC7252,mqtt-spec}, we also discover indications for TLS support in guidelines for two other protocols, i.e., Tridium Fox and Fox Platform---which is always offered in combination with Tridium Fox~\cite{niagara-2019,niagaraax-2012}.

A fundamental foundation for feasible active Internet measurements is a single port per protocol to perform scans on.
Hence, we filter out protocols where we are not aware of any (single) standard port which could be subject to our measurements.
As we are not able to find evidence for any standard port used by retrofitted BACnet protocol implementations, we refrain from analyzing this retrofitted protocol.
Furthermore, we focus our analysis in this paper on (D)TLS-based protocols.
Since we find no evidence for Zigbee IP secure relying on (D)TLS, we exclude it from our analysis.

As a result, we obtain a curated list of ten (D)TLS-based (I)IoT protocols subject to our Internet-wide analysis, i.e., we show for these protocols how large the TLS-adoption is and whether the deployments are configured securely.

\section{Ethical Considerations}
\label{sec:ethics}
As measurements of (I)IoT systems could have unintended implications, e.g., concerning information security, privacy, or safety, we take several ethical considerations as the basis for our research.

We follow widely recognized ethical research guidelines~\cite{dittrich_menlo-report_2012} as well as practices and procedures imposed by our institutions during the design, execution, and analysis of our research.
Therefore, we handle all collected data with care as well as responsibly contact operators of systems not implementing access control, whenever contact information is available~(cf.\ Appendix~\ref{subsec:ethics:handling}).
Furthermore, we adhere to accepted measurement guidelines~\cite{durumeric-zmap-2013} to reduce the impact of our measurements (cf.\ Appendix~\ref{subsec:ethics:impact}).

\subsection{Handling of Data \& Responsibilities}
\label{subsec:ethics:handling}

During our measurements, in agreement with our institution's data protection officer, we only request publicly available data.
Specifically, we never request any payload data when protected by access control mechanisms.
Besides for AMQP, where we check whether deployments accept a pair of default credentials, we never bypass any security mechanisms.
Most importantly, we immediately close connections to AMQP servers accepting the default credentials.
Furthermore, we never alter the server's state, i.e., we never send any write or function execution requests to the servers.
Still, our dataset might contain sensitive data of servers that do not implement access control~(cf.\ Section~\ref{subsec:applicationsecurity}).
We store all data on secured systems to keep potentially included sensitive data private and prevent attackers from finding open or insecure systems by using our dataset.
For the public release of our dataset, we removed retrieved payload data and replaced all identifiers, e.g., IP addresses and subject names in certificates, by consecutive numbers.
While this restriction prevents others from independently reproducing our certificate templating results, we consider this decision to constitute a reasonable trade-off to protect affected users.

\textbf{Responsible Disclosure:}
We analyzed the received data to identify server operators to inform them about their accessible systems whenever possible, i.e., we automatically searched the received payload data of MQTT brokers for email addresses.
We found \SI{\Remark{emailcntmqttpayload}}{}~email addresses on~\SI{\Remark{hostscntemailpayload}}{}~MQTT brokers as contact information.
To reduce the overhead of our information campaign, we filtered out email addresses using a domain without any DNS \texttt{MX} record behind.
Finally, we reached out to \SI{2985}{}~email addresses pointing out potential privacy, security, and safety issues.
Although we received \SI{877}{} automatically generated responses indicating that these collected email addresses did not exist, we received nine responses within one day.
While two operators indicate that affected systems are test systems, the seven others report that they secured their systems after our message or intend to secure them soon.
All in all, we received very positive feedback on our information campaign.

Unfortunately, other protocols subject to our study do not allow performing similar information campaigns.
Other than MQTT, AMQP does not allow duplicating data on request, i.e., data would be lost when our scanner requests it, and none of the legitimate clients would receive the data.
We refrain from requesting payload data because it is against our ethical considerations to change a system's state and alter ongoing communication.
Furthermore, traditional (I)IoT protocols with and w/o TLS only allow to request single production values but do not provide any function to request contact information.

\subsection{Reducing Impact of Measurements}
\label{subsec:ethics:impact}

To minimize the implication of our active Internet measurements, we follow well-established Internet measurement guidelines~\cite{durumeric-zmap-2013}.

\textbf{Measurement Coordination:}
We coordinate all of our measurements with our institutional Network Operation Center to reduce the impact on the Internet.
Specifically, we ensure to answer and handle inquiries or abuse requests as fast as possible.

\textbf{External Perception:}
We display the research intent of our scans to external operators by providing rDNS records for our scanning IP address and transmitting contact information in our client certificate as well as protocol messages where possible.
Additionally, we provide a website behind our scanning IP address with detailed information on the scope and purpose of our research.
Furthermore, we list opt-out instructions on the website to request exclusion of our scans.
Based on such requests~(also regarding other scanning projects of our institute), we exclude 5.8\,M IP addresses (0.14\% of the IPv4 address space).

\textbf{Limiting Load:}
To not overload any autonomous system, we spread our scans over a timeframe of approximately 24~hours per protocol and rely on \texttt{zmap}'s address randomization.
More importantly, to not overload potentially (I)IoT devices during our subsequent TLS handshakes, we instruct our scanner module to wait \SI{15}{\minute} between subsequent handshakes to one server.
For MQTT, we further set a scanning time (\SI{30}{\minute}) and outgoing traffic (\SI{10}{\mega\byte}) limit per host, i.e., our scanner disconnects whenever the limit exceeds.

While (I)IoT protocols do not realize security by default, we consider it essential to know whether today's deployments benefit from recently introduced security features.
To answer this question, we have taken sensible measures to reduce the risks introduced by active Internet measurements of industrial appliances, aiming to positively influence the security of (I)IoT deployments. %

\section{(D)TLS Cipher Suite Sets}
\label{sec:cipherclasses}

To reduce the load on deployments subject to our measurements, we refrained from performing a single (D)TLS handshake per possible cipher suite to analyze server support.
Instead, we curate four sets of (D)TLS cipher suites and analyze which suite servers choose from each set.

\begin{table*}[!t]
\scriptsize
\input{tables/tls-ciphers.tab}
\vspace{-0.5em}
\caption{%
Cipher suite sets we use to analyze whether servers support cipher suites implementing different levels of security.
}%
\vspace{-2.5em}
\label{tab:tls-cipher}
\end{table*}

Table~\ref{tab:tls-cipher} lists our four sets, i.e., \textit{recommended}, \textit{recommended without perfect forward secrecy}, \textit{compatibility}, and \textit{insecure}, as well as reports on the assigned (D)TLS ciphers.

\textbf{Recommended}:
To check whether servers implement (at least one) secure and recommended cipher suite, we collect suites following basic recommendations and guidelines~\cite{nistsp800-tls,rfc7525,tr2102-2}.
Servers denying a connection when our scanner presents this cipher suite set most likely do not implement any recommended suite indicating insecure or outdated configuration.
While fundamental for secure communication, all the ciphers included in this set implement perfect forward secrecy.

\textbf{Recommended w/o perfect forward secrecy}:
However, especially constrained (I)IoT devices might not be able to implement perfect forward secrecy.
Hence, we also check whether deployments support less resource-intensive, still secure, but non-perfect forward secret cipher suites.
We base the choice of ciphers in this set also on basic guidelines~\cite{nistsp800-tls,rfc7525,tr2102-2}.
While still secure, disclosure of cryptographic key material used for communication allows attackers to decrypt all the communicated data.

\textbf{Compatibility}:
Implementing a large variety of different combinations of ciphers and MAC algorithms, the standard ZGrab2 set has very high compatibility.
The cipher suites include RSA and ECDSA key exchange mechanisms in combination with AES and RC4 ciphers in GCM or CBC mode, as well as MAC algorithms relying on SHA1 or SHA256.
Thus, this set encompasses secure and weak cipher suites, depending on the included mechanism.
Hence, we use this set to evaluate the security policy of deployments, i.e., assess whether deployments choose one of the secure or one of the insecure cipher suites.

\textbf{Insecure}:
Finally, we curate a list of cipher suites, including insecure cryptographic primitives, e.g., DES, MD5, and NULL encryption, to check whether servers allow clients to connect without a proper level of security.
To this end, we manually selected cipher suites including unrecommended primitives~(according to~\cite{tr2102-1}) from the IANA registration list, which are officially standardized.
Whenever servers select a cipher out of this set, it enables clients to communicate insecurely.

\balance

\section{Certificate Clustering}
\label{sec:certclustering}
To research whether operators use configuration templates predefining preferences when generating certificates, we cluster certificates and analyze whether the certificates in each cluster were likely generated using the same template.

We chose the certificate's subject name to find coherent certificates after a manual revision of our dataset showed many certificates distributed over different hosts all over the Internet using similar naming schemes.
However, the name typically is a free text field with only a few conventions.
To find related subject names, we calculate a similarity matrix based on methods from text mining and subsequently create clusters using DBSCAN.
Since the subject name includes insignificant keys in front of every value and is comprised in every certificate, e.g., \texttt{CN=}, we first vectorize the subject name using single character (1,3)-grams and apply term frequency-inverse document frequency~(TF-IDF) building one vector per term.
This vector indicates the term's distribution in the subject name of all certificates~(the more subject names include a term, the smaller its TF-IDF).
Subsequently, using a sparse cosine similarity metric, we calculate a similarity matrix including the 500 most similar TF-IDF vectors~(with a threshold of \num{0.5} to make the calculation feasible, directly performing DBSCAN over the vectors of all terms would be infeasible).

To build clusters of certificates most likely generated with the same template, we perform DBSCAN on the similarity matrix~(with \(\epsilon=0.8\), cosine similarity as a metric, and a minimum cluster size of three).
Here, we manually revisited the results to ensure that no obvious false positives were included.

\end{document}